\documentclass[a4paper]{article} 

\author{Jacques Smulevici \footnote{University of Cambridge,
Department of Applied Mathematics
and Theoretical Physics, Wilberforce Road, Cambridge CB3 0WA United Kingdom}}
\title{On the area of the symmetry orbits of cosmological spacetimes with toroidal or hyperbolic symmetry}

\usepackage{amsmath,amsthm}
\usepackage{amssymb}
\usepackage{graphicx,color}
\usepackage{hyperref}
\newtheorem{definition}{Definition}
\newtheorem{conjecture}{Conjecture}
\newtheorem*{theoremaux}{Theorem}
\newtheorem{proposition}{Proposition}
\newtheorem{theorem}{Theorem}
\newtheorem{lemma}{Lemma}[section]
\newtheorem{corollary}{Corollary}

\newtheorem{remark}{Remark}[subsection]

\newcommand{\intr}[1]{\int_{\mathbb{R}^{#1}}}

\newcommand{\demi}{\frac{1}{2}}

\newcommand{\drond}[2]{\frac{\partial #1}{\partial #2}}

\newcommand{\comment}[1]{}

\begin{document}
\maketitle

\begin{abstract}
We prove several global existence theorems for spacetimes with toro\-idal or hyperbolic symmetry with respect to a geometrically defined time. More specifically, we prove that generically, the maximal Cauchy development of $T^2$-symmetric initial data with positive cosmological constant $\Lambda >0$, in the vacuum or with Vlasov matter, may be covered by a global areal foliation with the area of the symmetry orbits tending to zero in the contracting direction. We then prove the same result for surface symmetric spacetimes in the hyperbolic case with Vlasov matter and $\Lambda \ge 0$. In all cases, there is no restriction on the size of initial data.
\end{abstract}

\tableofcontents

\section{Introduction} The study of the global Cauchy problem constitutes one of the main areas of research in mathematical relativity and is one of the most natural problems to investigate in view of the hyperbolicity of the Einstein equations and of the theorems concerning the local Cauchy problem \cite{ycb:teedpn, bg:gacgr}. These theorems assert that given an appropriate initial data set, there exists a maximal solution of the Einstein equations
\begin{equation} \label{ee}
R_{\mu \nu}-\demi g_{\mu \nu}R =8 \pi T_{\mu \nu}-\Lambda g_{\mu \nu},
\end{equation}
coupled if necessary to appropriate matter equations\footnote{See \cite{ycb:usel} and \cite{ycb:pcel} for the case of Vlasov matter.}, which is unique up to diffeomorphism in the class of globally hyperbolic spacetimes. We call this solution the \emph{maximal Cauchy development} of the initial data. The global hyperbolicity assumption guarantees the domain of dependence property and is essential to the uniqueness statement.

The global Cauchy problem consists in understanding the global geometry of the maximal Cauchy development. A fundamental conjecture, known as \emph{strong cosmic censorship}\footnote{The conjecture was originally developed by R. Penrose \cite{rp:sta} and first formulated as a statement about the global geometry of the maximal Cauchy development by V. Moncrief and D. Eardley in \cite{me:gepcc}. See also the presentation of D. Christodoulou in \cite{ch:givp}.}, states that \emph{the maximal Cauchy development of generic compact or asymptotically flat initial data is inextendible as a regular\footnote{The regularity concerns here the degree of differentiability of the possible extensions and gives rise to different versions of the conjecture. For instance, the $C^2$ formulation of the conjecture is obtained by replacing ``regular'' with $C^2$ in the above statement of the conjecture.} Lorentzian manifold}. This can be thought of as a statement of uniqueness in a class of spacetimes not assumed a priori to be globally hyperbolic.

The expression ``generic initial data'' in the statement of the conjecture reflects the fact that there exist particular initial data for which the maximal Cauchy development fails to be inextendible. However, the extendibility property of the maximal Cauchy development for these particular initial data is expected to be dynamically unstable and, as we shall see below, this expectation has been verified in several cases. From the point of view of physics, uniqueness means predictability and thus, strong cosmic censorship asserts that, generically, general relativity is a deterministic theory in the same sense that Newtonian mechanics is deterministic.

\subsection{Areal foliations of $T^2$-symmetric and $k\le 0$ surface-sym\-metric spacetimes}
In full generality, the questions tied to the global Cauchy problem are not accessible with the current set of mathematical techniques. In order to make progress, one may try to look at simpler but connected problems, such as the study of the global Cauchy problem within certain classes of symmetries. 

Following this approach, two classes of solutions arising from compact initial data with symmetry have been given much attention recently, the so-called \emph{$T^2$-symmetric} and \emph{surface-symmetric} spacetimes. The $T^2$-symmetric spacetimes constitute a class of solutions arising from initial data with spatial topology $T^3$ and admitting a torus action. They contain as special subcases the $T^3$-Gowdy spacetimes and the polarized $T^2$-symmetric spacetimes\footnote{They also contain the even more special case of polarized $T^3$-Gowdy spacetimes.}. The surface-symmetric spacetimes constitute a class of solutions arising from initial data where the initial Riemannian $3$-manifold is given by a doubly warped product $S^1 \times \mathcal{S}$, where $\mathcal{S}$ is a compact $2$-surface of constant curvature $k$ and such that the rest of the initial data is invariant under the local isometries of $\mathcal{S}$. By rescaling, $k$ may be taken as being $-1,0$ or $+1$ and the different cases are known as hyperbolic, plane\footnote{Note that the plane symmetric case is a special case of $T^3$-Gowdy polarized solutions.} or spherical symmetry. 

In the case of $T^2$-symmetric or $k \le 0$ surface-symmetric spacetimes, the local geometry of the solution possesses the particular property that, unless the spacetime is flat, the symmetry orbits are either trapped or antitrapped, a feature which is shared by the spheres of symmetry in the black or white hole regions of a Schwarzschild solution with $m>0$. If we denote by $t$ the area of the symmetry orbits, this means that the gradient of $t$ is everywhere timelike and that $t$ may be used as a time coordinate\footnote{In particular, any non-flat $T^2$-symmetric or $k \le 0$ surface-symmetric spacetime can be oriented by $\nabla t$. With this choice of orientation, the future corresponds to the direction  where $t$ increases (expanding direction) and the past to the direction where $t$ decreases (contracting direction).}. For the vacuum $T^2$-symmetric case with zero cosmological constant ($\Lambda=0$), the existence of a global areal foliation where $t$ takes value in $(t_0,\infty)$ with $t_0 \ge 0$ was proven in \cite{bicm:gft2}. The proof was then extended to the Vlasov case \cite{ha:gfmsg, arw:caft2v} and to the case with $\Lambda >0$ \cite{ci:t2pcc}. Similarly, the existence of a global areal foliation for the surface-symmetric case with $k=-1$, $\Lambda=0$ and Vlasov matter\footnote{Note that, in the surface-symmetric case, a result analogous to Birkhoff's theorem applies, by which we mean that these spacetimes have no dynamical degree of freedom in the vacuum.} was proven in \cite{arr:evhs} and extended to the case with $\Lambda >0$ in \cite{tr:geabevpcc, tn:ssevcc}.

 It was soon realized that in the vacuum $T^3$-Gowdy case with $\Lambda=0$, one has $t_0=0$ unless the spacetime is flat \cite{vm:gpgs,pc:u1u1}. The natural question arose: \emph{Is $t_0=0$ generically for all the possible cases?} The proofs that $t_0=0$ generically for $T^2$-symmetric spacetimes with $\Lambda=0$, in the vacuum or with Vlasov matter, were given in \cite{iw:ast2} and \cite{mw:ast2v}. It has also been proven that $t_0=0$ in the case of plane symmetric initial data with $\Lambda=0$ and Vlasov matter as well as in the case of plane or hyperbolic symmetric initial data with $\Lambda \ge 0$ and Vlasov matter under an extra small data assumption \cite{gr:csves, sbt:isss}. Moreover, the results for $k \le 0$ surface-symmetric initial data have been extended to the Einstein-Vlasov-scalar field system \cite{tr:isevs}.

\subsection{Strong cosmic censorship for $T^2$-symmetric or surface-symmetric spacetimes}
One motivation for the study of the value of $t_0$ was the expectation that, in the cases were $t_0=0$, the curvature should in general blow up as $t$ goes to 0, thus providing a proof of inextendibility (and thus of strong cosmic censorship) for these cases. Indeed, for vacuum $T^3$-Gowdy spacetimes with $\Lambda=0$, first in the polarized case\footnote{Note that, in \cite{cim:sccp}, strong cosmic censorship was also proved for polarized Gowdy spacetimes arising from initial data given on $S^2 \times S^1$, $S^3$ and $L(p,q)$.} and then for the full class, detailed asymptotic expansions were obtained and used in this sense to establish a proof of the $C^2$ formulation of the strong cosmic censorship conjecture \cite{cim:sccp, hr:gs, hr:scct3}. 

While it seemed difficult to extend the analysis of the vacuum $T^3$-Gowdy spacetimes to the more general case of $T^2$-symmetric spacetimes, strong cosmic censorship was nonetheless proven for $T^2$-symmetric spacetimes with $\Lambda=0$ in the presence of Vlasov matter \cite{dr:t2}. The analysis starts with the remark that for $T^2$-symmetric or $k \le 0$ surface-symmetric spacetimes, with or without Vlasov matter and with $\Lambda \ge 0$, the fact that $t$ is unbounded implies inextendibility in the expanding direction because of the continuous extension of the Killing fields to possible Cauchy horizons \cite{dr:iecs}. Thus it is sufficient to study the contracting direction in order to complete the proof of strong cosmic censorship for these classes of spacetimes. The proof given in \cite{dr:t2} relied on a rigidity of the possible Cauchy horizon, linked with the fact that $t_0=0$, and on the particular properties of the Vlasov equation. The assumption that $\Lambda=0$ was necessary only as to ensure that $t_0=0$. Therefore the proof remained valid in the case where $\Lambda > 0$, if one added the extra assumption that $t_0=0$. In \cite{js:scct2pccm}, we studied the remaining cases, namely the $T^2$-symmetric spacetimes with $\Lambda > 0$ and Vlasov matter for which $t_0 > 0$, and proved their inextendibility, thus completing a proof of strong cosmic censorship for $T^2$-symmetric spacetimes with $\Lambda \ge 0$ and Vlasov matter. In the same article, we proved that vacuum $T^2$-symmetric spacetimes with $\Lambda > 0$ and $t_0 > 0$ were also generically inextendible. Finally, in the surface-symmetric case with Vlasov matter, strong cosmic censorship was resolved in the affimative for $k \le 0$ and $\Lambda \ge 0$ and for $k=1$ and $\Lambda=0$, some obstructions remaining in the spherical case with $\Lambda > 0$, in particular the possible formation of Schwarzschild-de-Sitter or, even worse, extremal Schwarzschild-de-Sitter black holes \cite{dr:ss}.

\subsection{The past asymptotic value of $t$ and the main theorems}
The results of \cite{js:scct2pccm}, as well as the proof of inextendibility for the $k \le 0$ surface-symmetric cases where $t_0 > 0$ contained in \cite{dr:ss}, gave satisfactory answers to the strong cosmic censorship conjecture.  However, they did not address the question of the value of $t_0$. It is the subject of this article to resolve this question. 

First, in Theorem \ref{th:t2vla} (see section \ref{se:tth}), we will extend the work of M. Weaver \cite{mw:ast2v} proving that \emph{the maximal Cauchy development of $T^2$-symmetric initial data with $\Lambda \ge 0$ and non-vanishing Vlasov matter can be covered by global areal foliations with $t$ going to zero in the contracting direction.} Thus $t_0=0$ for these spacetimes. 

As it often happens in these types of problems, the vacuum case is more difficult than the Vlasov case. This is already reflected in the fact that for vacuum $T^2$-symmetric spacetimes with $\Lambda > 0$, one can find special families of (non-flat) solutions for which $t_0 > 0$ . That these solutions are indeed special is the content of Theorem \ref{th:t2vac} which states that \emph{vacuum $T^2$-symmetric spacetimes with $\Lambda \ge 0$ for which $t_0 > 0$ are necessarily polarized}. Thus, generically, $t_0=0$ for vacuum $T^2$-symmetric spacetimes with $\Lambda \ge 0$. 

Finally, we will show that the proof given for the $T^2$ case with Vlasov matter may be adapted to the hyperbolic case. We will obtain Theorem \ref{th:hyvla} which asserts that \emph{$t_0=0$ for $k \le 0$ surface-symmetric spacetimes with $\Lambda \ge 0$ and non-vanishing Vlasov matter}. Thus Theorem \ref{th:hyvla} asserts that the results of \cite{gr:csves, sbt:isss} are true in general and do not require any smallness assumption. To summarize, we provide in Tables 1 and 2 a picture of the current status of the analysis of singularities for the $T^2$-symmetric and surface-symmetric spacetimes in the vacuum or with Vlasov matter. 

\begin{table} \label{ta:t0}
\[
\input{resultr0.pstex_t}
\]
\caption{Value of $t_0$ for generic $T^2$-symmetric and $k \le 0$ surface-symmetric spacetimes.}
\end{table}

\begin{table}  \label{ta:t1}
\[
\input{resultscc.pstex_t}
\]
\caption{Status of strong cosmic censorship for $T^2$-symmetric and surface-symmetric spacetimes.}\label{ta:scc}
\end{table}

\subsection{Outline}
The outline of this article is as follows. We start in section \ref{se:prel} with an introduction to the different classes of symmetry and present the classes of initial data that we will consider in the rest of the paper.
 In section \ref{se:mcd}, we recall the existence and uniqueness of the maximal Cauchy development and in section \ref{se:gaf}, we present the previous results concerning the global foliations of $T^2$-symmetric and $k \le 0$ surface-symmetric spacetimes that we shall use as a starting point for our analysis. The statements of the theorems proved in this article then follow in section \ref{se:tth}. Before giving the proofs of the three theorems in sections \ref{se:t2vla}, \ref{se:t2vac} and \ref{se:hypvla}, it will be useful to describe the approach that we will take, especially for the proof of Theorem \ref{th:t2vac}, and this is done in section \ref{se:rsp}. We end this paper by some comments and open questions in section \ref{se:coq}. In appendix \ref{ap:idev}, we provide some information on the initial data sets of the Einstein and Einstein-Vlasov systems for the reader not familiar with this. In appendix \ref{ap:ssac}, we very briefly describe a coordinate transformation for $k=-1$ surface symmetric spacetimes and finally in appendix \ref{ap:sid}, we recall the classical results that symmetric initial data lead to symmetric spacetimes.

\section{Preliminaries} \label{se:prel}

\subsection{$T^2$-symmetric spacetimes with spatial topology $T^3$} \label{se:t2spt2}
A spacetime $(\mathcal{M},g)$ is said to be $T^2$-symmetric if the metric is invariant under an effective action of the Lie group $T^2$ and the group orbits are spatial.
The Lie algebra of $T^2$ is spanned by two commuting Killing fields $X$ and $Y$ everywhere non-vanishing and we may normalise them so that the area element $t$ of the group orbits is given by:
\begin{displaymath}
g(X,X)g(Y,Y)-g(X,Y)^2=t^2.
\end{displaymath}

In the previous analysis of these spacetimes \cite{pc:u1u1, bicm:gft2, arw:caft2v, ci:t2pcc}, it has been shown that any globally hyperbolic $T^2$-symmetric spacetime with spatial topology $T^3$ which satisfies the Einstein equations in the vacuum or with Vlasov matter and with $\Lambda \ge0$ admits a metric in areal coordinates of the form:

\begin{align} \label{t2:metric}
ds^2=&e^{2 (\nu -U) } \left( -\alpha dt^2+d\theta^2 \right)+e^{2U} \left[ dx+Ady+(G+AH)d\theta \right]^2\nonumber \\&+e^{-2U}t^2[dy+Hd\theta]^2,
\end{align}
where all functions depend only on $t$ and $\theta$ and are periodic in the latter. Note that the form (\ref{t2:metric}) of the metric is unchanged under an $SL(2,\mathbb{R})$ transformation of the Killing vectors $X=\drond{}{x}$ and $Y=\drond{}{y}$.

As $T^2$-symmetric spacetimes contain several dynamical degrees of freedom, certain special cases have been introduced. A first simplification appears in the case where the Killing fields $X$ and $Y$ may be chosen such that their inner product, and thus the function $A$, vanishes. Such cases are called \emph{polarized} $T^2$-symmetric spacetimes.

Associated with $T^2$-symmetric spacetimes, are certain quantities called the twist quantities which are defined by:

\begin{eqnarray}
J=\epsilon_{abcd}X^a Y^b \nabla^c X^d, \\
K=\epsilon_{abcd}X^a Y^b \nabla^c Y^d.
\end{eqnarray}
These are related to the metric functions by:

\begin{eqnarray}
J&=&-\frac{te^{-2\nu+4U}}{\sqrt{\alpha}}(G_t+AH_t), \\
K&=&AJ-\frac{t^3e^{-2\nu}}{\sqrt{\alpha}}H_t.
\end{eqnarray}

Note that for any pair of commuting Killing vectors on a spacetime satisfying the vacuum Einstein equations, the associated twists quantities are constant  \cite{rg:mgns}. Thus for vacuum $T^2$-symmetric spacetimes, by an $SL(2, \mathbb{R})$ transformation of the Killing fields $X$ and $Y$, we may ensure that the form of the metric (\ref{t2:metric}) is unchanged while one of twist quantities vanishes. Therefore, in the vacuum, we shall always assume that $J=0$.

The cases where both $J=0$ and $K=0$ are called \emph{$T^3$-Gowdy} spacetimes. Note that by Frobenius's theorem, the conditions $J=K=0$ are equivalent to the integrability of the planes orthogonal to $dx$, $dy$.

\subsection{Spacetimes with a hyperbolic surface of symmetry} \label{se:shss}

A spacetime $(\mathcal{M},g)$ is said to be $k=-1$ surface-symmetric if it can be foliated by spacelike surfaces $\Sigma_t$ such that for all $t$, $\Sigma_t$ is isometric to a doubly warped product $(S^1 \times \mathcal{S},h_t)$ where $S$ is a fixed compact surface of constant curvature $-1$.

It follows easily from the previous analysis on these spacetimes \cite{ar:cssph} that any $k=-1$ surface symmetric spacetime which is globally hyperbolic and satisfies the Einstein equations with $\Lambda \ge0$, in the vacuum or with Vlasov matter, admits a metric in areal coordinates of the form\footnote{Compared to the usual metric for these spacetimes, we use the square of the radius function $t=r^2$ as the time coordinate rather than the radius function $r$ itself. Moreover, we have introduced the functions $\alpha$ and $\nu$ by analogy with the $T^2$ case, so as to ease the application of the method of the $T^2$ case to this class of spacetimes. See Appendix \ref{ap:ssac} for a description of the change of coordinates from the usual parametrization to this one.}: 

\begin{equation} \label{hy:metric}
ds^2=-\frac{e^{2\nu}}{t}(\alpha dt^2-d\theta^2)+t \gamma_{ab}dx^a dx^b,
\end{equation}
where the functions $\nu$ and $\alpha$ depend only on $t$ and $\theta$ and are periodic in the latter with period $1$ and $\gamma$ induces a metric of constant curvature $-1$ on the orbits of symmetry.

\subsection{The Einstein-Vlasov system} \label{ap:ve}
Apart from in the vacuum case, where we will set $T_{\mu \nu}=0$ in (\ref{ee}), we will couple the Einstein equations to the Vlasov matter model which we present in this section.

Let $\mathcal{P} \subset \mathcal{TM}$ denote the set of all future directed timelike vectors of length $-1$. $\mathcal{P}$ is classically called the $\emph{mass shell}$. Let $f$ denote a nonnegative function on the mass shell. The Vlasov equation equation for $f$ is derived from the condition that $f$ be preserved along geodesics. In coordinates, we therefore have:

\begin{equation} \label{eq:vla}
p^\alpha \partial_{x^\alpha}f-\Gamma^{\alpha}_{\beta \gamma}p^{\beta}p^{\gamma}\partial_{p^\alpha}f=0,
\end{equation}
where $p^{\alpha}$ denotes the momentum coordinates on the tangent bundle conjugate to $x^\alpha$. 

The energy-momentum tensor is defined by:

\begin{equation} \label{def:emt}
T_{\alpha \beta}(x)=\int_{\pi^{-1}(x)}p_\alpha p_\beta f,
\end{equation}
where $\pi : \mathcal{P} \rightarrow \mathcal{M}$ is the natural projection from the mass shell to the spacetime and the integral is with respect to the natural volume form on $\pi^{-1}(x)$.

\subsection{The classes of initial data} \label{se:cid}
After this introduction to the symmetry classes and the matter fields, we are ready to present the initial data sets that will be studied in this article. For convenience, we will require that the initial data are smooth and, in the non-vacuum case, that the support of the Vlasov field is compact. These assumptions may clearly be relaxed if necessary\footnote{For instance, we could have chosen the initial data to be compatible with the statement of Theorem 4.1 of \cite{dr:t2}. However, we decided to give preference to clarity and will therefore stick with compact data for the Vlasov field.}.
\begin{definition}A vacuum $T^2$-symmetric initial data set is a triplet $(\Sigma,h,K)$ such that
\begin{enumerate}
\item{ $\Sigma$ is a smooth differential $3$-manifold with topology $T^3$ (in particular, $\Sigma$ admits an effective action of $T^2$),} \label{cd:si}
\item{ $h$ is a smooth Riemannian metric on $\Sigma$ which is invariant under an effective action of the Lie group $T^2$,} \label{cd:h}
\item{ $K$ is a smooth symmetric $2$-tensor also invariant under the same $T^2$ action,} \label{cd:k}
\item{ $(\Sigma,h,K_{ab})$ satisfies the vacuum constraint equations of general relativity.}
\end{enumerate}
\end{definition}

We describe in appendix \ref{ap:idev} the constraint equations in the vacuum or in the presence of Vlasov matter for the reader not familiar with them.

\begin{definition} A $T^2$-symmetric initial data set with Vlasov matter is a quadruplet $(\Sigma,h,K,\hat{f})$ such that
\begin{enumerate}
\item{conditions \ref{cd:si}., \ref{cd:h}. and  \ref{cd:k}. of the above definition hold,}
\item{ $\hat{f}$ is a smooth, non-negative function of compact support defined on $T\Sigma$ which is invariant under the natural lift to $T\Sigma$ of the $T^2$ action,}
\item{ $(\Sigma,h,K_{ab},\hat{f})$ satisfies the constraint equations of the Einstein-Vlasov system.}
\end{enumerate}
\end{definition}

Let us also define the notion of polarized $T^2$-symmetric initial data and of Gowdy initial data as follows:

\begin{definition}
A vacuum $T^2$-symmetric initial data set $(\Sigma,h,K)$ (respectively a $T^2$-symmetric initial data set with Vlasov matter $(\Sigma,h,K,\hat{f})$ ) is said to be polarized if there exist two Killing fields $(X,Y)$ which generate the $T^2$ action such that $h(X,Y)=0$ and $K(X,Y)=0$ on $\Sigma$.
\end{definition}

\begin{definition}
A vacuum $T^2$-symmetric initial data set $(\Sigma,h,K)$ (respectively a $T^2$-symmetric initial data set with Vlasov matter) is said to be a Gowdy initial data set if there exist linearly independent, commuting vector fields $Z,X,Y$ on $\Sigma$ such that $X,Y$ are Killing fields which generate the $T^2$ action and such that $h(Z,X)=h(Z,Y)=K(Z,Y)=K(Z,X)=0$.
\end{definition}

We define $k=-1$ surface-symmetric initial data with Vlasov matter as follows:
\begin{definition}A $k=-1$ surface-symmetric initial data set with Vlasov matter is a quadruplet $(\Sigma,h,K,\hat{f})$ such that
\begin{enumerate}
\item{ $\Sigma=S^1 \times \mathcal{S}$ where $\mathcal{S}$ is a smooth compact surface,}
\item{ $h$ is a smooth Riemannian doubly-warped product metric on $\Sigma$ of the form $a(\theta)d\theta^2+b(\theta)\gamma_{\mathcal{S}}$, where $\gamma_{\mathcal{S}}$ is a metric of constant curvature $-1$,}
\item{$\hat{f}$ is a smooth, non-negative function of compact support defined on $T\Sigma$ and invariant under the natural lift of the local isometries of $\mathcal{S}$ to $T\Sigma$.}
\item{ $(\Sigma,h,K_{ab},\hat{f})$ satisfies the constraint equations of the Einstein-Vlasov system.}
\end{enumerate} \label{def:hyid}
\end{definition}

\section{The maximal Cauchy development} \label{se:mcd}
We will recall in this section the classical results concerning the existence and uniqueness of the maximal Cauchy development to which we will refer often in the rest of this article. We will state the theorem in the case of Vlasov matter. For the vacuum case, it suffices to replace all matter terms by zero.

\begin{theoremaux} \label{th:mcd}
Let $(\Sigma,h,K,\hat{f})$ be an initial data set for the Einstein-Vlasov system. Then there exists a triplet  $(\mathcal{M},g,f)$, called the maximal Cauchy development of $(\Sigma,h,K,\hat{f})$, such that:

\begin{enumerate}
\item{$(\mathcal{M},g)$ is a smooth globally hyperbolic spacetime and $f$ is a smooth, non-negative function of compact support defined on the mass shell $\mathcal{P}$, \label{mc1}}
\item{$(\mathcal{M},g,f)$ satisfies the Einstein-Vlasov system (\ref{ee}), (\ref{def:emt}), (\ref{eq:vla}),\label{mc2}}
\item{there exists a smooth embedding $\phi:\Sigma \rightarrow \mathcal{M}$ such that $\phi(\Sigma)$ is a Cauchy surface for $\mathcal{M}$ and if $h'$, $K'$, $f'$ denotes respectively the first and second fundamental form of $\phi(\Sigma)$ and the restriction of $f$ to the tangent bundle of $\phi(\Sigma)$ then $\phi^*(h')=h$, $\phi^*(K')=K$, $\phi^*(f')=\hat{f}$,\label{mc3}}
\item{if $(\mathcal{\bar{M}},\bar{g},\bar{f})$ is another triplet satisfying \ref{mc1}, \ref{mc2} and \ref{mc3} and if $\bar{\phi}$ denotes the corresponding embedding of $\Sigma$ in $\mathcal{\bar{M}}$ then there exists an smooth isometry $\psi$ from $(\mathcal{\bar{M}},\bar{g})$ onto a subset of $(\mathcal{M},g)$ such that $\psi^*{\bar{f}}=f$ and $\psi(\bar{\phi}(\Sigma))=\phi(\Sigma)$.}
\end{enumerate}
\end{theoremaux}See \cite{ycb:teedpn, bg:gacgr,ycb:usel, ycb:pcel} for the original proofs of these theorems.

\section[Global areal foliations]{Global areal foliations of $T^2$-symmetric or $k=-1$ surface-symmetric spacetimes} \label{se:gaf}
We present in this section certain previous results concerning areal foliations of $T^2$-symmetric or $k=-1$ surface-symmetric spacetimes.
Let us first recall that symmetries of the initial data are transmitted to the maximal Cauchy development. For the reader not familiar with these results, they are presented in Appendix \ref{ap:sid}. Thus, $T^2$-symmetric (respectively surface-symmetric) initial data lead to $T^2$-symmetric (respectively surface-symmetric) spacetimes. We have moreover the following:

\begin{proposition} \label{pro:gaf}
Let $(\mathcal{M},g,f)$ be the maximal Cauchy development of $T^2$-sym\-metric initial data (respectively $k=-1$ surface-symmetric initial data) with $\Lambda \ge 0$, either in the vacuum or with Vlasov matter. Then

\begin{enumerate}
\item{ $(\mathcal{M},g)$ is $T^2$-symmetric (respectively $k=-1$ surface-symmetric) and $f$ is invariant under the natural lift to $T\mathcal{M}$ of the $T^2$ action (respectively under the natural lift to $T\mathcal{M}$ of the local isometries of $\mathcal{S}$, with $\mathcal{S}$ as in Definition \ref{def:hyid}),}
\item{$(\mathcal{M},g)$ can be covered by areal coordinates $(t,\theta,x,y)$ where the metric takes the form (\ref{t2:metric}) (respectively (\ref{hy:metric})) and $t$ ranges from $t_0 \ge 0$ to $+ \infty$.}
\item{In the $T^2$ case, $(\mathcal{M},g)$ is a polarized $T^2$-symmetric spacetime (respectively a $T^3$-Gowdy spacetime) if and only if the initial data are polarized (respectively Gowdy).}
\end{enumerate}
\end{proposition}

Furthermore, we have the following continuation criterion:

\begin{proposition} \label{pro:cc}
Let $(\mathcal{M},g,f)$ be a past development\footnote{Here and everywhere else in the paper, we will consider that, by definition, a development of an initial data set for the Einstein equations is a globally hyperbolic spacetime which satisfies the Einstein equations and agree with the given data initially in the usual sense of general relativity.} of $T^2$-symmetric initial data (respectively $k=-1$ surface-symmetric initial data) with $\Lambda \ge 0$, either in the vacuum or with Vlasov matter and assume that $(\mathcal{M},g)$ can be covered by areal coordinates $(t,\theta,x,y)$, where $t$ ranges from $t_f > 0$ to $t_i$, $t_f < t_i$ and the metric takes the form (\ref{t2:metric}) (respectively (\ref{hy:metric})) . Assume that:
\begin{enumerate}
\item{all metric functions and their derivatives admit a continuous extension to $t=t_f$,}
\item{in the Vlasov case, $f$ and all its derivatives admit a continuous extension to $t=t_f$.}
\end{enumerate}
Then there exists a past development $(\tilde{\mathcal{M}},\tilde{g},\tilde{f})$ of the initial data and an isometric embedding $i$ of $\mathcal{M}$ into $\tilde{\mathcal{M}}$ satisfying $i^*(\tilde{f})=f$ and such that $i(\mathcal{M}) \neq \tilde{\mathcal{M}}$.
\end{proposition}

The first proposition follows from the results of \cite{pc:u1u1, bicm:gft2, arw:caft2v, ci:t2pcc} for the $T^2$ case and from the results of \cite{arr:evhs} for the hyperbolic case. The second proposition follows from the standard local well-posedness theory for the Einstein-Vlasov system as found in \cite{ycb:usel,ycb:pcel}.
\section{The Theorems} \label{se:tth}

\begin{theorem} \label{th:t2vla}
Let $(\mathcal{M},g,f)$ be the maximal development of $T^2$-symmetric initial data with Vlasov matter and $\Lambda \ge 0$. Suppose that the Vlasov field $f$ does not vanish identically. Then $(\mathcal{M},g)$ admits a global foliation by areal coordinates with the time coordinate $t$ taking all values in $(0,\infty)$, i.e.~$t_0=0$ in the notation of Proposition \ref{pro:gaf}.
\end{theorem}
Thus the presence of Vlasov matter forbids $t_0 >0$. In the vacuum case, we know that non-flat solutions with $t_0 > 0$ exist (see appendix E in \cite{js:scct2pccm}) which already indicates that this case is more difficult. We will prove the following:
\begin{theorem} \label{th:t2vac}
Let $(\mathcal{M},g)$ be the maximal Cauchy development of vacuum $T^2$-symmetric initial data with $\Lambda >0$ and suppose that the spacetime is not polarized. Then $(\mathcal{M},g)$ admits a global foliation by areal coordinates with the time coordinate $t$ taking all values in $(0,\infty)$, i.e.~$t_0=0$ in the notation of Propo\-sition \ref{pro:gaf}.
\end{theorem}

The last theorem is the analogue of Theorem \ref{th:t2vla} in the hyperbolic symmetric case:
\begin{theorem} \label{th:hyvla}
Let $(\mathcal{M},g,f)$ be the maximal development of $k=-1$ surface-symmetric initial data with Vlasov matter and $\Lambda \ge 0$. Suppose that the Vlasov field $f$ does not vanish identically. Then $(\mathcal{M},g)$ admits a global foliation by areal coordinates with the time coordinate $t$ taking all values in $(0,\infty)$, i.e.~$t_0=0$ in the notation of Proposition \ref{pro:gaf}.
\end{theorem}
Note that in the vacuum case, there exist solutions of the Einstein equations with hyperbolic symmetry such that $t_0>0$ \cite{ar:cssph}. Thus, the assumption on the Vlasov field is necessary.

\section{Remarks on the strategy of the proofs} \label{se:rsp}
We will present here the main ideas of the proofs of the theorems. We will place particular emphasis on the proof of Theorem \ref{th:t2vac} as it is the most difficult one. The reader might want to return to this section while reading the proof of Theorem \ref{th:t2vac} in order to better follow the arguments.

The proofs of Theorems \ref{th:t2vla}  and \ref{th:hyvla} are based on the strategies developed in \cite{iw:ast2, mw:ast2v}. However, some crucial arguments of these previous works fail in the case of Theorem \ref{th:t2vac} and we have thus been forced to introduce a different approach which we will present below.

 In order to explain these differences and before presenting this new approach, let us first briefly revisit some of the ideas of the proofs contained in \cite{iw:ast2} and \cite{mw:ast2v} for $T^2$-symmetric spacetimes with $\Lambda=0$, respectively in the vacuum and in the Vlasov case. 

\subsection{Previous work} \label{se:pw}
Let us thus assume that $(\mathcal{M},g,f)$ is a past development of $T^2$-symmetric initial data, with $\Lambda=0$, in the vacuum or with Vlasov matter. Suppose that $(\mathcal{M},g)$ is covered by areal coordinates with $t \in (t_f,t_i]$, where $t_f >0$. In view of Proposition \ref{pro:cc}, in order to obtain a statement analogue to that of Theorem \ref{th:t2vla}, it is sufficient to prove that for all such $(\mathcal{M},g)$, all metric functions, the Vlasov field $f$ and all their derivatives admit continuous extensions to $t=t_f$.

\subsubsection{The conformal coordinate system and the function $\alpha$}
 Let us first recall from \cite{bicm:gft2} that another coordinate system may be introduced in $(\mathcal{M},g)$, the so-called conformal coordinate system. In this coordinate system, the metric takes the form:

\begin{align} \label{t2:conf}
ds^2=&e^{2 (\nu -U) } \left( -d\tau^2+d\chi^2 \right)+e^{2U} \left[ dx+Ady+(G+AH)d\chi \right]^2\nonumber \\&+e^{-2U}t^2[dy+Hd\chi]^2.
\end{align}
In the coordinate system $(\tau,\chi,x,y)$, if one assumes that the area of the symmetry orbits $t$ is uniformly bounded from below by a strictly positive constant, one may obtain\footnote{The proof (in the vacuum case) is essentially based on energy and null cone estimates where the energies considered arised naturally from the wave map background structure of the equations. On the other hand, these estimates and the results obtained in conformal coordinates do not provide any information concerning the behaviour of the function $t$, apart from what is already contained in the statement of proposition \ref{pro:gaf}, see \cite{bicm:gft2}.} continuous extensions of all metric functions, the Vlasov field and their derivatives \cite{bicm:gft2}. Thus, it is clear that in order to obtain the same statement in areal coordinates, the key point is to control the function $\alpha$ which appears in (\ref{t2:metric}), as well as its derivatives, as it is this function which dictates the change of coordinates from conformal to areal coordinates. Moreover, it turns out that the function $\alpha$ is necessarily non-decreasing in the past, and in fact increasing if $K>0$ (i.e.~the spacetime is not of $T^3$-Gowdy type). This implies that, in essence, one only need to prove that $\alpha$ is bounded above.

\subsubsection{The energy estimates}
For this purpose, one introduces the energy density\footnote{In the Gowdy case, this energy quantity arises naturally from the wave map structure of the equations. For the $T^2$ case, the vacuum Einstein equations may be regarded as the equations of a wave map problem with source, for which the natural associated energy density is $g$.}
\begin{eqnarray}
g&=&U_t^2+\alpha U_\theta^2+ \frac{e^{2U}}{4t^2}\left( A_t^2+\alpha A_\theta^2 \right)
\end{eqnarray}
and the energy integral:
\begin{equation}
E_g=\int_{\theta \in [0,1]} \frac{g}{\sqrt{\alpha}}.
\end{equation}
This energy can be easily shown to be bounded from above. 

\subsubsection{The estimate on $\alpha$}
Moreover, one can obtain an estimate of the type: 
\begin{equation} \label{ineq:betaener}
\alpha e^{2 \nu}(t,\theta) \le C  \frac{E_g(t_i)}{E_g(t)},
\end{equation}
 for some positive constant $C$ which depends only on the initial data and the value of $t_f>0$. Thus, in order to obtain an upper bound on $\alpha e^{2\nu}$, it is sufficient to have a lower bound on $E_g$.
In the vacuum case with $\Lambda=0$, this lower bound follows easily from the Einstein equations, as $E_g$ is necessarily non-decreasing in the past direction. From the bound on $\alpha e^{2 \nu}$, the upper bound on $\alpha$ follows easily by integration of the evolution equation for $\alpha$ (see equation (\ref{ee:alphat2}) with $\Lambda=0$). The key points are thus the estimate (\ref{ineq:betaener}) and the monoticity of $E_g$.

In the Vlasov case, the monotonicity of $E_g$ is actually broken and thus one loses the easy upper bound on $\alpha e^{2\nu}$. In order to obtain a bound on $\alpha$, one introduces another energy integral, which we shall call here $E_{g,f}$, which can also be proven to be bounded from above. It turns out that $E_{g,f}$ controls $\rho$, an energy density associated with the energy-momentum tensor, and using the fact that $f$ does not vanish identically, Weaver proved in \cite{mw:ast2v} that one can extract enough information from $\rho$ to obtain the following estimate: 

\begin{equation} \label{es:minalpha}
\min_{\theta \in [0,1]}\alpha(t,.) < M,
\end{equation}
for some constant $M>0$ . Thus, using the fact that $f$ does not vanish, one obtains an estimate on the function $\alpha$. This estimate is not as strong as in the vacuum case but it turns out that, together with the upper bound on $E_g$, this control on $\alpha$ is sufficient to derive pointwise estimates on $g$ and bounds on the support of $f$, from which it is easy to derive all the remaining estimates.
\subsection{The proofs of Theorem \ref{th:t2vla} and Theorem \ref{th:hyvla}}

Assume now that we are in the setting of Theorem \ref{th:t2vla}, where we focus on the $T^2$-symmetric case with Vlasov matter and $\Lambda \ge 0$. In this case, as in the case where $\Lambda=0, f \neq 0$ discussed in section \ref{se:pw}, we do not have monotonicity of $E_g$. However, all other important monotonicity properties hold and the estimate concerning $\min_{\theta \in [0,1]}\alpha(t,.)$ still holds. This implies that the proof in the Vlasov case with $\Lambda=0$ can be extended without too much difficulty to the case where $\Lambda >0$. This is treated in detail in section \ref{se:t2vla}. 

\begin{remark} \label{rm:minalpha}
In particular, we note that the assumption of the non-vanishing of the Vlasov field is necessary only so as to establish the estimate (\ref{es:minalpha}). In other words, we have the following proposition:
\begin{proposition}
Let $(\mathcal{M},g,f)$ be a development of $T^2$-symmetric initial data in the vacuum or with Vlasov matter and with $\Lambda \ge 0$. Assume that $(\mathcal{M},g)$ admits a global areal foliation $(t,\theta,x,y)$ where $t$ ranges from $t_f>0$ to $t_i$, $t_f < t_i$.  Assume moreover that the estimate (\ref{es:minalpha}) holds. Then, all the metric functions, the Vlasov field $f$ and all their derivatives admits continuous extensions to $t=t_f$, i.e~the assumptions of Proposition \ref{pro:cc} are verified.
\end{proposition}This simple remark will be useful in the course of the proof of Theorem \ref{th:t2vac}.
\end{remark}

Let us also note that the important monotonicity properties used in the proof of Theorem \ref{th:t2vla} remain valid in the case of hyperbolic symmetry. We will prove Theorem \ref{th:hyvla} by adapting the strategy of the proof of Theorem \ref{th:t2vla} to the hyperbolic symmetric case. This is treated in detail in section \ref{se:hypvla}.

\subsection{The proof of Theorem \ref{th:t2vac}}
In the vacuum case with $\Lambda >0$, we lose again the monotonicity property of $E_g$. Thus, one does not have a priori the lower bound on $E_g$ required to apply (\ref{ineq:betaener}). Moreover, we cannot obtain an a priori estimate on $\min_{\theta \in [0,1]}\alpha(t,.)$ as in the Vlasov case as this required that certain matter terms do not vanish. However, estimates similar to (\ref{ineq:betaener}) hold and thus, we easily obtain that the statement that $\alpha$ is bounded above is equivalent to the statement that $E_g$ is bounded from below by a strictly positive constant. 
 
\subsubsection{Different parametrisations for the orbits of symmetry and explicit solutions of the equations} \label{spse:dpo}
The monotonicity of $E_g$ is linked with the homogeneity or inhomogeneity of the wave equation for the metric function $U$ defined in (\ref{t2:metric}). When $\Lambda >0$, an extra term arises in the time derivative of $E_g$ which has the wrong sign (see equation (\ref{eq:eht})). In fact, in the case where both twists quantities vanish, i.e.~in the $T^3$-Gowdy case ($K=0$), there is a way to recover the monotonicity argument. Indeed, one may apply a simple tranformation to the function $U$ such that the the wave equation for the resulting metric function $P$ is homogeneous (see equation (\ref{ee:P}) with $K=0$). Using $(U,A)$ or $(P,A)$ corresponds to a different choice of parametrisation for the extrinsic geometry of the orbits of symmetry. The system of wave equations sastisfied by $(P,A)$ has a similar structure to that of $(U,A)$ and one may introduce an energy $E_h$ associated with it, which plays a role similar to that of $E_g$.

The interpretation of the transformation is as follows. In the case $(K=0,\Lambda=0)$, all flat Kasner spacetimes corresponding to $U=k \ln t$, $A=const$ are possible solutions of the equations. In the case $(K>0,\Lambda=0)$, the only Kasner spacetimes of the form $U=k \ln t$, $A=const$ which satisfies the Einstein equations are those for which $U=0$ and $A=const$. Another characterisation of these solutions is that they correspond to $E_g=0$. In the case $K=0$, $\Lambda>0$, there can be naturally no flat Kasner solutions, but there are plane symmetric solutions which are characterized by $E_h=0$. We also remark that in both cases $(K>0,\Lambda=0)$ and $(K=0,\Lambda>0)$, there are solutions, with respectively $E_g=0$ or $E_h=0$, for which $t_0 >0$ (see \cite{iw:ast2} and Appendix E in \cite{js:scct2pccm}).

\subsubsection{The easy case $(K=0, \Lambda > 0)$}
In this case, as mentioned above, the system of wave equations for $(P,A)$ is homogeneous. Moreover, one can easily prove that $E_h$ is non-decreasing in the past direction (see Remark \ref{rm:k0c}). An estimate similar to (\ref{ineq:betaener}) can be derived, from which we obtain the desired upper bound on $\alpha$ under the assumption that $E_h \neq 0$ initially. This case can thus be treated separetely and we present it in Proposition \ref{prop:ck0} (see section \ref{se:tbee}).

\subsubsection{The general case. The contradiction setting}
When both $K > 0$ and $\Lambda >0$, there is no easy way to recover a monotonicity property on $E_h$ or $E_g$ and thus there are no a priori lower bounds on $E_g$ or $E_h$. 
We will prove Theorem \ref{th:t2vac} by recovering such a lower bound via other methods. The aim will therefore be to bound away from $0$ the energy integrals $E_h$ and $E_g$ associated with the non-linear system of wave equations describing the motion of the orbits of symmetry. For this, we will proceed by contradiction, assuming that $t_0 >0$ for the maximal Cauchy development. 

This will allow us to obtain two important facts: $\alpha$ is uniformly blowing up (section \ref{se:conts}) and the energy integrals $E_g$ and $E_h$ tend to $0$ as $t \rightarrow t_0$ (section \ref{se:lgeos}). (The uniform blow up of $\alpha$ is in fact an immediate consequence of the remark \ref{rm:minalpha}.) 

\subsubsection{Control on the spatial differences of some metric functions}
We will use the uniform blow up of $\alpha$ and the vanishing limit of $E_h$ and $E_g$ to obtain successively more and more control on the solutions and improve our understanding of the non-linear terms in the equations. First, the vanishing of $E_h$ and $E_g$ in the limit $t \rightarrow t_0$ will imply a strong control on the spatial differences of some the metric functions (section \ref{se:scsd}). In particular, control on $\max_{\theta \in [0,1]} \alpha e^{\nu}-\min_{\theta \in [0,1]}\alpha e^{\nu}$ and similar quantities will be used extensively in the null cone estimates and the analysis of the characteristics which we will pursue later.

\subsubsection{Some tools for the null cone estimates} \label{spse:acac}
In sections \ref{se:pnceeca} and \ref{se:pnceecb}, we will derive null cone estimates. In order to do so, it will be necessary to have at hand the following tools:

\begin{enumerate}
\item[-]{an estimate on $\drond{}{\theta}\left( \ln {\alpha}\right)$ (section \ref{se:lnalphatheta}),}
\item[-]{estimates for the integrals of small powers of $\alpha$ (section \ref{se:eispa}), which will essentially be used to control some error terms in the null cone estimates,}
\item[-]{a parametrisation of the null rays in areal coordinates (section \ref{se:acac}).}
\end{enumerate}
Moreover, to exploit these null cone estimates in the last step of the proof, we will need to control a change of coordinates from the coordinates adapted to the null rays to the areal system of coordinates. The required estimate is proved in section \ref{se:eacat}.

Finally, in order to prove the pointwise estimates from below of section \ref{se:pnceecb}, we will need to start with large data. The analysis of the polarization energy, which we describe below, will enable us to exhibit such large data.

\subsubsection{The polarization energy $E_A$}
In section \ref{se:eacat}, we will focus our attention on the polarization energy $E_A$ of the spacetime associated with the wave equation satisfied by the polarization function $A$ defined in (\ref{t2:metric}). Since by definition, $E_A \le E_h$, a lower bound on $E_A$ is sufficient to obtain a lower bound on $E_h$ and close the estimates. (Motivation for considering this energy comes from the fact that the evolution equation for $A$ stays homogeneous even with $\Lambda >0$ and the simple remark that one of the common features of all known cases with $t_0 >0$ is that all such spacetimes are polarized and thus have $E_A=0$.) From the contradiction setting, it follows that $E_A \rightarrow 0$ as $t \rightarrow t_0$. Using the assumption $E_A >0$ and the vanishing limit of $E_A$, we will exhibit a sequence of points in the spacetime where the energy density $h$ is of the order of $\alpha$. 

\subsubsection{The null cone estimates}
These points will be used in section \ref{se:pnceecb} as large initial data for some null cone estimates along the characterisitics of the spacetime. The aim of these null cone estimates will be to prove that not only is $h$ of order $\alpha$ at some points, but it is in fact blowing up at least like $\alpha^{1-\epsilon}$ along certain characteristics. However, in order to control the spatial derivatives and the non-linearity of the equations, we will also need an estimate from above for $h$. Thus, we will first prove that $h$ is blowing up at most like $\alpha^{1+\epsilon}$. To derive these pointwise estimates on $h$, we will use the tools developed in the previous sections and apply null cone estimates similar to those we introduced in \cite{js:scct2pccm}.

By a continuity argument, it will actually follow that $h$ necessarily blows up along a whole family of characteristics. 

\subsubsection{The contradiction}
In the previous step, we have obtained the blow up of $h$ as $\alpha^{1-\epsilon}$ along a strip of characteristics. This can be integrated in space but if we want to relate the resulting integral to $E_h$, we need to control the difference between the integral of $h$ over the spacelike foliation associated with the conformal coordinate system and its integral over the spacelike foliations associated with the areal coordinate system. Using the results of section \ref{se:eacat}, we will prove  in section \ref{se:contrad} that the two integrals differ at most by a factor of $\alpha^{\epsilon}$. It follows that $E_h=\int_{[0,1]} \frac{h}{\sqrt{\alpha}}d\theta$ is bounded from below by $\delta \min_{\theta \in [0,1]}\alpha^{1/2-2\epsilon}$ for some $\delta>0$ and thus, in particular, does not vanish as $t$ goes to $t_0$. This is a contradiction, which concludes the proof of theorem \ref{th:t2vac}.

\section{Proof of Theorem \ref{th:t2vla}} \label{se:t2vla}
We will prove Theorem \ref{th:t2vla} in this section. As discussed above, the method will follow \cite{mw:ast2v}. It would be sufficient to check that the extra terms arising from the introduction of $\Lambda>0$ do not spoil any of the monotonicity arguments and may be controlled when required, but in order to be self-contained, we will provide a full proof. Moreover, some of the estimates given here will be useful later in order to prove Theorem \ref{th:t2vac} in section \ref{se:t2vac}, in particular, to obtain the uniform blow up of $\alpha$ of lemma \ref{bup:alpha}.
We start by recalling the Einstein-Vlasov system for $T^2$-symmetric spacetimes in areal coordinates.

\subsection{Vlasov matter in $T^2$-symmetric spacetimes} \label{t2:ve}
Let $(\mathcal{M},g,f)$ be a past development of $T^2$-symmetric initial data with Vlasov matter as described in section \ref{se:cid} and assume that $(t,\theta,x,y)$ is a system of areal coordinates such that the metric takes the form (\ref{t2:metric}). Let $v_i$, for $i=0,1,2,3$ denote the components of the velocity vector in the untwisted set of $1$-forms:

\begin{equation}
\{dt, d\theta, dx+Gd\theta,dy+Hd\theta \}.
\end{equation}
which has for dual basis:
\begin{equation}
\left \{\frac{\partial}{\partial t}, \frac{\partial}{\partial \theta}-G\frac{\partial}{\partial x}-H\frac{\partial}{\partial y}, \frac{\partial}{\partial x}, \frac{\partial}{\partial y} \right \}
\end{equation}
In this new frame, the metric (\ref{t2:metric}) and its inverse are given by:

\begin{eqnarray}
\tilde{g}_{ij}= \left(
\begin{array}{cccc}
- \alpha e^{2(\nu-U)} & 0 & 0 & 0 \\
0 & e^{2(\nu-U)} & 0 & 0 \\
0 & 0 & e^{2U} & e^{2U}A \\
0 & 0 & e^{2U}A & e^{-2U}t^2+e^{2U}A^2
\end{array}
\right),
\end{eqnarray}

\begin{eqnarray}
\tilde{g}^{ij}= \left(
\begin{array}{cccc}
- \alpha^{-1} e^{-2(\nu-U)} & 0 & 0 & 0 \\
0 & e^{-2(\nu-U)}  & 0 & 0 \\
0 & 0 & e^{-2U}+e^{2U}A^2t^{-2} & -e^{2U}At^{-2} \\
0 & 0 & -e^{2U}At^{-2} & e^{2U}t^{-2}
\end{array}
\right).
\end{eqnarray}

Note that, along a geodesic, the components $v_2$ and $v_3$ of the velocity vector are constant, since if we denote by $V$ the tangent vector to a geodesic, we have:
\begin{eqnarray}
v_2&=&g\left(V, \drond{}{x}\right), \\
\nabla_V g\left(V, \drond{}{x}\right)&=&g\left(\nabla_V V, \drond{}{x}\right)+g\left(V,\nabla_V \drond{}{x}\right)=0,
\end{eqnarray}
using the geodesic and the Killing equations.

We will parametrize the mass shell $\mathcal{P}$ by the coordinates $(t,\theta,x,y,v_1,v_2,v_3)$, where by an abuse of notation, we denote the lift to $\mathcal{P}$ of the coordinates on $\mathcal{M}$ by the same symbols. The Vlasov field $f$ can then be identified with a function of $(t,\theta,x,y,v_1,v_2,v_3)$ or using the symmetry, with a function of $(t,\theta,v_1,v_2,v_3)$ only and we shall, by an abuse of notation, use both definitions and always denote it by $f$.

With these definitions, the mass shell relation $v_\mu v^\mu=-1$, which holds on the support of the Vlasov field, is given by\footnote{Note that $v_0 < 0$ since $v^0 > 0$.}:

\begin{eqnarray} \label{ee:ms}
v_0=-\sqrt{\alpha e^{2(\nu-U)}+\alpha v_1^2 + \alpha e^{2(\nu-2U)}v_2^2+\alpha t^{-2} e^{2\nu}(v_3-Av_2)^2}
\end{eqnarray}
and the Vlasov equation reads as:
\begin{eqnarray} \label{ee:vlasov}
\drond{f}{t}&=&\drond{v_0}{v_1}\drond{f}{\theta}- \bigg \{ \drond{v_0}{\theta}+\frac{\sqrt{\alpha}e^{\nu}}{t^3}(K-AJ)(v_3-Av_2) \nonumber \\
&&\hbox{}+\frac{\sqrt{\alpha}e^{2\nu-4U}}{t}Jv_2 \bigg\}\drond{f}{v_1}.
\end{eqnarray}

\subsection{The Einstein equations in areal coordinates} \label{t2:ee}
The Einstein equations (\ref{ee}) give rise in areal coordinates to the following system of equations:

Constraint equations:
\begin{eqnarray}
\frac{\nu_t}{t}&=&U_t^2+\alpha U_\theta^2+\frac{e^{4U}}{{4t^2}}(A_t^2+\alpha A_\theta^2) + \frac{\alpha e^{2\nu-4U}}{4t^2}J^2+\frac{\alpha e^{2\nu}  (K-AJ)^2}{4t^4}\nonumber \\&&\hbox{}+\alpha e^{2(\nu-U)} \Lambda+8\pi\frac{\sqrt{\alpha}}{t}\int_{\mathbb{R}^3}f|v_0|dv_1dv_2dv_3, \label{ee:nut_vla}\\
\frac{\alpha_t}{\alpha}&=&-\frac{\alpha e^{2\nu-4U}J^2}{t}-\frac{\alpha e^{2\nu}  (K-AJ)^2} {t^3}-4t\alpha e^{2(\nu-U)}\Lambda  \label{ee:alphat_vla} \\
&&\hbox{}-16\pi\alpha^{3/2}e^{2(\nu-U)} \int_{\mathbb{R}^3} \frac{f(1+e^{-2U}v_2^2+e^{2U}t^{-2}(v_3-Av_2)^2)}{|v_0|}dv_1dv_2dv_3, \nonumber\\
\frac{\nu_\theta}{t}&=&2U_tU_\theta+\frac{e^{4U}}{2t^2}A_tA_\theta-\frac{\alpha_\theta}{2t \alpha} -8\pi\frac{\sqrt{\alpha}}{t}\int_{\mathbb{R}^3}fv_1 dv_1dv_2dv_3. \label{ee:nutheta_vla}
\end{eqnarray}

Evolution equations:
\begin{eqnarray}
\nu_{tt}-\alpha \nu_{\theta \theta}&=&\frac{\alpha_\theta \nu_\theta}{2}+\frac{\alpha_t \nu_t}{2 \alpha}-\frac{\alpha_\theta^2}{4\alpha}+\frac{\alpha_{\theta \theta}}{2}-U_t^2+\alpha U_\theta^2+\frac{e^{4U}}{4t^2}(A_t^2-\alpha A_\theta^2)\nonumber \\
&&\hbox{}-\frac{\alpha e^{2\nu-4U}J^2}{4t^2}-\frac{3\alpha e^{2\nu}(K-AJ)^2}{4t^4} \nonumber \\&&\hbox{}+\alpha \Lambda e^{2(\nu-U)}-8\pi\frac{\alpha^{3/2}e^{2\nu}}{t^3}\int_{\mathbb{R}^3}\frac{f(v_3-Av_2)^2}{|v_0|}dv_1 dv_2 dv_3, \\
U_{tt}-\alpha U_{\theta \theta}&=& - \frac{U_t}{t}+\frac{\alpha_\theta U_\theta}{2}+\frac{\alpha_t U_t}{2 \alpha}+\frac{e^{4U}}{2t^2}(A_t^2-\alpha A_\theta^2) \nonumber \\&&\hbox{}+\alpha \Lambda e^{2(\nu-U)}+\frac{\alpha e^{2\nu-4U}J^2}{2t^2} \nonumber \\
&&\hbox{}+8\pi\frac{\alpha^{3/2}e^{2(\nu-U)}}{2t} \int_{\mathbb{R}^3} \frac{f(1+2e^{-2U}v_2^2)}{|v_0|}dv_1 dv_2 dv_3, \label{ee:uevol}\\
A_{tt}-\alpha A_{\theta \theta}&=& \frac{A_t}{t}+\frac{\alpha_\theta A_\theta}{2}+\frac{\alpha_t A_t}{2 \alpha}-4(A_tU_t- \alpha A_\theta U_\theta)\nonumber \\
&&\hbox{}+\frac{\alpha e^{2\nu-4U}J(K-AJ)}{t^2}\nonumber \\&&\hbox{}+16\pi\frac{\alpha^{3/2}e^{2\nu-4U}}{t} \int_{\mathbb{R}^3} \frac{fv_2(v_3-Av_2)}{|v_0|}dv_1 dv_2 dv_3. \label{ee:aevol}
\end{eqnarray}

Auxiliary equations:
\begin{eqnarray}
J_t&=&-16\pi\alpha \int_{\mathbb{R}^3}\frac{f v_1 v_2}{|v_0|}dv_1 dv_2 dv_3, \\
J_\theta &=&16\pi \int_{\mathbb{R}^3}f v_2dv_1 dv_2 dv_3, \label{ee:jtheta_vla}\\
K_t &=&-16\pi\alpha \int_{\mathbb{R}^3}\frac{f v_1 v_3}{|v_0|}dv_1 dv_2 dv_3, \\
K_\theta &=&16\pi \int_{\mathbb{R}^3}f v_3 dv_1 dv_2 dv_3. \label{ee:ktheta_vla}
\end{eqnarray}

We will now proceed to the proof of Theorem \ref{th:t2vla}. 

 In the rest of this section, $(\mathcal{M},g,f)$ will be a past development of $T^2$-symmetric initial data with Vlasov matter and $\Lambda \ge 0$. We will cover $(\mathcal{M},g)$ by areal coordinates $(t,\theta,x,y)$, where the range of the coordinates is $(t_f,t_i] \times [0,1]^{3}$ with $0 < t_f < t_i$. The metric is then given by (\ref{t2:metric}) where all functions depend on $t$ and $\theta$ and are periodic in $\theta$ with period $1$. The Einstein-Vlasov system implies that the system (\ref{ee:nut_vla})-(\ref{ee:ktheta_vla}) completed by (\ref{ee:vlasov}) holds for all $(t,\theta) \in (t_f,t_i] \times [0,1]$. Moreover, we will assume that $f$ does not vanish identically. From what has been said in section \ref{se:rsp}, we will prove that for all such $(\mathcal{M},g,f)$, the hypotheses of proposition \ref{pro:cc} are satisfied, from which Theorem \ref{th:t2vla} follows immediately.

First we recall some standard facts about the Vlasov field in such spacetimes.
\subsection{Conservation laws} \label{se:clt2v}
From the conservation of the Vlasov field $f$ along geodesics, if follows immediately that $f$ is bounded above by some constant $F>0$:

\begin{equation}
f \le F.
\end{equation}

Since $v_2$ and $v_3$ are constant along geodesics, it follows that the support of $f$ in $v_2$ and $v_3$ is conserved. By compactness of the initial Cauchy surface, we therefore have an upper bound on the support of $f$ in $v_2$ and $v_3$ in $(\mathcal{M},g)$. Let $X$ be such an upper bound:

\begin{equation} \label{eq:upv2v3}
X=\sup\{ \max(|v_2|, |v_3|) / \exists (t,\theta,v_1) / f(t,\theta,v_1,v_2,v_3) > 0 \} < \infty.
\end{equation}

The particle current is given by:

\begin{equation}
N^\mu=\frac{\sqrt{\alpha}}{t}\int_{\mathbb{R}^3} \frac{f}{|v_0|}v^\mu dv_1 dv_2 dv_3.
\end{equation}

From the Vlasov equation it follows that $N^\mu$ is divergence free: $\nabla_\mu N^\mu=0$. We therefore have the conservation law,  $\forall t$, 

\begin{equation} \label{eq:clv_vla}
\int_{[0,1]} N^0 t \sqrt{\alpha} e^{2(\nu-U)} d\theta=\int_{[0,1]}\left( \int_{\mathbb{R}^3}f dv_1 dv_2 dv_3 \right)d\theta=Q,
\end{equation}
for some positive constant $Q$. Moreover, since by assumption, the Vlasov field does not vanish identically, we have:

\begin{equation}
 Q > 0
\end{equation}

\subsection{Lower bound on the mean value of $|v_1|$.} \label{se:lbav1}
In this section, we prove a lower bound on the mean value of $|v_1|$ for the measure $fdvd\theta$. This lower bound is the important estimate which takes avantage of the assumption that $f \neq 0$. Coupled to the energy estimates which we will derive in the next section, this estimate will give us uniform control of $\min_{[0,1]}\alpha(t,.)$ (See section \ref{sec:basc}).

\begin{lemma} \label{lem:lbv_vla}
There exists $\delta > 0$ such that $\int_{[0,1]} \int_{\mathbb{R}^3} f |v_1| dv_1 dv_2 dv_3 d\theta> \delta$ for all $t \in (t_f,t_i]$.
\end{lemma}

\begin{proof}: 

Let $\epsilon=\frac{Q}{16X^2F}$, so that $Q-\epsilon 8 X^2 F = Q/2 > 0$. We have:

\begin{eqnarray}
\int_{[0,1]} \int_{\mathbb{R}^3} f|v_1| dv_1 dv_2 dv_3&=&\int_{[0,1]} \intr{2} \left(\int_{-\epsilon}^\epsilon f|v_1| \right)dv_2 dv_3 \nonumber
\\&&+\int_{[0,1]} \intr{2} \left( \int_{|v_1| > \epsilon}f|v_1| dv_1 \right)dv_2 dv_3, \nonumber 
\\
&\ge& \epsilon \int_{[0,1]} \intr{2} \left( \int_{|v_1| > \epsilon}f dv_1 \right) dv_2 dv_3, \nonumber \\
&\ge& \epsilon Q/2.
\end{eqnarray}
\end{proof}

\subsection{Energy estimates}

The following energy estimates take their origins in the underlying wave map structure of the equations, visible in the vacuum case \cite{bicm:gft2} and easily modifiable to suit the Vlasov case.

Define the energy integral $E_{g,K,\Lambda,f}(t)$ by\footnote{The motivation for the notation $E_{g,K,\Lambda,f}$ comes from the fact that this energy may be decomposed in four terms containing respectively $g$, $K$, $\Lambda$ and $f$. Later, we will introduce several other energy integrals and the notation will follow the same pattern.}:

\begin{equation} \label{def:egkf}
E_{g,K,\Lambda,f}=\int_{[0,1]} \frac{\nu_t}{\sqrt{\alpha}t}d\theta.
\end{equation}

From the constraint equation (\ref{ee:nut_vla}), it follows that:

\begin{align}
E_{g,K,\Lambda,f}=&\int_{[0,1]} \frac{1}{\sqrt{\alpha}} \bigg( U_t^2+\alpha U_\theta^2+\frac{e^{4U}}{4t^2}(A_t^2+\alpha A_\theta^2)+\frac{\alpha e^{2\nu-4U}J^2}{4t^2}\nonumber \\
&+\frac{\alpha e^{2\nu}(K-AJ)^2}{4t^2}+\alpha e^{2(\nu-U)}\Lambda+8\pi\frac{\sqrt{\alpha}}{t}\intr{3}f|v_0|dv_1dv_2dv_3 \bigg)d\theta.
\end{align}

Using the Einstein equations, we may compute the time deriva\-tive\\ of $E_{g,K,\Lambda,f}$:

\begin{eqnarray} \label{eq:dedt}
\frac{dE_{g,K,\Lambda,f} }{dt}&=&-\int_{[0,1]} \Bigg[ \frac{2}{t} \left(\frac{U_t^2}{\sqrt{\alpha}}+\frac{e^{4U}}{4t^2}\sqrt{\alpha}A_\theta^2\right) \nonumber \\
&&+\frac{\sqrt{\alpha}e^{2\nu-4U}J^2}{2t^3}+\frac{\sqrt{\alpha}e^{2\nu}(K-AJ)^2}{t^4} \nonumber \\
&&+8\pi\intr{3} \left( \frac{f |v_0|}{t^2}+\frac{\alpha e^{2\nu}f(v_3-Av_2)^2}{t^4|v_0|} \right)dv_1 dv_2 dv_3 \Bigg]d\theta.
\end{eqnarray}

Since the right-hand side is non-positive, $E_{g,K,\Lambda,f}$ is non-decreasing when $t$ is decreasing\footnote{Note that in contrast, $E_g=\int_{[0,1]} \frac{1}{\sqrt{\alpha}} \bigg( U_t^2+\alpha U_\theta^2+\frac{e^{4U}}{4t^2}(A_t^2+\alpha A_\theta^2)\bigg)d\theta$ is not necessarily monotonic.}. Moreover, we have:

\begin{lemma} \label{lem:energy} $E_{g,K,\Lambda,f}$ is bounded on $(t_f,t_i]$ and admits a continuous extension at $t_f$.
\end{lemma}

\begin{proof} From  (\ref{def:egkf}), (\ref{eq:dedt}) and the mass shell relation (\ref{ee:ms}), we obtain:

\begin{equation}
\frac{dE_{g,K,\Lambda,f} }{dt} \ge \frac{-4}{t}E_{g,K,\Lambda,f},
\end{equation}
where the factor of $4$ arises because of the terms containing $(K-AJ)^2$. Applying Gronwall's lemma and using the lower bound $t \ge t_f >0$, we then obtain a uniform bound on $E_{g,K,f}$.
\end{proof}

\subsection{Estimate for $\sqrt{\alpha}e^{2\nu+bU}$}
In this section, we exploit the monotonicity properties of the constraint equations. We have the following lemma:

\begin{lemma} \label{lem:firstcontrol}
For any real number b, $\sqrt{\alpha}e^{2\nu+bU}$ is uniformly bounded on \\$(t_f,t_i] \times [0,1]$.
\end{lemma}
\begin{proof}: Using equations  (\ref{ee:nut_vla}) and (\ref{ee:alphat_vla}), we see that $(t^{b^2/8}\sqrt{\alpha}e^{2\nu+bU})$ is decreasing with decreasing $t$: 

\begin{eqnarray}
\partial_t(t^{b^2/8}\sqrt{\alpha}e^{2\nu+bU})&=&b^2/8t^{b^2/8-1}\sqrt{\alpha} e^{2\nu+bU}+t^{b^2/8}\frac{\alpha_t}{2\sqrt{\alpha}}e^{2\nu+bU}\nonumber \\
&&+ t^{b^2/8}\sqrt{\alpha}e^{2\nu+bU}(2\nu_t+bU_t), \nonumber\\
&=& t^{b^2/8}\sqrt{\alpha}e^{2\nu+bU}\Bigg(2t \bigg [(U_t+\frac{b}{4t})^2+\alpha U_\theta^2 \nonumber \\
&&+\frac{e^{4U}}{4t^2}(A_t^2+\alpha A_\theta^2) \bigg ] \nonumber \\
&&+8\pi\sqrt{\alpha}\intr{3}f\left(|v_0|+\frac{\alpha v_1^2}{|v_0|}\right)dv_1dv_2dv_3\Bigg) \ge 0.
\end{eqnarray}
\end{proof}

Thanks to the freedom in the choice of the Killing fields, we also have:
\begin{lemma} \label{lem:controlA}
For any positive real number $r$ and any real number $\lambda$:

\begin{equation}
\alpha^{r/2}e^{2 r \nu+\lambda U}A^2
\end{equation}
is bounded on $(t_f,t_i] \times [0,1]$.
\end{lemma}
\begin{proof}
Consider inverting the role of $X$ and $Y$ in the metric:

\begin{eqnarray}
\tilde{X}&=&Y, \\
\tilde{Y}&=&-X.
\end{eqnarray}

 This is an $SL(2,\mathbb{R})$ transformation and therefore (see section \ref{se:t2spt2}), the form of the metric is unchanged if we relabel the metric functions as follows, using tilde notations for the new metric functions:

\begin{eqnarray}
e^{2 \tilde{U}}&=&e^{2U}A^2+t^2 e^{-2U}, \\
e^{2 \tilde{U}}\tilde{A}&=&-A e^{2U},\\
\tilde{\alpha}&=&\alpha, \\
\tilde{\alpha}e^{2(\tilde{\nu}-\tilde{U})}&=&\alpha e^{2(\nu-U)}.
\end{eqnarray}

Let $q < r$, using the previous equations, it follows that:

\begin{equation}
\tilde{\alpha}^{q/2}e^{2q\tilde{\nu}+2(1-q)\tilde{U}}=\alpha^{q/2}e^{2q\nu+2(1-q)U}A^2+\alpha^{q/2}t^2e^{2q\nu-2(1+q)U} \label{eq:controlA}
\end{equation}

Since the tilded metric functions satisfy the same equations with respect to the same $t$, the left-hand side of equation (30) is bounded on $(t_f,t_i] \times [0,1]$ from lemma \ref{lem:firstcontrol}. Since the second term on the right hand side is positive, the first term is bounded. By lemma \ref{lem:firstcontrol}, 

\begin{equation}
\alpha^{(r-q)/2}e^{2(r-q)\nu+\left( \lambda-2(1-q) \right)U}
\end{equation}
is bounded, and multiplying this by the first term on the right-hand side of (\ref{eq:controlA}), we obtain the desired estimate.
\end{proof}

The quantity $\sqrt{\alpha}e^{\nu}$ will play an important role in the analysis. To simplify some of the computations, let us define $\beta$ by:

\begin{equation} 
e^{\beta}=\sqrt{\alpha}e^{\nu}.
\end{equation}

\subsection{Estimates for the integrals of the spatial derivatives of the metric functions}
From equation (\ref{ee:nutheta_vla}), we derive:

\begin{equation} \label{ee:betatheta_vla}
\beta_\theta=2t\left(U_t U_\theta+\frac{e^{4U}}{4t^2}A_t A_\theta\right)-8\pi\sqrt{\alpha}\intr{3} f v_1 dv_1 dv_2 dv_3
\end{equation}

It follows from the energy estimates obtained in lemma \ref{lem:energy} and the above equation that we can uniformly control the variation in $\theta$ of the metric functions i.e.~we have the following:

\begin{lemma} $\int_{[0,1]}|\beta_\theta|d\theta$, $\int_{[0,1]}|U_\theta|d\theta$, $\int_{[0,1]} e^{2U}|A_\theta|d\theta$, $\int_{[0,1]}|J_\theta|d\theta$ and \\$\int_{[0,1]}|K_\theta|d\theta$ are uniformly bounded on $(t_f,t_i]$. \label{lem:thetacontrol}
\end{lemma}
\begin{proof}

From equation (\ref{ee:betatheta_vla}), we obtain:

\begin{equation}
\frac{|\beta_\theta|}{t} \le \frac{\nu_t}{\sqrt{\alpha}t}
\end{equation}
and by integration we obtain a bound on $\int_{[0,1]}|\beta_\theta|d\theta$ in view of (\ref{def:egkf}) and the bound on $E_{g,K,\Lambda,f}$. The bounds on $\int_{[0,1]}|J_\theta|d\theta$ and $\int_{[0,1]}|K_\theta|d\theta$ follow from the auxiliary equations (\ref{ee:jtheta_vla}), (\ref{ee:ktheta_vla}), and the conservation of the flux (\ref{eq:clv_vla}) together with (\ref{eq:upv2v3}). The bounds on the remaining quantities follow from the definition of $E_{g,K,f}$ and the monotonicity in $t$ of $\alpha$.
\end{proof}

\subsection{Control of $\alpha$ along special curves} \label{sec:basc}
 In this section, we obtain a bound on $\min_{\theta \in [0,1]}\alpha(t,.)$, using the lower bound on the mean value of $|v_1|$.

\begin{lemma} $\min_{\theta \in [0,1]} \alpha(t,.)$ is uniformly bounded on $(t_f,t_i]$.
\end{lemma}
\begin{proof} From the definition of $E_{g,K,\Lambda,f}$, we have:
\begin{equation}
8\pi \int_{[0,1]} \intr{3} f|v_0|dv_1 dv_2 dv_3 d\theta \le t E_{g,K,\Lambda,f}
\end{equation}
and from the mass shell relation (\ref{ee:ms}), we obtain:

\begin{eqnarray}
\int_{[0,1]} \intr{3} f \sqrt{\alpha} |v_1| dv_1 dv_2 dv_3 d\theta &\le& \frac{t E_{g,K,\Lambda,f}}{8\pi},\\
\sqrt{\min_{\theta \in [0,1]} \alpha(t,.)} \int_{[0,1]} \intr{3} f |v_1| dv_1 dv_2 dv_3 d\theta &\le& \frac{t E_{g,K,\Lambda,f}}{8\pi}, \\
\sqrt{\min_{\theta \in [0,1]} \alpha(t,.)} &\le& \frac{t E_{g,K,\Lambda,f}}{8\pi\delta},
\end{eqnarray}
where we have used the lower bound of lemma \ref{lem:lbv_vla} to obtain the last inequality.
\end{proof}

\begin{remark}
Note that this is the only step in the proof of Theorem \ref{th:t2vla} where we need the assumption that $f$ does not vanish. In particular, in the proof by contradiction of Theorem \ref{th:t2vac} given in section \ref{se:t2vac}, we will be able to assume that the above lemma does not hold (see section \ref{se:bupalpha}).
 \end{remark}

As a corollary, we have:
\begin{corollary} There exists $\bar{\theta} \in [0,1]$ such that $\alpha(t,\bar{\theta})$ is bounded on $(t_f,t_i]$. \label{cor:controlalpha}
\end{corollary}
\begin{proof}
Let $M$ be a bound for $\min_{[0,1]} \alpha$. Suppose that for every $\theta \in [0,1]$, $\alpha(t,\theta)$ is unbounded. By assumption, for every $\theta$, there exists a $t^*(\theta)$, for which $\alpha(t^*(\theta),\theta) > 2M $ and by continuity, there exists an open interval $I_\theta=(\theta-\delta_\theta,\theta+\delta_\theta)$ such that:

\begin{equation}
\forall \theta' \in I_\theta, \alpha(t^*(\theta),\theta')> M.
\end{equation}

Consider $\cup_{\theta \in [0,1]} I_\theta$. This is an open cover of $[0,1]$ and by compactness, there exists a finite subcover. Let $\theta_0, \theta_1, ..., \theta_n$ be such that $[0,1]=\cup_{0 \le k \le n}I_{\theta_k}$ and let $T=\min_{0 \le k \le n} t^*(\theta_k)$. Since $\alpha$ is increasing with decreasing time, it follows that $\alpha(T,\theta) > M$ for every $\theta \in [0,1]$ which contradicts the definition of $M$.
\end{proof}

\subsection{Estimate on  $\sqrt{\alpha}e^{\nu+bU}$} \label{se:eb}
We are now ready to prove:

\begin{lemma} For any real number $b$, $\sqrt{\alpha}e^{\nu+bU}=e^{\beta+bU}$ is uniformly bounded on $(t_f,t_i] \times [0,1]$. \label{lem:seccontrol}
\end{lemma}
\begin{proof} By lemma \ref{lem:firstcontrol} and corollary \ref{cor:controlalpha}, we have:

\begin{equation}
e^{\beta(t,\bar{\theta})+bU(t,\bar{\theta})}=\sqrt{\alpha(t,\bar{\theta})} \sqrt{\alpha(t,\bar{\theta})}e^{2\nu(t,\bar{\theta})+bU(t,\bar{\theta})}\le B,
\end{equation}
for some constant $B>0$.

The uniform bound on $e^{\beta+bU}$  then follows from lemma \ref{lem:thetacontrol} since we have that, for all $(t,\theta) \in (t_f,t_i] \times [0,1]$:

\begin{eqnarray} 
|\beta(t,\bar{\theta})-\beta(t,\theta)| \le B', \nonumber \\
|U(t,\bar{\theta})-U(t,\theta)| \le B',
\end{eqnarray}
for some constant $B'>0$, and thus

\begin{equation}
e^{\beta(t,\theta)+bU(t,\theta)} \le B e^{B'+|b|B'} \qedhere  
\end{equation}

\end{proof}

\subsection{Control of the polarization}
Corollary \ref{cor:controlalpha} also implies a sharper estimate on the inner product of the Killing fields:

\begin{lemma} \label{lem:seccontrolA}
For any real numbers $r$ and $b$, $e^{r\beta+bU}A$ is uniformly bounded on $(t_f,t_i] \times [0,1]$.
\end{lemma}
\begin{proof} It follows from corollary \ref{cor:controlalpha} and lemma \ref{lem:controlA}, that $e^{r \beta+bU}A$ is bounded on $(t_f,t_I] \times \{\bar{\theta}\}$.
Furthermore, we have:

\begin{eqnarray}
e^{r \beta+bU}A(t,\theta) &\le& e^{r\beta+bU}A(t,\bar{\theta})+\int_{\bar{\theta}}^\theta \bigg( e^{r\beta+bU}A(r \beta_\theta+b U_\theta) \nonumber \\
&&+e^{r\beta+(b-2)U}e^{2U}A_\theta \bigg) d\theta'.
\end{eqnarray}

Using the bound on $e^{r\beta+bU}A(t,\bar{\theta})$, we therefore obtain:

\begin{eqnarray}
\left|e^{r \beta+bU}A(t,\theta)\right| \le B+\left|\int_{\bar{\theta}}^\theta e^{r \beta+bU}|A|(|r\beta_\theta|+|bU_\theta|)d\theta \right|.
\end{eqnarray}
for some constant $B>0$ and we can conclude using Gronwall's inequality and lemma \ref{lem:thetacontrol}.
\end{proof}

\subsection{Estimates for the time integrals of the twist quantities}
In order to estimate the first derivatives of $U$ and $A$ in the next section, we will need the following estimates for the time integrals of the twist quantities:

\begin{lemma}\label{lem:intjcontrol} $\int_t^{t_i} \max_{\theta \in [0,1]}[e^{2\beta-4U}J^2 ](t',\theta)dt'$ is uniformly bounded on $(t_f,t_i]$.
\end{lemma}

\begin{proof}
From lemma \ref{lem:thetacontrol}, there exists a constant $M$ such that:

\begin{eqnarray}
|J(t',\theta)| \le M + |J(t',\bar{\theta})|, \nonumber \\
e^{2\beta-4U}(t',\theta) \le e^{6M}e^{2\beta-4U}\beta(t',\bar{\theta}). \nonumber \\
\end{eqnarray}

Thus, we have:
\begin{eqnarray}
\int_t^{t_i} \max_{\theta \in [0,1]}[e^{2\beta-4U}J^2](t',\theta) dt' &\le& \int_t^{t_i}e^{6M}\Big[e^{2\beta-4U} \big( J^2 \nonumber \\
&&+2M|J|+M^2 \big)\Big](t',\bar{\theta}) dt'
\end{eqnarray}
and using $2 |J| \le J^2+1$ as well as lemma \ref{lem:seccontrol}, we obtain:

\begin{eqnarray}
\int_t^{t_i} \max_{\theta \in [0,1]}[e^{2\beta-4U}J^2](t',\theta) dt' &\le&B+B'\int_t^{t_i} [e^{2\beta-4U} J^2](t',\bar{\theta})dt', \label{ineq:J}
\end{eqnarray}
for some constants $B$ and $B'$.

Since by integration of equation (\ref{ee:alphat_vla}), we have:

\begin{equation}
\int_t^{t_i}[e^{2\beta-4U}J^2](t', \bar{\theta}) dt' \le t_i \ln \frac{\alpha(t,\bar{\theta})}{\alpha(t_i,\bar{\theta})}, \label{ineq:Jbar}
\end{equation}
which is bounded from corollary \ref{cor:controlalpha}, the right-hand side of (\ref{ineq:J}) is uniformly bounded. 
\end{proof}
Similarly, we have:

\begin{lemma} $\int_t^{t_i} \max_{\theta \in [0,1]}[e^{2\beta}(K-AJ)^2 ](t',\theta) dt'$ is uniformly bounded \\on $(t_f,t_i]$. \label{lem:intkcontrol}
\end{lemma}
\begin{proof}
We first integrate the $\theta$ derivative of $e^{\beta}(K-AJ)$, using the auxiliary equations (\ref{ee:ktheta_vla}) and (\ref{ee:jtheta_vla}) to replace the derivatives of $K_\theta$ and $J_\theta$ by matter terms: 

\begin{eqnarray}
e^{\beta}|K-AJ|(t,\theta) &\le& e^\beta|K-AJ|(t,\bar{\theta}) \nonumber \\
&&+\bigg|\int_{\bar{\theta}}^\theta \bigg[ e^\beta|K-AJ||\beta_\theta|+e^{\beta-2U}|J|e^{2U}|A_\theta| \nonumber \\
&&+16\pi e^\beta \intr{3}f|v_3|dv_1 dv_2 dv_3 \nonumber \\
&&+16\pi e^\beta A \intr{3} f|v_2| dv_1 dv_2 dv_3 \bigg]d\theta' \bigg|.
\end{eqnarray}

Using lemmas \ref{lem:thetacontrol}, \ref{lem:seccontrol} and  \ref{lem:seccontrolA}, as well as the conservation law (\ref{eq:clv_vla}) and the uniform boundedness (\ref{eq:upv2v3}) of the support of $f$ in $v_3$ and $v_2$, we obtain:

\begin{eqnarray}
e^{\beta}|K-AJ|(t,\theta)&\le& e^\beta|K-AJ|(t,\bar{\theta}) \nonumber \\
&&+B\bigg|\int_{\bar{\theta}}^\theta e^\beta|K-AJ||\beta_\theta|d\theta' \bigg| \nonumber \\
&&+C\max_{\theta \in [0,1]}[e^{\beta-2U}|J|(t,.)] +D,
\end{eqnarray}
for some constants $B$, $C$ and $D$.
Applying Gronwall's lemma, we obtain:

\begin{eqnarray}
e^{\beta}|K-AJ|(t,\theta)&\le& \Big(e^\beta|K-AJ|(t,\bar{\theta}) \nonumber \\
&&+B\max_{\theta \in [0,1]}[e^{\beta-2U}|J|(t,.)] +C \Big)\left(1+e^{|\int_{\bar{\theta}}^\theta |\beta_\theta|d\theta'|}\right)
\end{eqnarray}
and therefore, using lemma \ref{lem:thetacontrol} again, we have:

\begin{eqnarray}
\max_{\theta \in [0,1]}e^{\beta}|K-AJ|(t,.)&\le& D\Big(e^\beta|K-AJ|(t,\bar{\theta}) \nonumber \\
&&+B\max_{\theta \in [0,1]}[e^{\beta-2U}|J|(t,.)] +C\Big).
\end{eqnarray}
We may now conclude by integration of equation (\ref{ee:alphat_vla}) and the application of lemma \ref{lem:intjcontrol} to bound the term containing $J$.
\end{proof}

\subsection{Null cone estimates for the first derivatives of $U$ and $A$ coupled to an estimate for the support of $f$} \label{sec:ncet2v}
We will perform null cone energy estimates to bound the first derivatives of $U$ and $A$. However, to close the estimates we will also need to estimate the support of $f$. 
 \comment{As a first attempt, it would seem reasonable to try to obtain higher order energy estimates by commuting equations (\ref{ee:uevol}) and (\ref{ee:aevol}) by $\partial_t$ and $\partial_\theta$ and then apply a Sobolev inequality to derive pointwise bounds on the first derivatives. However, commuting equations (\ref{ee:uevol}) and (\ref{ee:aevol}) by $\partial_t$ and $\partial_\theta$ generates many terms involving derivatives of other functions than $U$ and $A$ which are not yet controlled. Moreover, null cone estimates are particularly suitable to $1+1$ dimensional problems, hence the preference given here to this approach.}

Recall the definition of the energy density:

\begin{equation} \nonumber
g= U_t^2+\alpha U_\theta^2+\frac{e^{4U}}{4t^2}\left(A_t^2+\alpha A_\theta^2 \right). 
\end{equation}
We will prove the following lemma:

\begin{lemma}$g$ is uniformly bounded on $(t_f,t_i] \times [0,1]$ and the support of $f$ is uniformly bounded on $(t_f,t_i] \times [0,1] \times \mathbb{R}^3$.
\end{lemma}
\begin{proof}

We define $g^{\times}$ by:

\begin{eqnarray}
g^{\times}&=&2 \sqrt{\alpha} \left( U_t U_\theta+\frac{e^{4U}}{4t^2}A_tA_\theta\right).
\end{eqnarray}
We have $g\pm g^{\times} \ge 0$. Let $\partial_u=\partial_t-\sqrt{\alpha} \partial_\theta$ and $\partial_v=\partial_t+\sqrt{\alpha} \partial_\theta$.

Using the Einstein equations, we can compute the null derivatives of $g+g^{\times}$ and $g-g^{\times}$:

\begin{eqnarray}
\partial_u (g+g^{\times})&=&-\frac{2}{t} \left( U_t^2+\frac{e^{4U}}{4t^2}\alpha A_\theta^2 \right)+\frac{\alpha_t}{\alpha}(g+g^{\times})-\frac{g^{\times}}{t} \nonumber \\
&&+2(U_t+\sqrt{\alpha}U_{\theta}) \bigg( \frac{e^{\beta-4U}}{2t^2}J^2 \nonumber \\
&&+8\pi\frac{\sqrt{\alpha}e^{2\beta-2U}}{2t} \intr{3} \frac{f(1+2e^{-2U}v_2^2)}{|v_0|}dv_1 dv_2 dv_3 +e^{2\beta -2U}\Lambda \bigg)  \nonumber \\
&&+\frac{e^{4U}}{2t^2}(A_t+\sqrt{\alpha}A_\theta) \bigg( \frac{e^{2\beta-4U}}{t^2}J(K-AJ) \nonumber \\
&&+16\pi\sqrt{\alpha}\frac{e^{2\beta-4U}}{t}\intr{3}\frac{fv_2(v_3-Av_2)}{|v_0|}dv_1dv_2dv_3 \bigg) \label{eq:derG+H}
\end{eqnarray}
and
\begin{eqnarray}
\partial_ v(g-g^{\times})&=&-\frac{2}{t} \left( U_t^2+\frac{e^{4U}}{4t^2}\alpha A_\theta^2 \right)+\frac{\alpha_t}{\alpha}(g-g^{\times})+\frac{g^{\times}}{t} \nonumber \\
&&+2(U_t-\sqrt{\alpha}U_\theta) \bigg( \frac{e^{\beta-4U}}{2t^2}J^2 \nonumber \\
&&+8\pi\frac{\sqrt{\alpha}e^{2\beta-2U}}{2t} \intr{3} \frac{f(1+2e^{-2U}v_2^2)}{|v_0|}dv_1 dv_2 dv_3 +e^{2\beta -2U}\Lambda \bigg)  \nonumber \\
&&+\frac{e^{4U}}{2t^2}(A_t-\sqrt{\alpha}A_\theta) \bigg( \frac{e^{2\beta-4U}}{t^2}J(K-AJ) \nonumber \\
&&+16\pi\sqrt{\alpha}\frac{e^{2\beta-4U}}{t}\intr{3}\frac{fv_2(v_3-Av_2)}{|v_0|}dv_1dv_2dv_3 \bigg). \label{eq:derG-H}
\end{eqnarray}

Define $T_1$ and $T_2$ by:
\begin{eqnarray}
T_1&=&\frac{e^{\beta-4U}}{2t^2}J^2+8\pi\frac{\sqrt{\alpha}e^{2\beta-2U}}{2t} \intr{3} \frac{f(1+2e^{-2U}v_2^2)}{|v_0|}dv_1 dv_2 dv_3\nonumber \\
&&+e^{2\beta -2U}\Lambda,  \\
T_2&=& \frac{e^{2\beta-2U}}{t^3}J(K-AJ)\nonumber\\
&&+16\pi\sqrt{\alpha}\frac{e^{2\beta-2U}}{t^2}\intr{3}\frac{fv_2(v_3-Av_2)}{|v_0|}dv_1dv_2dv_3.
\end{eqnarray}

$T_1$ and $T_2$ can be estimated using equation (\ref{ee:alphat_vla}):
\begin{eqnarray}
|T_1| &\le& \left| \frac{\alpha_t}{2t \alpha} \right|, \label{ineq:t1} \\ 
|T_2| &\le& \left| \frac{\alpha_t}{2 \alpha} \right|. 
\end{eqnarray}

We therefore obtain:

\begin{eqnarray}
|\partial_u(g+g^{\times})| &\le& \left| \frac{\alpha_t}{ \alpha} \right| \left( \frac{g}{t}+\frac{1}{2t}+2g \right)+\frac{3g}{t},\label{es:ghuder} \\
|\partial_v(g-g^{\times})| &\le& \left| \frac{\alpha_t}{ \alpha} \right| \left( \frac{g}{t}+\frac{1}{2t}+2g \right)+\frac{3g}{t}. \label{es:ghvder}
\end{eqnarray}
To perform null cone estimates, in view of the last two inequalities, we need to control the time integral of $\frac{\alpha_t}{\alpha}$, that is to say, we need to control $\ln \alpha$. Consider  the right-hand side of equation (\ref{ee:alphat_vla}). The time integral of the two terms containing the twist quantities are bounded from lemmas \ref{lem:intjcontrol}, \ref{lem:intkcontrol} and the term containing the cosmological constant is bounded from lemma \ref{lem:seccontrol}.
Therefore to control $\left|\frac{\alpha_t}{\alpha}\right|$, we only need to control the last term, which is the term containing the Vlasov field. While we already have a bound on the support of the Vlasov field in $v_2$ and $v_3$, we still cannot estimate the support of $f$ in $v_1$. Therefore, the best we can obtain from equation (\ref{ee:alphat_vla}) is an estimate for $\left|\frac{\alpha_t}{\alpha}\right|$ which depends on the support of $f$ in $v_1$ and quantities which have been shown to be bounded. On the other hand, using the characteristic equations associated with the Vlasov equation, i.e.~using the geodesic equations, we can obtain a bound on the support of $v_1$ in terms of $g$ and quantities which have been shown to be bounded. The strategy, which was originally developed by Andreasson \cite{ha:gfmsg}, is therefore to combine the two.
For this, we define the following functions:

\begin{eqnarray}
u_1&=&\sqrt{\alpha}v_1\\
\bar{u}_1(t)&=&\sup \bigg\{ \sqrt{\alpha}|v_1| / \exists (t',\theta,v_1,v_2,v_3)  \nonumber \\
&& \in [t,t_i] \times [0,1] \times \mathbb{R}^3 /f(t',\theta,v_1,v_2,v_3) \neq 0\bigg\}, \\
\psi(t)&=& \max\left( \sup_{\theta \in [0,1]} g(t,.)+\bar{u}_1^2(t),2\right).
\end{eqnarray}

We start by estimating $\left|\frac{\alpha_t}{\alpha}\right|=-\frac{\alpha_t}{\alpha}$ in terms of $\bar{u}_1$:
\begin{eqnarray} \label{es:alphatalphainter}
-\frac{\alpha_t}{\alpha}(t,\theta) &\le& C(t)+B(t, \theta),
\end{eqnarray}
for some non-negative function $C(t)$ whose integral in time is bounded and where $B(t,\theta)$ is given by:
\begin{equation}
B(t,\theta)=16\pi e^{2\beta-2U}\intr{3}\frac{f(1+e^{-2U}v_2^2+t^{-2}e^{2U}(v_3-Av_2)^2)}{|v_0|}du_1 dv_2 dv_3. \nonumber
\end{equation}
We have:

\begin{eqnarray}
B(t,\theta) &\le& 16\pi e^{2\beta-2U}\intr{3}\frac{f(1+e^{-2U}v_2^2+t^{-2}e^{2U}{(v_3-Av_2)}^2)}{e^{\beta-U}\sqrt{1+e^{-2\beta+2U}u_1^2}}du_1 dv_2 dv_3, \nonumber \\
&\le& 16\pi  e^{\beta-U}F\Big(1+e^{-2U}X^2+\frac{e^{2U}}{t^2}(X \nonumber \\
&&+|A|X)^2 \Big) 4X^2 \int^{\bar{u}_1}_{-\bar{u}_1}\frac{du_1}{\sqrt{1+e^{-2\beta+2U}u_1^2}}, \nonumber \\
&\le& 16\pi  e^{\beta-U}F\Big(1+e^{-2U}X^2+\frac{e^{2U}}{t^2}(X \nonumber \\
&&+|A|X)^2 \Big) 4X^2 2 \left( e^{\beta-U}\ln\left(\bar{u}_1+\sqrt{e^{2\beta-2U}+\bar{u_1}^2}\right)+e^{-1}\right).
\end{eqnarray}

Therefore, using lemmas \ref{lem:seccontrol} and \ref{lem:seccontrolA}, it follows from (\ref{es:alphatalphainter}) that there exist a non-negative function $C(t)$ whose integral in time is bounded and a constant $D>0$ such that we have the following estimate:

\begin{equation}
-\frac{\alpha_t}{\alpha}(t,\theta) \le C(t)+D \ln(1+\bar{u}_1^2). \label{es:alphat_vla} 
\end{equation}

On the other hand, from the characteristic equation of the Vlasov equation (\ref{ee:vlasov}), it follows that:

\begin{eqnarray}
\frac{d u_1^2}{ds}&=& \frac{\alpha_t}{\alpha}u_1^2+\frac{2\sqrt{\alpha}u_1}{v_0} \bigg( e^{2\beta-2U}(\beta_\theta-U_\theta)  \nonumber \\
&&\hbox{}+e^{2\beta-4U}(\beta_\theta-2U_\theta)v_2^2  \nonumber \\
&&\hbox{}+\frac{e^{2\beta}}{t^2}(v_3-Av_2)[(v_3-Av_2)\beta_\theta-A_\theta v_2] \bigg) \nonumber \\
&&\hbox{}+\frac{2e^{2\beta}u_1}{t}\bigg( \frac{(K-AJ)(v_3-Av_2)}{t^2}+e^{-4U}Jv_2 \bigg)
\end{eqnarray}
and therefore, we have by integration:

\begin{eqnarray} \label{ee:u1int}
|u_1^2(s)-u_1^2(t_i)|&=&\Bigg|\int_{t_i}^s \Bigg( \frac{\alpha_t}{\alpha}u_1^2+\frac{2\sqrt{\alpha}u_1}{v_0} \bigg( e^{2\beta-2U}(\beta_\theta-U_\theta) \nonumber \\
&&\hbox{}+e^{2\beta-4U}(\beta_\theta-2U_\theta)v_2^2 \nonumber \\
&&\hbox{}+\frac{e^{2\beta}}{t^2}(v_3-Av_2)[(v_3-Av_2)\beta_\theta-A_\theta v_2] \bigg) \nonumber \\
&&\hbox{}+\frac{2e^{2\beta}u_1}{t}\bigg( \frac{(K-AJ)(v_3-Av_2)}{t^2}\nonumber \\
&&\hbox{}+e^{-4U}Jv_2 \bigg) \Bigg)ds'\Bigg|.
\end{eqnarray}
Let us estimate one by one the terms on the right-hand side of (\ref{ee:u1int}). 

The first term can be estimated using (\ref{es:alphat_vla}) as follows\footnote{Note the importance of the independence in $\theta$ of the right-hand side of (\ref{es:alphat_vla}) to perform the estimate along the characteritics.}, for $s < t_i$, we have

\begin{eqnarray}
\left| \int_{t_i}^s \frac{\alpha_t}{\alpha}u_1^2 ds'\right| &\le& \bigg|\int_{t_i}^s \left|\frac{\alpha_t}{\alpha} \bar{u}_1^2(s')\right| ds'\bigg|, \nonumber \\
 &\le& \bigg|\int_{t_i}^s \left(C(s)+D \ln(1+\bar{u}_1^2(s')\right)\bar{u}^2_1(s') ds' \bigg|  \label{es:term1},
\end{eqnarray}
where $C(s)$ is a non-negative function whose integral is uniformly bounded and $D$ is a non-negative constant.

To estimate the second term on the right-hand side of (\ref{ee:u1int}), we use equation (\ref{ee:betatheta_vla}) to obtain:

\begin{equation}
\sqrt{\alpha}|\beta_\theta| \le B g+D\bar{u}_1^2,
\end{equation}
for some constants $B$ and $D$ which depend on the bounds on $t$, $f$ and the support of $f$ in $v_2$ and $v_3$. Moreover, from the definition of $g$, we have:

\begin{eqnarray}
\sqrt{\alpha}|U_\theta| &\le& \frac{g}{2}+\frac{1}{2}, \\
\sqrt{\alpha}\frac{e^{2U}|A_\theta|}{t} &\le& 2g+\frac{1}{2}.
\end{eqnarray}
From the uniform bounds on $e^{2\beta-2U}$, $e^{2\beta}A$, the support of $f$ in $v_2$ and $v_3$, and from the estimate for $\beta_\theta$, we have, using  $|u_1| \le |v_0|$, that along a characteristic for which $f$ does not uniformly vanish:

\begin{eqnarray} \label{es:term2}
&&\bigg|\int_{t_i}^s \frac{2\sqrt{\alpha}u_1}{v_0} \bigg( e^{2\beta-2U}(\beta_\theta-U_\theta) \nonumber \\
&&+e^{2\beta-4U}(\beta_\theta-2U_\theta)v_2^2+\frac{e^{2\beta}}{t^2}(v_3-Av_2)[(v_3-Av_2)\beta_\theta-A_\theta v_2] \bigg)ds'\bigg| \nonumber \\
&\le& B+\bigg|\int_{t_i}^s \left( Dg+E\bar{u}_1^2 \right) ds'\bigg|.
\end{eqnarray}
for some constants $B$, $D$ and $E$.

Consider the last term on the right-hand side of (\ref{ee:u1int}). We have

\begin{eqnarray} 
\lefteqn{\bigg|\int_{t_i}^s \frac{2e^{2\beta}u_1}{t^3} \left( (K-AJ)(v_3-Av_2) +e^{-4U}Jv_2 \right) ds' \bigg|} \nonumber \\
&\le& \bigg|\int_{t_i}^s \frac{2e^\beta\bar{u}_1}{t_f^3}\bigg( \max_{\theta \in [0,1]}(e^{\beta}|K-AJ|)(t,.)|X+|A|X| \nonumber \\
&&\hbox{}+e^{\beta-2U}X \max_{\theta \in [0,1]}(e^{\beta-2U}|J|)(t,.) \bigg) ds'\bigg|
\end{eqnarray}
and thus, using lemmas \ref{lem:seccontrol}, \ref{lem:seccontrolA}, \ref{lem:intjcontrol}, \ref{lem:intkcontrol} and the inequality $2a \le a^2+1$ to replace $\bar{u}_1$, $\max_{\theta \in [0,1]}(e^{\beta}|K-AJ|)(t,.)$  and  $\max_{\theta \in [0,1]}(e^{\beta-2U}|J|)(t,.)$ by their respective squares, we obtain:

\begin{eqnarray} \label{es:term3}
&&\bigg| \int_{t_i}^s \frac{2e^{2\beta}u_1}{t^3} \left( (K-AJ)(v_3-Av_2) +e^{-4U}Jv_2 \right) ds'\bigg|  \nonumber  \\
&&\phantom{a}\quad \le B +\bigg|\int_{t_i}^s \bar{u}_1^2 F(s) ds'\bigg|,
\end{eqnarray}
where $B$ is a constant and $F(s)$ is a non-negative function whose integral is uniformly bounded.
Using (\ref{es:term1}), (\ref{es:term2}) and (\ref{es:term3}), we therefore obtain the following estimate for $\bar{u}_1$:

\begin{eqnarray} \label{es:baru1}
\bar{u}_1^2(t) &\le& B+\int_{t_i}^s \left(C(s)+B\ln(1+\bar{u}_1^2(s'))\right)\bar{u}_1^2 ds' \nonumber \\
&&+\int_{t_i}^s \left( Bg+B\bar{u}_1^2 \right) ds' +\int_{t_i}^s \bar{u}_1^2 F(s) ds'.
\end{eqnarray}
where $B$ is a non-negative constant and $C(s)$, $F(s)$ are non-negative function whose integrals are uniformly bounded.

These estimates are sufficient to obtain an upper bound on $\psi$. 
We first use equations (\ref{es:ghuder}) and (\ref{es:ghvder}) to do a null cone estimate for $g(t,\theta)$. For this let $(t,\theta)$ be in $(t_f,t_i] \times [0,1]$ and integrate (\ref{es:ghuder}) and (\ref{es:ghvder}) along the integral curves of $\partial_u$, $\partial_v$ ending at $(t,\theta)$. Adding the obtained equations, we have:

\begin{eqnarray}
2g(t, \theta) &\le& B+\int_u \left|\frac{\alpha_t}{\alpha}\right| \left(  \left( \frac{2g}{t}+1+2g \right)+\frac{3g}{t} \right)du' \nonumber \\
&&+\int_v  \left|\frac{\alpha_t}{\alpha}\right| \left(  \left( \frac{2g}{t}+1+2g \right)+\frac{3g}{t} \right)dv'.
\end{eqnarray}
where $B$ is a constant which depends on the maximum of $g$ on the initial hypersurface and is finite by compactness.
Using the estimate (\ref{es:alphat_vla}) and taking the maximum for $\theta$ in $[0,1]$, we obtain, for $t \in (t_f,t_i]$:

\begin{eqnarray} \label{es:maxG}
  \max_{\theta \in [0,1]}g(t,.) &\le& B+  \int^{t_i}_t \left(C(t')+B\ln(1+\bar{u}_1^2(t')\right)\max_{\theta \in [0,1]}g(t',.)dt'
\end{eqnarray}
where $B$ is a non-negative constant and $C(t)$ is a non-negative function whose integral is uniformly bounded.
Combining this with (\ref{es:baru1}), we derive the following estimate for $\psi$:

\begin{equation} \label{es:flnf}
\psi(t) \le B+\int^{t_i}_t F(s) \ln(\psi)(s)\psi(s) ds,
\end{equation}
where $B$ is non-negative constant and $F(s)$ is a non-negative function whose integral is uniformly bounded. From the last line it follows that:

\begin{eqnarray}
F \psi \ln \psi \bigg( B &+&\int_{t}^{t_i} F(s) \ln(\psi)(s)\psi(s) ds \bigg)^{-1} \nonumber \\
&&\cdot \left(\ln  \left( B + \int_{t}^{t_i} F(s) \ln(\psi)(s)\psi(s) ds \right)\right)^{-1}\le F(s)
\end{eqnarray}
and by integration of the last line, we obtain:
\begin{equation} \label{es:flnfi}
\psi(t) \le B^{\exp{ \int^{t_i}_t F(s)ds}}
\end{equation}
and since the integral is uniformly bounded, it follows that $\psi$ is uniformly bounded.
\end{proof}

\subsection{Continuous extension of the metric functions}
Now that $g$ and the support of $f$ have been proven to be uniformly bounded, it follows easily that:

\begin{lemma}
The first derivatives of $U$, $A$, $J$, $K$, together with  $\nu_t$, $\alpha_t$ are uniformly bounded on $(t_f,t_i] \times [0,1]$ and $U$, $A$, $\nu$, $\alpha$, $J$, $K$ admit continuous extension to $t=t_f$.
\end{lemma}

\subsection{Estimates for the derivatives of $f$, $\nu_\theta$, $\alpha_\theta$ and higher order estimates} \label{se:efho}
Such estimates follow by standard methods which can be found for instance in \cite{mw:ast2v}.

\subsection{The conclusion}
Since all metric functions, the Vlasov field and all their derivatives have been shown to be uniformly bounded, the assumptions of Proposition \ref{pro:cc} have been retrieved. In particular, the maximal Cauchy development cannot have $t_f>0$ which concludes the proof of Theorem \ref{th:t2vla}.

\section{Proof of Theorem \ref{th:t2vac}} \label{se:t2vac}
We will now proceed to the proof of Theorem \ref{th:t2vac}.

\subsection[The Einstein equations in areal coordinates]{The Einstein equations in areal coordinates for vacuum $T^2$-symmetric spacetimes}
The Einstein equations (\ref{ee}) for vacuum $T^2$-symmetric solutions reduce in areal coordinates to the following system of equations:

Constraint equations:
\begin{eqnarray}
\frac{\nu_t}{t}&=&U_t^2+\alpha U_\theta^2+\frac{e^{4U}}{{4t^2}}(A_t^2+\alpha A_\theta^2)+\frac{\alpha e^{2\nu}  K^2}{4t^4} +\alpha e^{2(\nu-U)} \Lambda, \label{ee:nut2}\\
\frac{\nu_\theta}{t}&=&2U_tU_\theta+\frac{e^{4U}}{2t^2}A_tA_\theta-\frac{\alpha_\theta}{2t \alpha}, \label{ee:nuthetav}\\
\frac{\alpha_t}{\alpha}&=& -4t\alpha e^{2(\nu-U)}\Lambda -\frac{\alpha e^{2\nu}  K^2} {t^3}. \label{ee:alphat2}
\end{eqnarray}
Evolution equations:
\begin{eqnarray}
\nu_{tt}-\alpha \nu_{\theta \theta}&=&\frac{\alpha_\theta \nu_\theta}{2}+\frac{\alpha_t \nu_t}{2 \alpha}-\frac{\alpha_\theta^2}{4\alpha}+\frac{\alpha_{\theta \theta}}{2}-U_t^2+\alpha U_\theta^2+\frac{e^{4U}}{4t^2}(A_t^2-A_\theta^2)\nonumber \\
&&-\frac{3\alpha e^{2\nu} K^2}{4t^4} +\alpha \Lambda e^{2(\nu-U)}, \\
U_{tt}-\alpha U_{\theta \theta}&=& - \frac{U_t}{t}+\frac{\alpha_\theta U_\theta}{2}+\frac{\alpha_t U_t}{2 \alpha}+\frac{e^{4U}}{2t^2}(A_t^2-\alpha A_\theta^2)+\alpha \Lambda e^{2(\nu-U)}, \label{ee:U}\\
A_{tt}-\alpha A_{\theta \theta}&=& \frac{A_t}{t}+\frac{\alpha_\theta A_\theta}{2}+\frac{\alpha_t A_t}{2 \alpha}-4(A_tU_t- \alpha A_\theta U_\theta). \label{ee:A}
\end{eqnarray}
Auxiliary equations:
\begin{eqnarray} \label{ee:aev}
0&=&G_t+AH_t, \\
0&=&H_t-\frac{\sqrt{\alpha}e^{2\nu}K}{t^3}. \label{ee:aev2}
\end{eqnarray}
Note that the Killing fields have been chosen such that the twist quantity $J$ vanishes and note that $K$ is a non-negative constant (see section \ref{se:t2spt2}).

Let us define the following replacement for the function $U$: 

\begin{eqnarray}
P&=&2U-\ln t.
\end{eqnarray}
We refer to the discussion in section \ref{spse:dpo} for an exposition of the motivation for the introduction of the quantity $P$.

The evolution equation for $U$ leads to the following equation for $P$:
\begin{eqnarray} \label{ee:P}
P_{tt}-\alpha P_{\theta \theta}= \left(-\frac{1}{t}+\frac{1}{2}\frac{\alpha_t}{\alpha}\right)P_t+\frac{\alpha_\theta P_\theta}{2}+e^{2 P}( A_t^2-\alpha A_\theta^2)-\frac{1}{2t^4}\alpha e^{2 \nu} K^2.
\end{eqnarray}
As mentioned in section \ref{spse:dpo}, we note that in the Gowdy case $K=0$, this equation is homogeneous, since there are no terms containing $\Lambda$ compared to equation (\ref{ee:U}). In the following, it will be useful to work both with $P$ and $U$ and to use two energy densities, one associated with the system of wave equations for $(U,A)$ and one associated with the system of wave equations for $(P,A)$.

\subsection{The universal cover of $\mathcal{M}/T^2$}
In section \ref{se:acac}, we will study the characteristic equation which defines null rays in areal coordinates. It will be easier to address this problem in the universal cover of the quotient of the spacetime. For any $T^2$-symmetric spacetimes $(\mathcal{M},g)$, we introduce $\mathcal{Q}=\mathcal{M}/T^2$, the quotient of the spacetime by the orbits of symmetry, and then define $\tilde{\mathcal{Q}}$ as the universal cover of $\mathcal{Q}$. Let $\pi_1: \mathcal{M} \rightarrow \mathcal{Q}$ be the natural projection from $\mathcal{M}$ to $\mathcal{Q}$. 

Suppose $(\mathcal{M},g)$ is foliated by areal coordinates with the metric taking the form (\ref{t2:metric}). Let $\alpha_Q$ be such that $\alpha$ is the pull-back of $\alpha_Q$ by $\pi_1^{*}$. We then define $\tilde{\alpha}$ to be the lift to $\tilde{\mathcal{Q}}$ of $\alpha_Q$. We may define similarly tilded functions for all metric functions, such as $\tilde{\nu}$, $\tilde{U}$, etc. Note that $\tilde{\mathcal{Q}}$ has topology $\mathbb{R} \times \mathbb{R}$ and admits areal coordinates $(\tilde{t},\tilde{\theta}) \in (t_f.t_i] \times \mathbb{R}$ and Lorentzian metric:
\begin{equation}
ds^2=-e^{2(\tilde{\nu}-\tilde{U})}(\tilde{\alpha}d\tilde{t}^2-d\tilde{\theta}^2).
\end{equation}
Note also that all tilde functions $\tilde{\nu}$, $\tilde{U}$, etc. are periodic in $\theta$ with period $1$ and that they satisfy the system of equations (\ref{ee:nut2})-(\ref{ee:aev2}) on $(t_f.t_i] \times \mathbb{R}$.

In the following, we will often\footnote{That is to say, we shall use the same symbol for a function defined on $\mathcal{M}$ and for its associated tilde function.}, by an abuse of notation\footnote{Note that strictly speaking, in the analysis of section \ref{se:t2vla}, all metric functions were also defined on $\mathcal{Q}$ rather than $\mathcal{M}$ since we had considered them to be function of $(t,\theta)$. The same remark applies for the analysis carried in section \ref{se:hypvla}.}, drop the tildes on the functions defined on $\tilde{\mathcal{Q}}$.

\subsection{The contradiction setting} \label{se:conts}
As explained in section \ref{se:rsp}, the proof will follow by contradiction. Let us thus assume that $(\mathcal{M},g)$ is the past maximal development of vacuum $T^2$-symmetric spacetimes with $\Lambda >0$ such that $t_0 > 0$. By proposition \ref{pro:gaf}, there exist a global areal foliation where the metric takes the form (\ref{t2:metric}) and such that $t$ lies in $(t_0,t_i]$. Thus, there exists functions $\alpha$, $\nu$, $U$, $A$ defined on $(t_0,t_i] \times [0,1]$ which are periodic in $\theta$ with period $1$, and a constant $K$ such that $\alpha$, $\nu$, $U$, $A$ and $K$ satisfy the system of equations (\ref{ee:nut2}), (\ref{ee:A}). 
Moreover, since the cases where $\Lambda=0$ have already been treated, and since the cases where $K=0$, $\Lambda>0$ may be treated by similar methods as we explained in the previous section, we will suppose that we are in the case where $K >0$ and $\Lambda >0$. Finally, let us assume that the assumptions of Theorem \ref{th:t2vla} hold, i.e.~the spacetime is not polarized.

\subsection{Uniform blow up of $\alpha$} \label{se:bupalpha}
The contradiction setting immediately implies the following:
\begin{lemma} \label{bup:alpha} Under the assumptions of section \ref{se:conts}, $\forall \theta \in [0,1]$, $\alpha(t,\theta) \rightarrow \infty$ as $t \rightarrow t_0$ and $\min_{\theta \in [0,1]} \alpha(t, \theta) \rightarrow \infty$ as $t \rightarrow t_0$.
\end{lemma}
\begin{proof}Suppose the lemma does not hold. Because of the monotonicity of $\alpha$, it follows that $\min_{\theta \in [0,1]} \alpha(t, \theta)$ is uniformly bounded, i.e.~results similar to those of section \ref{sec:basc} hold. We may then apply similar estimates as the estimates of sections \ref{se:eb} to \ref{se:efho}, replacing $f$ by $0$ everywhere. Indeed, the presence of the Vlasov matter was necessary only so as to ensure that the content of section \ref{sec:basc} is valid. Proposition \ref{pro:cc} then applies, and thus $(\mathcal{M},g)$ is not maximal, a contradiction.
\end{proof}
\begin{remark}
Since the rest of the proof of Theorem \ref{th:t2vac} will rely on the assumptions of section \ref{se:conts}, it will be from now on assumed that they hold.
\end{remark}

\subsection{The basic energy estimates} \label{se:tbee}
We will need to work with several energy densities and several energy integrals.
Let us thus define:
\begin{eqnarray}
g&=&U_t^2+\alpha U_\theta^2+ \frac{e^{4U}}{4t^2}\left( A_t^2+\alpha A_\theta^2 \right), \\
h&=&P_t^2+\alpha P_\theta^2+ e^{2P}\left( A_t^2+\alpha A_\theta^2 \right). \\
E_g(t)&=&\int_{[0,1]} \frac{g}{\sqrt{\alpha}} d\theta, \\
E_h(t)&=&\int_{[0,1]} \frac{h}{\sqrt{\alpha}} d\theta, \\
E_{h,K}(t)&=&E_h(t)+\int_{[0,1]}\frac{\sqrt{\alpha}e^{2 \nu}K^2}{t^4}d\theta, \\
E_{h,K,\Lambda}(t)&=&E_h(t)+\int_{[0,1]}\left(\frac{\sqrt{\alpha}e^{2 \nu}K^2}{t^4}+4\Lambda \frac{\sqrt{\alpha}e^{2 \nu-P}}{t}\right)d\theta.
\end{eqnarray}

Several computations will also be useful for the rest of the analysis. First, using the constraint equations (\ref{ee:nut2}) and (\ref{ee:alphat2}), we have the identities:
\begin{eqnarray}
\drond{}{t} \left(\frac{\sqrt{\alpha} e^{2 \nu-P}}{t} \right)&=& \frac{1}{2} \sqrt{\alpha}e^{2\nu-P}\left(h-\frac{1}{t^2}\right), \label{id:2}\\
\drond{}{t}\left( \sqrt{\alpha}e^{2 \nu} \right) &=&2 t \sqrt{\alpha}e^{2 \nu}g .\label{id:1}
\end{eqnarray}

Taking the time derivative of $E_h$ and using the Einstein equations, we obtain:

\begin{eqnarray}
\frac{d E_h}{dt}&=&-\frac{2}{t}\int_{[0,1]}\frac{P_t^2}{\sqrt{\alpha}}+e^{2P}\sqrt{\alpha} A_t^2 \nonumber\\
&&\hbox{}-2\Lambda \int_{[0,1]}\sqrt{\alpha}e^{2 \nu-P}h \nonumber\\
&&\hbox{}-\frac{2}{t^3}\int_{[0,1]}\sqrt{\alpha}e^{2 \nu}K^2g \nonumber\\
&&\hbox{}+\frac{1}{2t^5}\int_{[0,1]}\sqrt{\alpha}e^{2\nu}K^2 \nonumber\\
&&\hbox{}-\frac{2}{t}\int_{[0,1]}\sqrt{\alpha}P_{\theta \theta}+\frac{\alpha_\theta}{2 \sqrt{\alpha}}P_\theta.
\end{eqnarray}
The terms on the last line vanish thanks to the $\theta$ periodicity so we obtain\footnote{The fact that the terms involving derivatives in $\theta$ add up to an exact derivative is due to the wave map background structure of the equations. See \cite{bicm:gft2}.}:
\begin{eqnarray}
\frac{d E_h}{dt}&=&-\frac{2}{t}\int_{[0,1]}\frac{P_t^2}{\sqrt{\alpha}}+\frac{e^{2P}}{\sqrt{\alpha}} A_t^2 \nonumber\\
&&\hbox{}-2\Lambda \int_{[0,1]}\sqrt{\alpha}e^{2 \nu-P}h \nonumber\\
&&\hbox{}-\frac{2}{t^3}\int_{[0,1]}\sqrt{\alpha}e^{2 \nu}K^2g \nonumber\\
&&\hbox{}+\frac{1}{2t^5}\int_{[0,1]}\sqrt{\alpha}e^{2\nu}K^2. \label{eq:eht}
\end{eqnarray}
or written only in terms of $h$ and $P_t$, we have:

\begin{eqnarray}
\frac{d E_h}{dt}&=&-\frac{2}{t}\int_{[0,1]}\frac{P_t^2}{\sqrt{\alpha}}+\frac{e^{2P}}{\sqrt{\alpha}} A_t^2 \nonumber\\
&&\hbox{}-\int_{[0,1]}\frac{P_t}{\sqrt{\alpha}t^4}\alpha e^{2\nu}K^2 \nonumber\\
&&\hbox{}+\int_{[0,1]}\frac{1}{2}\frac{\alpha_t}{\alpha^{3/2}}h.
\end{eqnarray}
We see that the last term on the right-hand side of (\ref{eq:eht}) is competing against the others. 

\begin{remark} \label{rm:k0c}
In the case where $K=0$, the last term vanishes, thus, we obtain the desired monotonicity\footnote{Note that the parallelism between the cases $(K>0,\Lambda=0)$ and $(K=0,\Lambda>0)$ does not extend beyond the issue of the value of $t_0$. Indeed, once we know that $t_0=0$, the different powers of $t$ for the terms containing $\Lambda$ and $K$ in equation (\ref{ee:alphat2}) are likely to yield different asymptotics for the solutions.}  on $E_h$ and we could conclude as in \cite{iw:ast2}. Thus, we obtain the following proposition:

\begin{proposition} \label{prop:ck0}
Let $(\mathcal{M},g)$ be the maximal development of $T^2$-symmetric initial data in the vacuum with $\Lambda \ge 0$ and $K=0$. Suppose that $E_h$ does not vanish identically. Then $(\mathcal{M},g)$ admits a global foliation by areal coordinates with the time coordinate $t$ taking all values in $(0,\infty)$, i.e.~$t_0=0$ in the notation of Proposition \ref{pro:gaf}.
\end{proposition}
 \end{remark}
Unfortunately, in the general case, we lose this monotonicity and the analysis is, as we will see, more complex.

We may also compute the time derivative of $E_{h,K}$ and $E_{h,K,\Lambda}$:

\begin{eqnarray} \label{eq:ekt}
\frac{d E_{h,K}}{dt}&=&-\frac{2}{t}\int_{[0,1]}\frac{P_t^2}{\sqrt{\alpha}}+e^{2P}\sqrt{\alpha} A_t^2 \nonumber\\
&&\hbox{}-2\Lambda \int_{[0,1]}\sqrt{\alpha}e^{2 \nu-P}h \nonumber\\
&&\hbox{}-\frac{7}{2 t^5} \int_{[0,1]}\sqrt{\alpha}e^{2\nu}K^2
\end{eqnarray}
and 
\begin{eqnarray} \label{ee:dekl}
\frac{d E_{h,K,\Lambda}}{dt}&=&-\frac{2}{t}\int_{S^1[0,1]}\frac{P_t^2}{\sqrt{\alpha}}+e^{2P}\sqrt{\alpha} A_t^2 \nonumber\\
&&\hbox{}-\frac{7}{2t^5}\int_{[0,1]} \sqrt{\alpha}e^{2\nu}K^2 \nonumber\\
&&\hbox{}-2\Lambda \int_{[0,1]}\frac{\sqrt{\alpha}e^{2\nu-P}}{t^2}.
\end{eqnarray}
We see in particular that $E_{h,K}$ and $E_{h,K,\Lambda}$ are non-decreasing with decreasing time\footnote{Note that this monotonicity cannot be used as a replacement of the monotonocity of $E_g$ or $E_h$, since no estimate similar to (\ref{ineq:betaener}) can hold when $E_g$ is replaced by $E_{h,K}$ or $E_{h,K,\Lambda}$, as can be seen by studying homogeneous plane symmetric solutions.}.

We prove moreover the following:

\begin{lemma} \label{lem:bes}
$E_g$, $E_h$, $E_K$ and $E_{K, \Lambda}$ are uniformly bounded on $(t_0,t_i]$ and the last two quantities can be continuously extended to $t_0$.
\end{lemma}
\begin{proof}

From (\ref{ee:dekl}), we have in particular that:

\begin{equation}
\frac{d E_{h,K,\Lambda}}{dt} \ge -\frac{7}{2t} E_{K,\Lambda},
\end{equation}
which implies by application of Gronwall's lemma that $E_{h,K,\Lambda}$ is bounded uniformly if $t_0 > 0$. However, since:

\begin{equation}
E_h \le E_{h,K} \le E_{h,K,\Lambda},
\end{equation}
we also obtain a uniform bound on $E_h$ and $E_K$. Since moreover, $E_{h,K}$ and $E_{h,K,\Lambda}$ are monotonically increasing they admit strictly positive limits at $t=t_0$. A similar analysis implies the uniform bound on $E_g$.
\end{proof}

\subsection{Continuous extensions of the twist and cosmological energies}
In order to extract some information from the continuous extensions of $E_{h,K}$ and  $E_{h,K,\Lambda}$, we will need the following:

\begin{lemma} \label{lem:bdtce}
$\sqrt{\alpha}e^{2 \nu}$, $\sqrt{\alpha}e^{2 \nu-P}$ and therefore $\frac{\sqrt{\alpha}e^{2 \nu}K^2}{t^4}$ and $\Lambda \frac{\sqrt{\alpha}e^{2 \nu-P}}{t}$ admit continuous extensions to $t=t_0$ and are uniformly bounded in $(t_0,t_i] \times [0,1]$.
\end{lemma}
\begin{proof}The derivatives with respect to $t$ of $\sqrt{\alpha}e^{2 \nu}$ and $t^{-1/2}\sqrt{\alpha}e^{2 \nu-P}$ are positive, as can be verified by direction computation. Therefore, they are monotonically decreasing in the past direction and admit continuous extensions to $t=t_0$. Moreover, they are bounded by the maximum of their values on the initial data surface, which is finite by compactness.
\end{proof}

Since $\sqrt{\alpha}e^{2 \nu}$ and $t^{-1/2}\sqrt{\alpha}e^{2 \nu-P}$ are pointwise decreasing with $t$ in the past direction and are positive, their integrals over $\theta$ at fix $t$ are positive functions which are decreasing in the past direction and therefore, they admit a limit as $t$ goes to $t_0$. Thus we have the following:

\begin{lemma} \label{lem:linttce}
$\int_{[0,1]} \frac{\sqrt{\alpha}e^{2 \nu}K^2}{t^4}d\theta$ and $\int_{[0,1]} \Lambda \frac{\sqrt{\alpha}e^{2 \nu-P}}{t}d\theta$ admit continuous extensions to $t=t_0$. 
\end{lemma}

\subsection{Estimate for the the spatial derivatives of $\beta$ and $\beta-\frac{P}{2}$}
We define $\beta$ as in the Vlasov case by:

\begin{equation}
e^{2\beta}= \alpha e^{2\nu}.
\end{equation}
It follows as in lemma \ref{lem:thetacontrol} that $\beta_\theta$ is bounded by $\frac{g}{\sqrt{\alpha}}$:
\begin{equation} \label{ineq:betatheta}
|\beta_\theta| \le t \frac{g}{\sqrt{\alpha}}
\end{equation}
and by integration, we obtain:

\begin{lemma} \label{lem:bintsdbeta}
For all $t \in (t_0,t_i]$, 

\begin{equation}
\max_{[0,1]} \beta(t,.)-\min_{[0,1]} \beta(t,.) \le t_i E_g.
\end{equation}
In particular, $\max_{[0,1]} \beta(t,.)-\min_{[0,1]} \beta(t,.)$ is uniformly bounded.
\end{lemma}

We may do the same analysis using $h$ and $P$. First, we rewrite equation (\ref{ee:nuthetav}) as:

\begin{equation}
\beta_\theta-\frac{P_\theta}{2}=\frac{t}{2}\left( P_t P_\theta+e^{2P}A_t A_\theta\right),
\end{equation}
from which we obtain that

\begin{equation} \label{ineq:betaptheta}
\left|\beta_\theta-\frac{P_\theta}{2}\right| \le \frac{t}{4} \frac{h}{\sqrt{\alpha}}
\end{equation}
and therefore, using the bounds on $E_h$, we have the following lemma:

\begin{lemma} \label{lem:bintsdbetap}
For all $t \in (t_0,t_i]$, 

\begin{equation} 
\max_{[0,1]} ( 2\beta-P)(t,.)-\min_{[0,1]} ( 2\beta-P)(t,.) \le \frac{t_i}{2} E_h.
\end{equation}
In particular, $\max_{[0,1]} ( 2\beta-P)(t,.)-\min_{[0,1]} ( 2\beta-P)(t,.)$ is uniformly bounded.
\end{lemma}

\subsection{Limit of the gravitational energy of the orbits of symmetry} \label{se:lgeos}
We may then prove the following lemma:
\begin{lemma} \label{lem:nbfbe}
$\forall \epsilon > 0$, $\exists t_\epsilon > t_0$, such that either $E_h(t_\epsilon) \le \epsilon$ or  $E_g(t_\epsilon) \le \epsilon$.
\end{lemma}
\begin{proof}
Suppose the lemma does not hold. Then there exists an $\epsilon > 0$ such that $\min(E_h,E_g) > \epsilon$, $\forall t > t_0$. 

By integration of equation (\ref{eq:eht}), we have $\forall t \in (t_0,t_i]$:

\begin{eqnarray} \label{ineq:beta}
&&\int_{t}^{t_i} 2 \Lambda \int_{[0,1]} \sqrt{\alpha} e^{2\nu-P}h d\theta dt'+\int_{t}^{t_i}\frac{2K^2}{t^3}\sqrt{\alpha}e^{2\nu}g d\theta dt' \nonumber \\
&&\le E_h(t)-E_h(t_i)+\int^{t_i}_t\frac{1}{2t^5}\int_{[0,1]}\sqrt{\alpha}e^{2\nu}K^2.
\end{eqnarray}
Since all terms on the right-hand side are bounded by lemmas \ref{lem:bes} and \ref{lem:linttce}, we have in particular, that, there exits some constant $D>0$ such that:

\begin{equation}
\int_{t}^{t_i} 2 \Lambda \int_{[0,1]} \sqrt{\alpha} e^{2\nu-P}h d\theta dt \le D.
\end{equation}
Using the control on the spatial derivatives of $2\beta-P$ obtained in lemma \ref{lem:bintsdbetap}, we obtain the following, for all $(t,\theta) \in (t_0,t_i] \times [0,1]$:

\begin{eqnarray}
\int_{t}^{t_i} \int_{[0,1]} e^{2\beta-P} \frac{h}{\sqrt{\alpha}} d\theta ds &\le& B, \\
\int_{t}^{t_i} \min_{\theta' \in [0,1]} e^{2\beta-P}(s,.) E_h(s) ds &\le& B ,\\
\int_{t}^{t_i} \min_{\theta' \in [0,1]} e^{2\beta-P}(s,.) ds &\le& \frac{B}{\epsilon},\\
\int_{t}^{t_i} e^{2\beta-P}(s,\theta) ds &\le& \frac{B}{\epsilon}+B'(t_i-t), \\
\int_{t}^{t_i} e^{2 \beta-P}(s,\theta) ds &\le& B'', \label{bd:intbetap}
\end{eqnarray}
for some constants $B> 0$, $B'> 0$ and $B''> 0$.

Similarly, one obtain from inequality (\ref{ineq:beta}) and lemma \ref{lem:bintsdbeta} that their exists a constant $B'''>0$ such that, for all $(t,\theta) \in (t_0,t_i]$:

\begin{equation} \label{bd:intbeta}
\int_{t}^{t_i} e^{2 \beta}(s,\theta) ds \le B'''.
\end{equation}

It follows from (\ref{bd:intbetap}) and (\ref{bd:intbeta}) that the right-hand side of (\ref{ee:alphat2}) is bounded and by integration, $\ln \alpha$ and therefore $\alpha$ are uniformly bounded above, which contradicts lemma \ref{bup:alpha}.
\end{proof}
We may now prove a stronger version of the above result:

\begin{lemma}
$E_h \rightarrow 0$ as $t \rightarrow 0$ and $E_g \rightarrow 0$ as $t \rightarrow 0$.
\end{lemma}
\begin{proof} $E_h=E_K-\int_{[0,1]}\frac{\sqrt{\alpha}e^{2 \nu}K^2}{t^4}$. In view of lemmas \ref{lem:bes} and \ref{lem:linttce}, both terms on the right-hand side have a limit, thus $E_h$ has a limit. Similarly, $E_g$ has a limit. In view of the last lemma, both limits cannot be strictly positive and therefore at least one of them has to be zero. Suppose for instance, that $E_h$ tends to $0$ as $t$ tends to $t_0$. From the definition of $h$, $g$, $P$ and $U$ it follows that $g=\frac{h}{4}+\frac{P_t}{2t}+\frac{1}{4t^2}$ and therefore:
\begin{equation}
g \le \frac{h}{2}+\frac{1}{2t^2}.
\end{equation}
Since on the other hand, $\sqrt{\alpha}$ tends to infinity uniformly in $\theta$ by lemma \ref{bup:alpha}, it follows from the last inequality that $E_g$ also tends to $0$ as $t$ tends to $t_0$. The case where we know a priori that $E_g$ tends to $0$ and we need to deduce that $E_h$ tends to $0$ may be treated similarly.
\end{proof}

\subsection{Strong control on the spatial derivative of $\beta$} \label{se:scsd}
An immediate application of these limits is an improvement of lemmas \ref{lem:bintsdbeta} and  \ref{lem:bintsdbetap}:

\begin{lemma}
\begin{eqnarray}
\lim_{t \rightarrow t_0}\left( \max_{[0,1]} \beta(t,.)-\min_{[0,1]} \beta(t,.)\right)=0, \nonumber \\
\lim_{t \rightarrow t_0}\left( \max_{[0,1]} (2 \beta-P)(t,.)-\min_{[0,1]} (2\beta-P)(t,.) \right) =0.
\end{eqnarray}
\end{lemma}
From which it follows that:

\begin{lemma} \label{lem:scbeta}
$\forall \epsilon > 0$, there exists $t' >t_0$, such that for all $t \in (t_0, t']$:

\begin{eqnarray}
\max_{\theta \in [0,1]}e^{2\beta}(t,.) &\le&  e^{\epsilon}\min_{[0,1]} e^{2\beta}(t,.), \\
\max_{\theta \in [0,1]}e^{2\beta-P}(t,.)&\le&  e^{\epsilon}\min_{[0,1]} e^{2\beta-P}(t,.), \\
\max_{\theta \in [0,1]}\left(-\frac{\alpha_t}{\alpha}(t,.)\right) &\le&  e^{\epsilon} \min_{\theta \in [0,1]}\left(-\frac{\alpha_t}{\alpha}(t,.)\right).
\end{eqnarray}
\end{lemma}
\begin{proof}The first two inequalities follows directly from the last lemma. The last inequality follows from the first two, since the Einstein equation for $\alpha$, equation (\ref{ee:alphat2}) can be rewritten in terms of $\beta$ and $P$ as follows:
\begin{equation} \label{ee:alphatbp}
\frac{\alpha_t}{\alpha}=-4 e^{2\beta-P}\Lambda-\frac{e^{2\beta}K^2}{t^3}
\end{equation}
\end{proof}

Note that by integration, we could easily obtain from the last line that $\forall \epsilon > 0$, there exists $t' >t_0$ and a constant $C>0$ such that for all $t \in (t_0, t']$:
\begin{equation}
\max_{\theta \in [0,1]} \alpha(t,.) \le C \min_{\theta \in [0,1]} \alpha(t,.)^{1+\epsilon}.
\end{equation}
Unfortunately, the exponent of the right-hand side is not $1$ and this will not be sufficient for our analysis. Thus, we need a stronger estimate than this one, which we provide in the next section.

\subsection{An estimate for $\drond{}{\theta}\left( \ln(\alpha)\right)$}  \label{se:lnalphatheta}
The estimates on $\beta_\theta$ and $2\beta_\theta-P_\theta$ coming from the inequalities (\ref{ineq:betatheta}) and (\ref{ineq:betaptheta}) were based on previously known estimates for $T^2$-symmetric spacetimes written in areal coordinates. Here, we will derive a stronger estimate from these inequalities, using the identities (\ref{id:1}) and (\ref{id:2}) and the equation (\ref{ee:alphat2}). The estimate that we obtain is the following:

\begin{lemma}
There exists a constant $C>0$ such that $\forall (t,\theta) \in (t_0,t_i] \times [0,1]$, 

\begin{equation}
\left|\drond{}{\theta}\left( \ln(\alpha)\right)(t,\theta)\right| \le C.
\end{equation}
\end{lemma}
\begin{proof}
Multiplying (\ref{ineq:betatheta}) and (\ref{ineq:betaptheta}) by $e^{2\beta}$ and $e^{2\beta-P}$, we obtain:

\begin{eqnarray}
|\beta_\theta|e^{2\beta} &&\le t\frac{g}{\sqrt{\alpha}}e^{2\beta}=\frac{1}{2}\partial_t(\sqrt{\alpha}e^{2\nu}), \label{ineq:ebetatheta}\\
\left|\beta_\theta-\frac{P_\theta}{2}\right|e^{2\beta-P} &&\le \frac{t}{4}\frac{h}{\sqrt{\alpha}}e^{2\beta-P}=\frac{t^{1/2}}{2} \partial_t \left(t^{-1/2}\sqrt{\alpha}e^{2\nu-P} \right), \label{ineq:ebetaptheta}
\end{eqnarray}
where we have used the identities (\ref{id:1}) and (\ref{id:2}) arising from the constraints to rewrite the right-hand sides of the equations.

On the other end, from equation (\ref{ee:alphat2}), we have:

\begin{eqnarray}
-\drond{}{t}\left(\ln \alpha \right)=4\Lambda e^{2\beta-P}+\frac{K^2e^{2\beta}}{t^3}.
\end{eqnarray}
Thus, taking the $\theta$ derivative of the last equation, we obtain:

\begin{eqnarray}
-\drond{}{\theta}\left( \drond{}{t}\left(\ln \alpha \right) \right) =4\Lambda ( 2\beta_\theta-P_\theta)e^{2\beta-P}+ 2\beta_\theta \frac{K^2e^{2\beta}}{t^3}.
\end{eqnarray}

We now integrate the last line and commute the $\theta$ and $t$ partial derivatives in the integrand of the left-hand side to obtain, $\forall (t,\theta) \in (t_0,t_i] \times [0,1]$:

\begin{eqnarray}
\partial_\theta \ln \alpha (t,\theta)&=&\partial_\theta \ln \alpha (t_i,\theta)+\int^{t_i}_t 4\Lambda (2\beta_\theta-P_\theta)e^{2\beta-P}(s,\theta)ds \nonumber \\
&&+\int^{t_i}_t 2\beta_\theta \frac{K^2e^{2\beta}}{t^3}(s,\theta) ds.
\end{eqnarray}
Using (\ref{ineq:ebetatheta}) and (\ref{ineq:ebetaptheta}), we have:

\begin{eqnarray}
|\partial_\theta \ln \alpha(t,\theta)| &\le& \sup_{\theta \in [0,1]} |\partial_\theta \ln \alpha (t_i,.)| +\int^{t_i}_t 4\Lambda |(2\beta_\theta-P_\theta) e^{2\beta-P}|(s,\theta)ds \nonumber \\
&&+\int^{t_i}_t \left|2\beta_\theta \frac{K^2e^{2\beta}}{t^3}\right|(s,\theta) ds,\nonumber \\
 &\le& \sup_{\theta \in [0,1]} \left|\partial_\theta \ln \alpha (t_i,.)\right|+ 4\Lambda t_i^{1/2} \int^{t_i}_t \partial_t \left(t^{-1/2}\sqrt{\alpha}e^{2\nu-P} \right) \nonumber \\
&& \hbox{}+\frac{K^2}{t_0^3}\int^{t_i}_t \partial_t(\sqrt{\alpha}e^{2\nu}),
\end{eqnarray}
and the lemma follows from the uniform bounds on $\sqrt{\alpha}e^{2\nu}$ and $\sqrt{\alpha}e^{2\nu-P}$.
\end{proof}

By integration, we immediately obtain:
\begin{corollary} \label{cor:alphatheta}
There exists a constant $C>0$ such that, for all $t \in (t_0,t_i]$, we have:

\begin{equation}
\max_{\theta \in [0,1]} \alpha(t,\theta) \le C \min_{\theta \in [0,1]} \alpha(t,\theta).
\end{equation}
\end{corollary}

Combining this with lemma \ref{lem:bintsdbeta}, we may obtain:

\begin{corollary} \label{cor:alcobe}
There exist constants $M_1$ and $M_2$ such that for all $(t,\theta) \in (t_0,t_i]$, we have:

\begin{equation} \label{ineq:alcobe1}
M_1 \sqrt{\alpha}(t,\theta) \ge e^{2\beta}(t,\theta) \ge M_2\sqrt{\alpha} (t,\theta).
\end{equation}
Similarly, there exist constants $M_1'$ and $M_2'$ such that for all for all $(t,\theta) \in (t_0,t_i]$, we have:

\begin{equation}  \label{ineq:alcobe2}
M'_1 \sqrt{\alpha}(t,\theta) \ge e^{2\beta-P}(t,\theta) \ge M'_2\sqrt{\alpha} (t,\theta).
\end{equation}

\end{corollary}
\begin{proof}
Given that $E_{h,K}$ is non-decreasing in the past direction, that $E_h$ tends to zero as $t$ tends to $t_0$ and that $K >0$, it follows that the limit of $\int_{[0,1]}\sqrt{\alpha}e^{2\nu}d\theta$ is non-zero. This implies, using the monotonicity of $\sqrt{\alpha}e^{2\nu}$ as a function of $t$ and the monotone convergence theorem that there exists a $\theta_0$ and a constant $M>0$ such that $\sqrt{\alpha}e^{2\nu}(t,\theta_0) \ge M$ for all $t \in (t_0,t_i]$. Let $M'$ be an upper bound for $\sqrt{\alpha}e^{2\nu}(t,\theta_0)$. By lemma \ref{lem:bintsdbeta}, there exits a constant $M''$ such that, for all $(t,\theta) \in (t_0,t_i] \times [0,1]$:

\begin{equation}
e^{M''}e^{2\beta}(t,\theta_0) \ge e^{2\beta}(t,\theta) \ge  e^{-M''}e^{2\beta}(t,\theta_0)
\end{equation}
and thus:
\begin{equation}
M' e^{M''} \sqrt{\alpha}(t,\theta_0) \ge e^{2\beta}(t,\theta) \ge M  e^{-M''}\sqrt{\alpha}(t,\theta_0)
\end{equation}
Let $M'''$ be such that, for all $(t,\theta) \in t \in (t_i,t_0] \times [0,1]$

\begin{equation}
e^{M'''}\sqrt{\alpha}(t,\theta) \ge \sqrt{\alpha}(t,\theta_0) \ge e^{-M'''}\sqrt{\alpha}(t,\theta)
\end{equation}
then we have:
\begin{equation}
M'e^{M'''} e^{M''} \sqrt{\alpha}(t,\theta) \ge e^{2\beta}(t,\theta) \ge M e^{-M'''} e^{-M''}\sqrt{\alpha}(t,\theta)
\end{equation}
This proves the inequalities (\ref{ineq:alcobe2}). The second set of inequalities can be treated similarly, using $E_{g}$ and another energy integral

\begin{equation}
E_{g,\Lambda}=\int_{[0,1]} \left(\frac{g}{\sqrt{\alpha}}\\+\alpha e^{2(\nu-U)}\Lambda \right)d\theta,
\end{equation}
which may be easily proven to be non-decreasing in the past direction and uniformly bounded.
\end{proof}

The aim of the next two sections will be to describe the characteristics curves and the establish several estimates about their behaviour for $t$ close to $t_0$. 
We will actually not need to analyse all null curves, but only null curves orthogonal to the orbits of symmetry. Note that in the next sections, we will often, by an abuse of notation, denote by the same name functions defined on $\mathcal{M}$ or $\mathcal{Q}$ together with their lifts to $\tilde{\mathcal{Q}}$, the universal cover of $\mathcal{Q}$.

\subsection{An analysis of the characteristics in areal coordinates} \label{se:acac}
Consider a null curve $\gamma$ in $\mathcal{M}$ which is orthogonal to the orbits of symmetry and let $\tilde{\gamma}$ be the lift to $\tilde{\mathcal{Q}}$ of the projection to $\mathcal{Q}$ of $\gamma$. In null coordinates as those used in \cite{js:scct2pccm}, $\gamma$ is given by $u=constant$ or $v=constant$. In areal coordinates, we obtain $\gamma$ by solving the characteristic equation:
\begin{equation} \label{ee:char}
\Theta'(s)=\pm \sqrt{\alpha(s,\Theta(s))},
\end{equation}
with appropriate initial conditions. If $\Theta(t)$ is a solution to the above equation, then $\gamma$ is given in areal coordinates by $\left(t,\Theta(t)\right)$.

By standard arguments, solutions of $(\ref {ee:char})$ exist are smooth and unique on $(t_0,t]$ for any $t \in (t_0,t_i]$ once initial conditions have been fixed.

Consider now the characteristics parallel to the constant $v$ lines. They are parametrized by $\left(s,\Theta(s,\theta,t)\right)$, where $\Theta(s,\theta,t)$ satisfies:

\begin{equation}
\Theta(s,\theta,t)=\theta-\int^{s}_{t}\sqrt{\alpha(s',\Theta(s',\theta,t))}ds'.
\end{equation}
Take the $\theta$ derivative of the last line:

\begin{equation}
\Theta_\theta(s,\theta,t)=1-\int^{s}_{t}\frac{1}{2}\left(\frac{\alpha_\theta}{\sqrt{\alpha}}\right)\left(s',\Theta(s',\theta,t)\right)\Theta_\theta(s',\theta,t)ds'.
\end{equation}
Solving this equation implicitly, we see that:

\begin{equation} \label{eq:thetatheta}
\Theta_\theta(s,\theta,t)=\exp{\int^{t}_{s}\frac{1}{2}\left(\frac{\alpha_\theta}{\sqrt{\alpha}}\right)(s',\Theta(s,\theta,t))ds'}.
\end{equation}
We are naturally lead to estimate $\int^{t}_{s}\frac{1}{2}\left(\frac{\alpha_\theta}{\sqrt{\alpha}}\right)(s',\Theta(s',\theta,t))ds'$. This is the subject of the next section.

\subsection{Estimates for the integral along the characteristics of $\frac{\alpha_{\theta}}{\sqrt{\alpha}}$} \label{se:eacat}

\begin{lemma} \label{es:sqalphathetaint}
$\forall \epsilon >0$, there exists a $\bar{t}>t_0$, such that for all $t' \in (t_0,\bar{t}]$ there exists a negative constant $M_1$ and a positive constant $M_2$ such that, for all $(t,\theta) \in (t_0,t'] \times [0,1]$:
\begin{equation}
M_1-\epsilon \ln \alpha (t,\Theta(t,\theta,t')) \le  \int^{t'}_t -\frac{\alpha_{\theta}}{\sqrt{\alpha}}(s,\Theta(s,\theta,t')) ds \le M_2+\epsilon \ln \alpha (t,\Theta(t,\theta,t')).
\end{equation}
\end{lemma}
\begin{proof}
Let $\epsilon > 0$ and let $\bar{t} \in (t_0,t_i]$ be such that lemma \ref{lem:scbeta} holds in the following way: for all $(t,\theta,\theta') \in (t_0,\bar{t}] \times [0,1]^2$, 

\begin{equation} \label{ineq:scbeta2}
-(1-\epsilon)\frac{\alpha_t}{\alpha}(t,\theta') \le -\frac{\alpha_t}{\alpha}(t,\theta) \le -(1+\epsilon)\frac{\alpha_t}{\alpha}(t,\theta').
\end{equation}

Let $t' \in (t_0,\bar{t}]$ and let $\Theta(t,\theta,t')$ be a characteristic such that:

\begin{equation}
\Theta(t,\theta,t')=\theta-\int^t_{t'}\sqrt{\alpha}(s,\Theta(s,\theta,t'))ds.
\end{equation}

We have, for all $(t,\theta) \in (t_0,t'] \times [0,1]$, 

\begin{eqnarray} 
\lefteqn{\int^{t'}_t -\frac{\alpha_{\theta}}{\sqrt{\alpha}}(s,\Theta(s,\theta,t')) ds} \nonumber \\
&=&\int^{t'}_t \left( \frac{\alpha_t}{\alpha} -\frac{\alpha_{\theta}}{\sqrt{\alpha}}- \frac{\alpha_t}{\alpha}\right)(s,\Theta(s,\theta,t'))ds, \\
&=&\int^{t'}_t \frac{d}{ds}\left(\ln\alpha(s,\Theta(s,\theta,t') \right)ds \nonumber\\
&&\hbox{}-\int^{t'}_{t} \frac{\alpha_t}{\alpha}(s,\Theta(s,\theta,t'))ds. \label{es:intalphathetaalpha}
\end{eqnarray}
We now use (\ref{ineq:scbeta2}) to estimate the second integral on the right-hand side. Let $\theta_0$ be in $[0,1]$. Then we have:

\begin{eqnarray}
&&-\int^{t'}_{t} \frac{\alpha_t}{\alpha}(s,\Theta(s,\theta,t'))ds \ge -(1-\epsilon)\int^{t'}_{t} \frac{\alpha_t}{\alpha}(s,\theta_0)ds, \\
&&-\int^{t'}_{t} \frac{\alpha_t}{\alpha}(s,\Theta(s,\theta,t'))ds \ge -(1-\epsilon) \left(\ln \alpha(t',\theta_0)-\ln(\alpha(t,\theta_0)) \right).
\end{eqnarray}
Using Corollary  \ref{cor:alphatheta}, there exists a constant $M>0$ such that:

\begin{eqnarray} \label{es:intalphatalpha}
-\int^{t'}_{t} \frac{\alpha_t}{\alpha}(s,\Theta(s,\theta))ds \ge& -(1-\epsilon) \big(\ln \alpha(t',\Theta(t',\theta,t')) \nonumber \\
&\hbox{}-\ln\alpha(t,\Theta(t,\theta,t')) \big)-M.
\end{eqnarray}
Similarly, we obtain:

\begin{eqnarray}
-\int^{t'}_{t} \frac{\alpha_t}{\alpha}(s,\Theta(s,\theta,t'))ds \le& -(1+\epsilon) \big(\ln \alpha(t',\Theta(t',\theta,t')) \nonumber \\
&\hbox{}-\ln\alpha(t,\Theta(t,\theta,t')) \big)+M.
\end{eqnarray}

Thus we have, from (\ref{es:intalphathetaalpha}) and (\ref{es:intalphatalpha}):
\begin{eqnarray}
&&\ln\alpha(t',\Theta(t',\theta,t'))-\ln\alpha(t,\Theta(t,\theta,t'))-(1-\epsilon)\Big(\ln\alpha(t',\Theta(t',\theta,t')) \nonumber \\
&&-\ln\alpha(t,\Theta(t,\theta,t'))\Big)-M \le \int^{t'}_t -\frac{\alpha_{\theta}}{\sqrt{\alpha}}(s,\Theta(s,\theta,t')) ds \label{ineq:atcm}
\end{eqnarray}
and similarly
\begin{eqnarray}
&&\int^{t'}_t -\frac{\alpha_{\theta}}{\sqrt{\alpha}}(s,\Theta(s,\theta,t')) ds  \le\ln\alpha(t',\Theta(t',\theta,t'))-\ln\alpha(t,\Theta(t,\theta,t'))\nonumber \\
&&\hbox{}-(1+\epsilon)\Big(\ln\alpha(t',\Theta(t',\theta,t'))-\ln\alpha(t,\Theta(t,\theta,t')\Big)+M. \label{ineq:atcp}
\end{eqnarray}
The lemma follows by simplifying the terms containing $\alpha(t,\Theta(t,\theta))$ in (\ref{ineq:atcm}) and (\ref{ineq:atcp}).
\end{proof}

\subsection{Estimates for the integrals of small powers of $\alpha$} \label{se:eispa}
It will be useful for the derivation of pointwise energy estimates to have some control over the integral of $\alpha^p$ for small enough $p$. We first need the following result:

\begin{lemma}
There exists $\theta \in [0,1]$, such that:

\begin{equation}
 \lim_{t \rightarrow  t_0}\sqrt{\alpha}e^{2\nu}(t,\theta) > 0
\end{equation}
\end{lemma}
\begin{proof} Suppose that the lemma does not hold. Since $\sqrt{\alpha}e^{2\nu}(t,\theta)$ is a decreasing function of $t$, it must then tend to $0$ as $t$ tends to $t_0$ for any $\theta$. From the compactness of $[0,1]$ and using again the fact that $\sqrt{\alpha}e^{2\nu}$ is decreasing in $t$, it follows that $\int_{[0,1]} \sqrt{\alpha}e^{2\nu} d\theta $ tends to $0$ as $t$ tends to $t_0$. This contradicts the facts that $E_K$ tends to a strictly positive value by monotonicity and $E_h$ has limit $0$.
\end{proof}
We may then obtain the following:

\begin{lemma} \label{lem:intspa}
For all $p <1/2$, there exists a function $B(t')$ such that $B(t') \rightarrow 0$ as $t' \rightarrow t_0$ and such that for all $(t,\theta) \in (t_0,t'] \times [0,1]$, with $t'>t_0$, we have

\begin{equation}
\int^{t'}_{t} \alpha^p(s,\theta)ds \le B(t').
\end{equation}
\end{lemma}
\begin{proof}
Let $\theta_0 \in [0,1]$ be such that the previous lemma holds, and thus such that $\sqrt{\alpha}e^{2\nu}(.,\theta_0)$ is bounded from below by a stricly positive constant on $(t_0,t']$.

We then rewrite equation (\ref{ee:alphat2}) as follows:

\begin{equation}
-(1/2-p)\frac{\alpha_t}{\alpha^{3/2-p}}=(1/2-p)\alpha^p f(t,\theta_0),
\end{equation}
where $f(t,\theta_0)=4 \Lambda\sqrt{\alpha} e^{2\nu-P}+\frac{\sqrt{\alpha}e^{2\nu} K^2}{t^3}$ is a function bounded from below by a strictly positive constant.
Integrating the last equation, we obtain:

\begin{equation}
\int_t^{t'}\frac{d }{dt}\left( \alpha^{p-1/2}\right)ds=\int^{t'}_t (1/2-p)\alpha^p f(s,\theta_0) ds.
\end{equation}
Using the lower bound on $f(s,\theta)$, we therefore obtain that:

\begin{equation}
\int^{t'}_t \alpha^p(s,\theta_0) ds \le \frac{C}{\alpha^{1/2-p}(t',\theta_0)},
\end{equation}
for some constant $C>0$.
The lemma then follows by application of Corollary \ref{cor:alphatheta} of section \ref{se:lnalphatheta} and the fact that $\lim_{t' \rightarrow t_0} \alpha(t',\theta_0)=+\infty$.
\end{proof}

From equation (\ref{eq:eht}), we have seen that $E_h$ is a priori not monotonic. In the next section, we will analyse an energy integral associated with the polarization function $A$. The advantage of this energy integral over $E_h$ is that, as the wave equation for $A$ is homogeneous, we will be able to extract useful information from the sign of $\frac{dE_A}{dt}$.
\subsection{Analysis of the polarization energy} \label{se:ape}

Define the energy associated with the wave equation for $A$ as:

\begin{equation}
E_A=\int_{[0,1]} \frac{e^{2P}}{\sqrt{\alpha}}\left( A_t^2+\alpha A_\theta^2\right) d\theta.
\end{equation}
Since by definition $E_A \le E_h$, we immediately obtain that $E_A \rightarrow 0$, when $t \rightarrow t_0$. The aim of this section is to extract some information from this remark.
Note that the wave equation for $A$, equation (\ref{ee:A}), may also be written as:

\begin{eqnarray}
\partial_t \left( \frac{t e^{2P} A_t}{\sqrt{\alpha}}\right)-\partial_\theta \left( t e^{2P} \sqrt{\alpha} A_\theta \right)=0
\end{eqnarray}

Let us  first compute the time derivative of $E_A$:

\begin{eqnarray}
\frac{dE_A}{dt}&=&\int_{[0,1]} \drond{}{t}\left( \frac{e^{2P}}{\sqrt{\alpha} }A_t^2 \right)+\drond{}{t}\left( e^{2P}\sqrt{\alpha}A_\theta^2 \right), \nonumber\\
&=&\int_{[0,1]} A_t \partial_t \left( \frac{e^{2P}}{\sqrt{\alpha}}A_t \right)+A_{tt} \frac{e^{2P}}{\sqrt{\alpha}}A_t \nonumber \\
&&\hbox{}+A_\theta^2 \partial_t(e^{2P}\sqrt{\alpha})+2 A_{\theta t} A_\theta e^{2P}\sqrt{\alpha}, \nonumber \\
&=&\int_{[0,1]} A_t \left( \partial_\theta( e^{2P}\sqrt{\alpha}A_\theta)-\frac{e^{2P}A_t}{t\sqrt{\alpha}}\right)\nonumber \\
&&+\Big( \alpha A_{\theta \theta}+(-\frac{1}{t}+\frac{\alpha_t}{2\alpha} )A_t+\demi \alpha_\theta A_\theta \nonumber\\
&&\hbox{}-2(A_t P_t-\alpha A_\theta P_\theta) \Big) \frac{e^{2P}}{\sqrt{\alpha}}A_t \nonumber\\
&&\hbox{}+2P_t e^{2P} \sqrt{\alpha}A_\theta^2+\demi\frac{\alpha_t}{\sqrt{\alpha}}e^{2P} A_\theta^2+2 A_{\theta t} A_\theta e^{2P}\sqrt{\alpha}, \nonumber\\
&=&\int_{[0,1]} -\frac{1}{t}\frac{e^{2P}A_t^2}{\sqrt{\alpha}}+A_t \partial_\theta ( e^{2P} \sqrt{\alpha}A_\theta ) \nonumber\\
&&\hbox{}+\frac{e^{2P}}{\sqrt{\alpha}}A_t \alpha A_{\theta \theta}\nonumber \\
&&-\frac{1}{t}\frac{e^{2P}A_t^2}{\sqrt{\alpha}}+\frac{\alpha_t}{2\alpha}\frac{e^{2P}A_t^2}{\sqrt{\alpha}} \nonumber\\
&&\hbox{}+\demi \alpha_\theta A_\theta \frac{e^{2P}A_t}{\sqrt{\alpha}} \nonumber\\
&&\hbox{}-2 A_t^2P_t \frac{e^{2P}}{\sqrt{\alpha}}+2\alpha P_\theta A_\theta\frac{e^{2P}A_t}{\sqrt{\alpha}} \nonumber\\
&&\hbox{}+2P_t e^{2P}\sqrt{\alpha}A_\theta^2 \nonumber\\
&&\hbox{}+\demi \frac{\alpha_t}{\sqrt{\alpha}} e^{2P}A_\theta^2+2A_{\theta t}A_\theta e^{2P} \sqrt{\alpha}, \nonumber\\
&=&\int_{[0,1]} -2\frac{e^{2P} A_t^2}{\sqrt{\alpha}} \nonumber\\
&&\hbox{}+2\partial_\theta (A_t e^{2P} \sqrt{\alpha} A_\theta ) \nonumber\\
&&\hbox{}+\left(\frac{\alpha_t}{2\alpha}-2P_t\right)\frac{e^{2P}A_t^2}{\sqrt{\alpha}} \nonumber\\
&&\hbox{}+\left(\frac{\alpha_t}{2\alpha}+2 P_t\right) e^{2P}\sqrt{\alpha} A_\theta^2.
\end{eqnarray}

Since the second term vanishes due to the periodicity, we obtain:
\begin{eqnarray} \label{ee:tdea}
\frac{dE_A}{dt}&=&\int_{[0,1]} -2\frac{e^{2P} A_t^2}{\sqrt{\alpha}} \nonumber\\
&&+\left(\frac{\alpha_t}{2\alpha}-2P_t\right)\frac{e^{2P}A_t^2}{\sqrt{\alpha}} \nonumber\\
&&+\left(\frac{\alpha_t}{2\alpha}+2 P_t\right) e^{2P}\sqrt{\alpha} A_\theta^2 \label{eq:eat}.
\end{eqnarray}

Note that by assumption, the spacetime is not polarized and thus $E_A$ cannot identically vanish on any Cauchy surface, in particular, on any surface of constant $t$. Now, if there exists $t' \in (t_0,t_i]$, such that for all $(t,\theta) \in (t_0,t'] \times [0,1]$, both $\frac{\alpha_t}{2\alpha}\pm P_t \le 0$, it follows that $E_A$ is increasing in the past direction, which contradicts the fact that $E_A \rightarrow 0$ as $t \rightarrow t_0$. We are lead to the following:

\begin{lemma} \label{lem:ubpt}
There exists a constant $C >0 $ and a sequence of points $(t_n,\theta_n)$ in $(t_0,t_i] \times [0,1]$, with $t_n \rightarrow t_0$, as $n \rightarrow +\infty$ such that $\frac{|P_t|}{\sqrt{\alpha}}(t_n,\theta_n)\ge C $.
\end{lemma}

\begin{proof}
As explain above, we have a sequence of points $(t_n,\theta_n)$ such that $|P_t|+\frac{\alpha_t}{2\alpha} \ge 0$ otherwise $E_A$ is increasing for $t$ close to $t_0$. From Corollary \ref{cor:alcobe} of section \ref{se:lnalphatheta} and equation (\ref{ee:alphatbp}), there exists a constant $M>0$ such that, for all $(t,\theta) \in (t_0,t_i] \times [0,1]$:

\begin{equation}
\frac{\alpha_t}{\alpha}(t,\theta) \le - M \sqrt{\alpha}(t,\theta),
\end{equation}
from which we obtain:

\begin{equation}
\left(|P_t|-\frac{M}{2}\sqrt{\alpha}\right)(t_n,\theta_n) \ge 0,
\end{equation}
which proves the lemma.
\end{proof}

The set of points we have just obtained will be used as initial data for some null cone estimates, where the aim will be to estimate from below the energy density $h$. However, we will need to treat some of the non-linear terms as error terms, and for this, it will be necessary to first control $h$ from above, which is the subject of the next section.

\subsection{Pointwise null cone energy estimates: control from above} \label{se:pnceeca}

We introduce the energy density:
\begin{eqnarray}
h^{\times}&=& 2 \sqrt{\alpha} P_t P_\theta+2 e^{2P} \sqrt{\alpha} A_t A_\theta.
\end{eqnarray}

Let us compute the sum and the difference of $h$ and $h^{\times}$:

\begin{eqnarray}
h+h^{\times}&=& (P_t+\sqrt{\alpha} P_\theta)^2+e^{2P}(A_t+\sqrt{\alpha}A_\theta)^2, \\
h-h^{\times}&=& (P_t-\sqrt{\alpha} P_\theta)^2+e^{2P}(A_t-\sqrt{\alpha}A_\theta)^2. 
\end{eqnarray}

Define:

\begin{eqnarray}
D_u&=&\partial_t -\sqrt{\alpha} \partial_\theta, \\
D_v&=&\partial_t +\sqrt{\alpha} \partial_\theta, \\
P_u&=&D_u P, \quad P_v=D_v P, \\
A_u&=&D_u A, \quad A_v=D_v A.
\end{eqnarray}

With this notation, we have:

\begin{eqnarray}
h+h^{\times}&=&P_v^2+e^{2P}A_v^2, \label{eq:h+hcrossv}\\
h-h^{\times}&=&P_u^2+e^{2P}A_u^2.
\end{eqnarray}

We may also rewrite the wave equations (\ref{ee:P}) and (\ref{ee:A}) for $P$ and $A$ as follows\footnote{Note that $D_u D_v=D_v D_u+\frac{\alpha_t}{\sqrt{\alpha}}\partial_\theta$}:

\begin{eqnarray}
D_u D_v P &=&\frac{\alpha_t}{2 \alpha}P_v-\frac{1}{2t}(P_u +P_v)+e^{2P}A_u A_v-\frac{1}{2t^4}\alpha e^{2\nu}K^2, \\
D_v D_u P &=&\frac{\alpha_t}{2 \alpha}P_u-\frac{1}{2t}(P_u +P_v)+e^{2P}A_u A_v-\frac{1}{2t^4}\alpha e^{2\nu}K^2, \\
D_u D_v A &=&\frac{\alpha_t}{2 \alpha}A_v-\frac{1}{2t}(A_u +A_v)-A_uP_v-A_vP_u, \\
D_v D_u A &=&\frac{\alpha_t}{2 \alpha}A_u-\frac{1}{2t}(A_u +A_v)-A_uP_v-A_vP_u. 
\end{eqnarray}

We have:

\begin{eqnarray}
D_u(h+h^{\times})&=& \left(-\frac{1}{t}+\frac{\alpha_t}{\alpha}\right)(P_v^2+e^{2P}A_v^2)-\frac{1}{t}(P_u P_v+e^{2P}A_v A_u)\nonumber\\
&&-\frac{P_v}{t^4}\alpha e^{2\nu}K^2, \\
D_v(h-h^{\times})&=& \left(-\frac{1}{t}+\frac{\alpha_t}{\alpha}\right)(P_u^2+e^{2P}A_u^2)-\frac{1}{t}(P_u P_v+e^{2P}A_v A_u)\nonumber\\
&&-\frac{P_u}{t^4}\alpha e^{2\nu}K^2,
\end{eqnarray}
i.e.~we have:
\begin{eqnarray}
D_u(h+h^{\times})&=& \left(-\frac{1}{t}+\frac{\alpha_t}{\alpha}\right)(h+h^{\times})-\frac{1}{t}(P_u P_v+e^{2P}A_v A_u) \label{eq:g+hu} \nonumber\\
&&-\frac{P_v}{t^4}\alpha e^{2\nu}K^2,\\
D_v(h-h^{\times})&=& \left(-\frac{1}{t}+\frac{\alpha_t}{\alpha}\right)(h-h^{\times})-\frac{1}{t}(P_u P_v+e^{2P}A_v A_u))\nonumber\\
&&-\frac{P_u}{t^4}\alpha e^{2\nu}K^2. \label{eq:g-hv}
\end{eqnarray}

We will prove, using null cone estimates, the following:
\begin{lemma} \label{lem:cah}
$\forall \epsilon > 0$, there exists a constant $B > 0$, a $t' > t_0$ and a $\theta_0 \in [0,1]$ such that for all $t' \ge t > t_0$, 

\begin{equation}
\sup_{\theta \in [0,1]} h(t,.) \le B \alpha^{1+\epsilon}(t,\theta_0).
\end{equation}
\end{lemma}
\begin{proof}
Let $t \in (t_0,t_i]$ and let $\Theta(s,\theta,t)$ denote a solution of the characteristic equation with initial conditions $\Theta(t,\theta,t)=\theta$ such that $(s,\Theta(s,\theta,t))$ correponds to a constant $v$ line in null coordinates, as introduced in section \ref{se:acac}. 

We have:
\begin{eqnarray}
\frac{\partial}{\partial s}\left( (h+h^{\times})(s,\Theta(s,\theta,t)) \right)&=&\drond{(h+h^{\times})}{t}-\sqrt{\alpha}\drond{(h+h^{\times})}{\theta}, \nonumber \\
&=&D_u (h+h^{\times})(s,\Theta(s,\theta,t))
\end{eqnarray}
and therefore equation (\ref{eq:g+hu}) can be rewritten as follows, for any $t' >t_0$:

\begin{eqnarray}
&&\frac{\partial}{\partial s} \left( (h+h^{\times})(s,\Theta(s,\theta,t))\exp{\int^{t'}_s\left(-\frac{1}{s'}+\frac{\alpha_t}{\alpha} \right)(s',\Theta(s',\theta,t))ds'}\right)=\nonumber \\
&& \left( \exp{\int^{t'}_{s}\left(-\frac{1}{s'}+\frac{\alpha_t}{\alpha} \right)(s',\Theta(s',\theta,t))ds'} \right)\phi(s,\Theta(s,\theta,t)),
\end{eqnarray}
where 
\begin{displaymath}
\phi=-\frac{1}{s}(P_u P_v+e^{2P}A_v A_u)-\frac{P_v}{s^4}\alpha e^{2\nu}K^2.
\end{displaymath}

Let $t' \ge t > t_0$ and integrate the last line between $t'$ and $t$ to obtain:
\begin{eqnarray}
&&(h+h^{\times})(t',\Theta(t',\theta,t))\nonumber \\
&&\hbox{}-(h+h^{\times})(t,\theta)\exp \int^{t'}_t\left(-\frac{1}{s'}+\frac{\alpha_t}{\alpha} \right)(s',\Theta(s',\theta,t))ds'=\nonumber \\
&&\int^{t'}_t \Bigg[ \left( \exp{\int^{t'}_{s}\left(-\frac{1}{s'}+\frac{\alpha_t}{\alpha} \right)(s',\Theta(s',\theta,t))ds'} \right) \nonumber \\
&&\phantom{a} \quad  \phantom{a} \quad\cdot \phi(s,\Theta(s,\theta,t)) \Bigg]ds. \label{eq:g+hint}
\end{eqnarray} 
Let $\epsilon > 0$ and fix a $\theta_0$ in $[0,1]$. Assume $t'$ is such that lemma $\ref{lem:scbeta}$ holds in the following sense: for all $(t,\theta) \in (t_0,t'] \times [0,1]$, 

\begin{eqnarray} \label{ineq:betascontrolrw}
(1+\epsilon)\frac{\alpha_t}{\alpha}(t,\theta_0) \le \frac{\alpha_t}{\alpha}(t,\theta) \le (1-\epsilon)\frac{\alpha_t}{\alpha}(t,\theta_0) 
\end{eqnarray}
which implies the following estimates:

\begin{eqnarray} \label{es:intalphat1}
&&\frac{t}{t'} \left( \frac{\alpha(t',\theta_0)}{\alpha(t,\theta_0)} \right)^{1+\epsilon} \le \exp \int^{t'}_t\left(-\frac{1}{s'}+\frac{\alpha_t}{\alpha} \right)(s',\Theta(s',\theta,t))ds' 
\end{eqnarray}
and 
\begin{eqnarray} \label{es:intalphat2}
&& \exp \int^{t'}_t\left(-\frac{1}{s'}+\frac{\alpha_t}{\alpha} \right)(s',\Theta(s',\theta,t))ds' \le \frac{t}{t'} \left( \frac{\alpha(t',\theta_0)}{\alpha(t,\theta_0)} \right)^{1-\epsilon}.
\end{eqnarray}

Define $F(s,\theta,t)$ by:

\begin{equation}
F(s,\theta,t)=(h+h^{\times})(s,\Theta(s,\theta,t))\frac{s}{t'} \left( \frac{\alpha(t',\theta_0)}{\alpha(s,\theta_0)} \right)^{1+\epsilon}.
\end{equation}
From (\ref{eq:g+hint}), (\ref{es:intalphat1}) and (\ref{es:intalphat2}), we have:

\begin{eqnarray} \label{ineq:F}
F(t,\theta,t) \le F(t',\theta,t)+\int^{t'}_t \frac{s}{t'} \left( \frac{\alpha(t',\theta_0)}{\alpha(s,\theta_0)} \right)^{1-\epsilon}|\phi(s,\Theta(s,\theta,t))|ds.
\end{eqnarray}
We will now estimate the second term on the right-hand side of the last inequality. First note that:

\begin{eqnarray}
|\phi(s,\Theta(s,\theta,t))|&&=\left|-\frac{1}{s}(P_u P_v+e^{2P}A_v A_u)-\frac{P_v}{s^4}\alpha e^{2\nu}K^2\right|, \\
 &&\le \frac{h}{2s}+\sqrt{h+h^{\times}}\frac{e^{2\beta}K^2}{s^4}.
\end{eqnarray}

Thus we have:

\begin{eqnarray}
 \left( \frac{\alpha(t',\theta_0)}{\alpha(s,\theta_0)} \right)^{1-\epsilon}\phi(s, \Theta(s,\theta)) \le  \left( \frac{\alpha(t',\theta_0)}{\alpha(s,\theta_0)} \right)^{1-\epsilon}\frac{1}{2s}h \nonumber \\
+ \left( \frac{\alpha(t',\theta_0)}{\alpha(s,\theta_0)} \right)^{1-\epsilon}\sqrt{h+h^{\times}}\frac{e^{2\beta}K^2}{s^4}.\label{es:phialpha}
\end{eqnarray}
The second term on the right-hand side of this last line may then be rewritten in terms of $F(s,\theta,t)$:

\begin{eqnarray}
\lefteqn{ \left( \frac{\alpha(t',\theta_0)}{\alpha(s,\theta_0)} \right)^{1-\epsilon}\sqrt{h+h^\times}\frac{e^{2\beta}K^2}{s^4}=} \nonumber \\
&&\left( \frac{\alpha(t',\theta_0)}{\alpha(s,\theta_0)} \right)^{1/2-(3\epsilon)/2}\frac{e^{2\beta}K^2}{s^4}\left(\frac{t'}{s}\right)^{1/2}\sqrt{F(s,\theta,t)}.
\end{eqnarray}
Moreover, from lemmas \ref{lem:bintsdbeta} and \ref{lem:intspa} and corollary \ref{cor:alcobe}, for $\epsilon$ small enough we have:

\begin{equation} \label{es:stphi}
 \int^{t'}_t \left( \frac{\alpha(t',\theta_0)}{\alpha(s,\theta_0)} \right)^{1/2-(3\epsilon)/2}\frac{e^{2\beta}K^2}{s^4}\left(\frac{t'}{s}\right)^{1/2} ds \le  \int^{t'}_t C \alpha^{(3 \epsilon)/2}(s,\theta_0)ds \le M,
\end{equation}
for some constant $M >0$.
We now use estimates of the type found in \cite{js:scct2pccm}. Let $t_m$ be such that $F(t_m,\theta,t_m)$ is a maximum of $F(s,\theta,s)$ with $s \in [t,t']$. Note the trivial fact that $\sup_{[t,t']}F(.,\theta,.)=F(t_m,\theta,t_m)=\sup_{[t_m,t']}F(.,\theta,.)$. It follows from (\ref{ineq:F}), (\ref{es:phialpha}) and (\ref{es:stphi}) that:

\begin{eqnarray} \label{ineq:ftm}
F(t_m,\theta,t_m) &\le& F(t',\theta,t_m)+\sqrt{F(t_m,\theta,t_m)}M+\nonumber \\
&&\int^{t'}_{t_m}\frac{s}{t'}\left( \frac{\alpha(t',\theta_0)}{\alpha(s,\theta_0)}\right)^{1-\epsilon}\frac{h(s,\Theta(s,\theta,t_m))}{2s} ds,
\end{eqnarray}
for some constant $M>0$. Note that $F(t',\theta,t_m)$ is uniformly bounded since by definition:

\begin{equation}
F(t',\theta,t_m)=(h+h^\times)(t',\Theta(t',\theta,t_m)) \le \sup_{\theta \in [0,1]} (h+h^\times)(t',.) \le C
\end{equation}
for some constant $C>0$. Thus, we have from (\ref{ineq:ftm}), 

\begin{eqnarray}
F(t_m,\theta,t_m) &\le& C+\sqrt{F(t_m,\theta,t_m)}M\nonumber \\
&&\hbox{}+\int^{t'}_{t_m}\frac{s}{t'}\left( \frac{\alpha(t',\theta_0)}{\alpha(s,\theta_0)}\right)^{1-\epsilon}\frac{h(s,\Theta(s,\theta,t_m))}{2s} ds,
\end{eqnarray}
We interpret the last line as an inequality for a second order polynomial equation in $\sqrt{F(t_m,\theta,t_m)}$. Thus $\sqrt{F(t_m,\theta,t_m)}$ must lie between the roots of this polynomial and we obtain easily that:

\begin{eqnarray}
F(t_m,\theta,t_m) \le B+C\int^{t'}_{t_m}\left( \frac{\alpha(t',\theta_0)}{\alpha(s,\theta_0)}\right)^{1-\epsilon}h(s,\Theta(s,\theta,t_m)) ds,
\end{eqnarray}
for some constants $B>0$ and $C>0$ independent of $\theta$. Since $F(t,\theta,t) \le F(t_m,\theta,t_m)$ and since $t \le t_m$, we have:

\begin{eqnarray}
F(t,\theta,t) \le B+C\int^{t'}_{t}\left( \frac{\alpha(t',\theta_0)}{\alpha(s,\theta_0)}\right)^{1-\epsilon}h(s,\Theta(s,\theta,t_m)) ds.
\end{eqnarray}
Taking the maximum over all $\theta \in [0,1]$, it follows that:
\begin{eqnarray} \label{es:h+hcross}
&&\left( \sup_{\theta \in [0,1]}(h+h^{\times})(t,.)\right) \frac{t}{t'} \left( \frac{\alpha(t',\theta_0)}{\alpha(t,\theta_0)}\right)^{1+\epsilon} \le B \nonumber \\
&&+ C\int^{t'}_{t}\left( \frac{\alpha(t',\theta_0)}{\alpha(s,\theta_0)}\right)^{1-\epsilon}\sup_{\theta \in [0,1]}(h(s,.)) ds.
\end{eqnarray}
A similar estimate may be obtained using $h-h^{\times}$ and equation (\ref{eq:g-hv}). Adding the estimate for $h-h^{\times}$ to (\ref{es:h+hcross}), we obtain easily that:
\begin{eqnarray}
&&\left( \sup_{\theta \in [0,1]}h(t,.)\right) \frac{t}{t'} \left( \frac{\alpha(t',\theta_0)}{\alpha(t,\theta_0)}\right)^{1+\epsilon} \le B \nonumber  \\
&&\hbox{}+ C\int^{t'}_{t}\left( \frac{\alpha(t',\theta_0)}{\alpha(t,\theta_0)}\right)^{1-\epsilon}\sup_{\theta \in [0,1]}(h(s,.)) ds.
\end{eqnarray}
Applying Gronwall's lemma to the last line, together with lemma \ref{lem:intspa}, completes the proof of the lemma.
\end{proof}

\subsection{Pointwise null cone energy estimates: control from below} \label{se:pnceecb}
With the control from above for $h$ that we have just obtained, we may now prove an estimate from below for $h$ if we are given appropriate initial data:

\begin{lemma} \label{lem:cfb}
Suppose that there exists a constant $B>0$ and a sequence of points $(t_{n},\theta_n)$ with $t_n \rightarrow t_0$ as $n \rightarrow +\infty$, such that, for all $n$, $\frac{|P_{v}|}{\sqrt{\alpha}}(t_n,\theta_n) > B$. Then for all $\epsilon>0$, there exists $C>0$, $t' >t_0$, $\theta' \in [0,1]$ and an interval $[\theta'-\delta,\theta'+\delta]$ with $\delta>0$ such that, for all $(t,\theta) \in (t_0,t']\times [\theta'-\delta,\theta'+\delta]$:

\begin{equation}
h(t,\Theta(t,\theta,t')) \ge C \alpha^{1-\epsilon}(t,\Theta(t,\theta,t')),
\end{equation}
where $(s,\Theta(s,\theta,t'))$ denote the parametrizations of the null lines parallel to the constant $v$ lines starting at $(t',\theta)$ which were introduced in section \ref{se:acac}.

\end{lemma}
\begin{proof}

Let $\epsilon >0$ and $n_0 \in \mathbb{N}$ be such that lemma \ref{lem:cah} holds and lemma \ref{lem:scbeta} holds as in (\ref{ineq:betascontrolrw}), with $t'$ replaced by $t_{n_0}$ in both lemmas. Let $n \ge n_0$. We will integrate equation (\ref{eq:g+hu}) in a way similar to the proof of the last lemma. Let us denote by $\Theta_n(t,\theta)$ the null lines parallel to the constant $v$ lines starting at $(t_n,\theta)$, i.e.~$\Theta_n(t,\theta)=\Theta(t,\theta,t_n)$. Equation (\ref{eq:g+hu}) can then be integrated as:
\begin{eqnarray}
&&(h+h^{\times})(t_n,\theta)\nonumber \\
&&-(h+h^{\times})(t,\Theta_{n}(t,\theta))\exp \int^{t_n}_t\left(-\frac{1}{s'}+\frac{\alpha_t}{\alpha} \right)(s',\Theta_{n}(s',\theta))ds'=\nonumber \\
&&\int^{t_{n}}_t \Bigg( \left( \exp{\int^{t_{n}}_{s}\left(-\frac{1}{s'}+\frac{\alpha_t}{\alpha} \right)(s',\Theta_n(s',\theta))ds'} \right) \nonumber \\ 
&& \phi(s,\Theta_n(s,\theta)) \Bigg)ds,  \label{eq:intg+hb}
\end{eqnarray} 
where $\phi=-\frac{1}{s}(P_u P_v+e^{2P}A_v A_u)-\frac{P_v}{s^4}\alpha e^{2\nu}K^2$.

Fix a $\theta_0 \in [0,1]$. Since lemma \ref{lem:scbeta} holds in the sense of (\ref{ineq:betascontrolrw}) for $t \in (t_0,t_{n}]$, we have again the following estimates:

\begin{eqnarray}
\frac{t}{t_n}\left(\frac{\alpha(t_n,\theta_0)}{\alpha(t,\theta_0)}\right)^{1+\epsilon} &\le& \exp \int^{t_n}_t \left(\frac{-1}{s}+\frac{\alpha_t}{\alpha}\right)(s,\Theta_n(s,\theta))ds
\end{eqnarray}
and
\begin{eqnarray}
 \exp \int^{t_n}_t \left(\frac{-1}{s}+\frac{\alpha_t}{\alpha}\right)(s,\Theta_n(s,\theta))ds &\le& \frac{t}{t_n}\left(\frac{\alpha(t_n,\theta_0)}{\alpha(t,\theta_0)}\right)^{1-\epsilon}. \label{es:inteaat}
\end{eqnarray}

Using this, we may estimate the last term on the right-hand side of  (\ref{eq:intg+hb}):

\begin{eqnarray}
&&\int^{t_n}_t \left( \left( \exp{\int^{t_n}_{s}\left(-\frac{1}{s'}+\frac{\alpha_t}{\alpha} \right)(s',\Theta_n(s',\theta))ds'} \right)\phi(s,\Theta_n(s,\theta)) \right)ds  \nonumber \\
&&\le \int^{t_n}_t  \frac{t}{t_n}\left( \frac{\alpha(t_n,\theta_0)}{\alpha(t,\theta_0)}\right)^{1-\epsilon}|\phi|(s,\Theta_n(s,\theta))ds,  \nonumber \\
&&\le \int^{t_n}_t  \frac{t}{t_n}\left( \frac{\alpha(t_n,\theta_0)}{\alpha(t,\theta_0)}\right)^{1-\epsilon}\left(\frac{h}{2s}+\sqrt{h+h^{\times}}\frac{e^{2\beta}K^2}{s^4}\right)ds.
\end{eqnarray}
We now use the estimates $h+h^{\times} \le 2h$, $h \le C\alpha^{1+\epsilon}$,  $\sqrt{h} \le  \sqrt{C}\sqrt{\alpha^{1+\epsilon}}$, $e^{2\beta} \le C \sqrt{\alpha}$ for some constant $C>0$ independent of $n$, as well as lemma \ref{lem:intspa} to obtain:

\begin{eqnarray}
&&\int^{t_n}_t \left( \left( \exp{\int^{t_n}_{s}\left(-\frac{1}{s'}+\frac{\alpha_t}{\alpha} \right)(s',\Theta_n(s',\theta))ds'} \right)\phi(s,\Theta_n(s,\theta)) \right)ds  \nonumber \\
&&\le \alpha(t_n,\theta_0)^{1-\epsilon}\int^{t_n}_t C\alpha^{2\epsilon} \le C'_n \alpha(t_n,\theta_0),
\end{eqnarray}
where $C'_n  \rightarrow 0$ as $n \rightarrow \infty$. 

We obtain from equation (\ref{eq:intg+hb}) that:

\begin{eqnarray}
\lefteqn{(h+h^\times)(t,\Theta_n(t,\theta)) \ge \bigg((h+h^\times)(t_n,\theta))} \nonumber \\
&&\hbox{}-C'_n(t)\alpha(t_n,\theta_0)\bigg)\exp \int^{t_n}_t\left(+\frac{1}{s'}-\frac{\alpha_t}{\alpha} \right)(s',\Theta_n(s',\theta))ds', \\
\lefteqn{(h+h^\times)(t,\Theta_n(t,\theta)) \ge  \bigg((h+h^\times)(t_n,\theta))} \nonumber \\
&&\hbox{}-C'_n(t)\alpha(t_n,\theta_0)\bigg)\frac{t_n}{t}\left(\frac{\alpha(t,\theta_0)}{\alpha(t_n,\theta_0)}\right)^{1-\epsilon}.\label{h+hcrossbelow}
\end{eqnarray}
By assumption, we have for all $n \in \mathbb{N}$, $\frac{|P_{v}|}{\sqrt{\alpha}}(t_n,\theta_n) > B$, and thus from equation (\ref{eq:h+hcrossv}), $\frac{(h+h^{\times})(t_n,\theta_n)}{\alpha(t_n,\theta_n)} \ge A$, for some $A>0$. By application of corollary \ref{cor:alphatheta}, we obtain  $\frac{(h+h^{\times})(t_n,\theta_n)}{\alpha(t_n,\theta_0)} \ge A'$, for some constant $A'>0$. Thus, for all $n \in \mathbb{N}$, there exists an interval around $\theta_n$, $[\theta_n-\delta_n,\theta_n+\delta_n]$ with $\delta_n>0$, such that for all $\theta \in  [\theta_n-\delta_n,\theta_n+\delta_n]$, 
\begin{equation} \label{ineq:h+hcrossalphanbelow}
\frac{(h+h^\times)(t_n,\theta)}{\alpha(t_n,\theta_0)} \ge \frac{A'}{2}.
\end{equation}
Let $n_1$ be such that for all $n \ge n_1$  and all $t \in (t_0,t_n]$, $C'_n \le \frac{A'}{4}$. Let $n_2= \max (n_0, n_1)$. Then, we have, from (\ref{h+hcrossbelow}) and (\ref{ineq:h+hcrossalphanbelow}), for all $(t,\theta) \in (t_0,t_{n_2}]\times [\theta_{n_2}-\delta_{n_2},\theta_{n_2}+\delta_{n_2}]$:

\begin{eqnarray}
(h+h^{\times})(t,\Theta_{n_2}(t,\theta)) \ge \frac{A'}{4}\alpha(t_{n_2},\theta_0)\frac{t_{n_2}}{t}\left(\frac{\alpha(t,\theta_0)}{\alpha(t_{n_2},\theta_0)}\right)^{1-\epsilon}.
\end{eqnarray}
Moreover, we have $\alpha(t,\Theta_{n_2}(t,\theta)) \le M \alpha(t,\theta_0)$ for some constant $M>0$, thus we obtain:

\begin{eqnarray}
(h+h^{\times})(t,\Theta_{n_2}(t,\theta)) \ge \frac{A'}{4M^{1-\epsilon}}\alpha(t_{n_2},\theta_0)\frac{t_{n_2}}{t_0}\left(\frac{\alpha(t,\Theta_{n_2}(t,\theta))}{\alpha(t_{n_2},\theta_0)}\right)^{1-\epsilon},
\end{eqnarray}
which proves the lemma.
\end{proof}
\begin{remark} \label{rm:bartlem}
With the notation of lemma \ref{lem:cfb}, it is possible to choose $t'$ so that $t' \in (t_0,\bar{t}]$, where $\bar{t}$ is such that lemma \ref{es:sqalphathetaint} holds. To see this, just replace in the above proof $n_0$ by $n_0' \ge n_0$ such that  $t_{n_0'} \in (t_0,\bar{t}]$.
\end{remark}

\subsection{The contradiction} \label{se:contrad}
From lemma \ref{lem:ubpt}, there exists a sequence of points $(t_n,\theta_n)$ and a constant $A>0$ such that $\frac{|P_{t}|}{\sqrt{\alpha}}(t_n,\theta_n) \ge A$. Thus, without of generality, we may assume that there exists a sequence of points $(t_n',\theta_n')$ and a constant $A>0$ such $\frac{|P_{v}|}{\sqrt{\alpha}}(t_n',\theta_n') \ge \frac{A}{2}$, exchanging the role of $u$ and $v$ if necessary. Therefore, lemma \ref{lem:cfb} applies and $\forall \epsilon>0$, there exists a $C>0$, a $t' >t_0$, a $\theta' \in [0,1]$ and an interval $[\theta'-\delta,\theta'+\delta]$ with $\delta >0$ such that, for all $(t,\theta) \in (t_0,t']\times [\theta'-\delta,\theta'+\delta]$:

\begin{equation}
h(t,\Theta(t,\theta,t')) \ge C \alpha^{1-\epsilon}(t,\Theta(t,\theta,t')),
\end{equation}
where $(s,\Theta(s,\theta,t'))$ denote the parametrizations of the null lines parallel to the constant $v$ lines, starting at $(t',\theta)$. Moreover, let us choose $t'$ so that $t' \in (t_0,\bar{t}]$, where $\bar{t}$ is such that lemma \ref{es:sqalphathetaint} holds, as in remark \ref{rm:bartlem}.

Consider the integral in $\theta$ of $h(t,\Theta(t,\theta,t')) $ and fix a $\theta_{0} \in [0,1]$. We have:

\begin{equation} \label{es:inthb}
\int_{[0,1]} h(t,\Theta(t,\theta,t')) d\theta \ge  2\delta C \alpha^{1-\epsilon}(t,\theta_{0}),
\end{equation}
using corollary \ref{cor:alphatheta}.
On the other hand, we have, by the change of variable $\theta'=\Theta(t,\theta,t')$,

\begin{equation}
\int_{[0,1]} h (t,\Theta(t,\theta,t')) d\theta= \int_{[0,1]}h (t,\theta') \Theta^{-1}_\theta d\theta'.
\end{equation}
From equation (\ref{eq:thetatheta}), we therefore have:

\begin{equation}
\int_{[0,1]} h(t,\Theta(t,\theta,t')) d\theta= \int_{[0,1]}h(t,\theta') \left(\exp{\int^{t}_{t'}\frac{1}{2}\left(\frac{\alpha_\theta}{\sqrt{\alpha}}\right)ds}\right) d\theta',
\end{equation}
where the integral in the exponential is taken along the characteristics.

Since lemma \ref{es:sqalphathetaint} holds, we have:

\begin{equation}
 \exp{\int^{t}_{t'}\frac{1}{2}\left(\frac{\alpha_\theta}{\sqrt{\alpha}}\right)ds} \le M \alpha^{\epsilon}.
\end{equation}
Thus, we obtain:

\begin{equation}
\int_{[0,1]} h(t,\Theta(t,\theta,t')) d\theta \le \int_{[0,1]} h M \alpha^{\epsilon} (t,\theta') d\theta'.
\end{equation}
Using again  Corollary \ref{cor:alphatheta} as well as the fact that $E_h=\int_{[0,1]} \frac{h}{\sqrt{\alpha}}d\theta$ is bounded, we see that the right-hand of the last inequality is bounded by $M' \alpha^{1/2+\epsilon}(t,\theta_0)$ for some constant $M'$. Choosing $\epsilon$ small enough, this contradicts (\ref{es:inthb}) since $\alpha \rightarrow \infty$ as $t \rightarrow t_0$. Thus theorem \ref{th:t2vac} is proved.\footnote{We see that the margin of error is, up to $\epsilon$, $\alpha^{1/2}$. This margin follows from our estimates because, up to $\alpha^{\epsilon}$, we have $h \sim \alpha$ along certain characteristics. On the other hand, if we did not have this margin, i.e.~if we had $h \sim \alpha^{1/2}$, then it would follow that for $t'$ close enough to $t_0$, $\frac{\alpha_t}{2\alpha}\pm P_t \le 0$ for all $\theta \in [0,1]$ and from equation (\ref{ee:tdea}), that $E_A$ is increasing the past. This would contradict the fact that $E_A \rightarrow 0$ as $t \rightarrow t_0$.}

\section{Proof of Theorem \ref{th:hyvla}} \label{se:hypvla}
We will prove Theorem \ref{th:hyvla} in this section. For this, we will adapt the proof found in \cite{iw:ast2} to the case of $k=-1$ surface-symmetric spacetimes. To exploit the methods of \cite{iw:ast2}, we have rewritten the metric in a form similar to the $T^2$ case (see (\ref{hy:metric}) in section \ref{se:shss}). In particular,  the coordinate $t$ used in (\ref{hy:metric}) denotes the square of the usual areal time used for these spacetimes, as found for instance in \cite{sbt:isss}.

We start by recalling the Einstein-Vlasov system for spacetimes with a hyperbolic surface of symmetry.

\subsection{Vlasov matter in $k=-1$ surface-symmetric spacetimes}
Let $(\mathcal{M},g,f)$ be a past development of $k=-1$ surface-symmetric initial data with Vlasov matter as described in section \ref{se:cid} and assume that $(t,\theta,x,y)$ is a system of areal coordinates such that the metric in $\mathcal{M}$ takes the form (\ref{hy:metric}). Let $v_i$, $i=0,1,2,3$ denotes the components of the velocity vector in the canonical basis of $1$-forms associated with the coordinate system $(t,\theta,x,y)$. We will parametrize the mass shell $\mathcal{P}$ by the coordinates $(t,\theta,x,y,v_1,v_2,v_3)$, where by an abuse of notation, we denote the lift to $\mathcal{P}$ of the coordinates on $\mathcal{M}$ by the same symbols.
The Vlasov field $f$ can be seen as a function of $(t,\theta,x,y,v_1,v_2,v_3)$ or, using the symmetry, as a function depending only on $t$, $\theta$, $w=\frac{\sqrt{t}}{e^{\nu}} v_1$ and $L=\gamma^{ab}v_a v_b$, and we will, by an abuse of notation, use both definitions and always denote it by $f$ \footnote{Note that indices on the velocities $v_i$ are raised or lowered using the metric (\ref{hy:metric}), not using $\gamma_{ab}$. This implies that if $p^a$ denotes the canonical momentum associated with the coordinates system $(t.\theta.x.y)$, then $L=t^2 \gamma_{ab} p^a p^b$}.

With these definitions, the mass shell relation $v_\mu v^\mu=-1$ is given by:

\begin{equation}
v_0=-\sqrt{\frac{\alpha}{t}e^{2\nu}+\alpha v_1^2+\frac{\alpha e^{2\nu}}{t^2} \gamma^{ab}v_a v_b}=-\frac{\sqrt{\alpha}e^{\nu}}{\sqrt{t}}\sqrt{1+w^2+\frac{L}{t}}
\end{equation}
and the Vlasov equation for $f(t,\theta,w)$ reads as:
\begin{eqnarray} \label{eq:vlasov}
\lefteqn{2\sqrt{t} \partial_{t}f + \frac{2\sqrt{t \alpha}w}{\sqrt{1+w^{2}+L/t}} 
\partial_{\theta}f - \bigg(\sqrt{t}(2\nu_t-1/t) w }\nonumber \\ 
&&\hbox{}+(\nu_\theta+\frac{\alpha_\theta}{2 \alpha})2\sqrt{t \alpha}\sqrt{1+w^{2}+L/t} \bigg)\partial_{w}f = 0.
\end{eqnarray}

\subsection{The Einstein equations}
The Einstein equations (\ref{ee}) reduce to the following system of equations:

Constraint equations:
\begin{eqnarray}
\nu_t &=& \frac{1}{4t}+\alpha e^{2\nu}\Lambda-\frac{k\alpha e^{2\nu}}{t} \nonumber \\
&&\hbox{}+8 \pi \sqrt{\alpha} \intr{3}f|v_0| \sqrt{\gamma^{-1}}dv_1 dv_2 dv_3, \label{ee:nut}\\
\frac{\alpha_t}{\alpha} &=& -4 \Lambda \alpha e^{2\nu}+\frac{4k\alpha e^{2\nu}}{t} \label{ee:alphat} \nonumber \\ 
&&\hbox{}-16 \pi \alpha^{3/2} e^{2\nu} \intr{3} \frac{ f \left(\frac{1}{t}+\frac{L}{t^2} \right)}{|v_0|}\sqrt{\gamma^{-1}}dv_1 dv_2 dv_3,  \\
\nu_{\theta}+\frac{1}{2}\frac{\alpha_\theta}{\alpha}&=&-8 \pi \sqrt{\alpha} \intr{3} f v_1 \sqrt{\gamma^{-1}}dv_1 dv_2 dv_3. \label{ee:nutheta}
\end{eqnarray}

Evolution equation:

\begin{eqnarray}
\nu_{tt}-\alpha \nu_{\theta \theta}&=&\frac{1}{2}\alpha_{\theta \theta}-\frac{1}{4}\frac{\alpha_\theta^2}{\alpha}+\frac{\nu_\theta \alpha_\theta}{2}-\frac{1}{4t^2}+\frac{\alpha_t \nu_t}{2 \alpha}+\frac{\alpha e^{2\nu} \Lambda}{t} \label{ee:evol} \nonumber \\
&&\hbox{}-4 \pi \frac{\alpha^{3/2} e^{2\nu}}{t^3} \intr{3} \frac{f L}{|v_0|}\sqrt{\gamma^{-1}}dv_1 dv_2 dv_3.
\end{eqnarray}
Here $k$ denotes the curvature of the surface of symmetry and will therefore be $-1$ in the case of hyperbolic symmetry. $\gamma$ denotes the determinant of the metric $\gamma_{ab}$.

In the rest of this section, $(\mathcal{M},g,f)$ will be a past development of $k=-1$ surface-symmetric initial data with Vlasov matter and $\Lambda \ge 0$. We will cover $(\mathcal{M},g)$ by areal coordinates $(t,\theta,x,y)$, where the range of the coordinates $(t,\theta)$ is $(t_f,t_i] \times [0,1]$ with $0<t_f<t_i$. The metric will be given by (\ref{hy:metric}) with the functions $\alpha$ and $\nu$ depending only on $(t,\theta)$ and being periodic in $\theta$ with period $1$. The Einstein-Vlasov system implies that the system (\ref{ee:nut})-(\ref{ee:evol}) completed with (\ref{eq:vlasov}) holds for all $(t,\theta) \in (t_f,t_i] \times [0,1]$. Moreover, we will assume that $f$ does not vanish identically. Following what has been said in section \ref{se:rsp}, we will prove that for all such $(\mathcal{M},g,f)$, the hypotheses of Proposition \ref{pro:cc} are satisfied, from which Theorem \ref{th:hyvla} follows immediately.

First, we recall some properties of the Vlasov field for such spacetimes.

\subsection{Conservation laws}
As in section \ref{se:clt2v}, since $f$ is conserved along geodesics, we have an immediate upper bound on $f$:

\begin{equation}
f \le F,
\end{equation}
for some $F>0$.
Since by assumption, $f$ has compact support, conservation of angular momentum along geodesics implies an upper bound on the support of $f$ in $L$, i.e.~we have:

\begin{equation}
X= \sup_{L \in \mathrm{supp}(f)} L < \infty.
\end{equation}

The particle current is given by:

\begin{equation}
N^\mu=\frac{\sqrt{\alpha}}{t}\int_{\mathbb{R}^3} \frac{f}{|v_0|}v^\mu \sqrt{\gamma^{-1}}dv_1 dv_2 dv_3.
\end{equation}

From the Vlasov equation it follows that $N^\mu$ is divergence free $\nabla_\mu N^\mu=0$ and therefore, we have the conservation law,  $\forall t$, 

\begin{equation} \label{eq:clv}
\int_{[0,1]} N^0 \sqrt{\alpha} e^{2\nu} d\theta  =\int_{[0,1]}\left( \int_{\mathbb{R}^3} f\sqrt{\gamma^{-1}}dv_1 dv_2 dv_3\right) d\theta=Q,
\end{equation}
for some non-negative constant $Q$. Moreover, since by assumption, the Vlasov field does not vanish identically, we have:

\begin{equation}
 Q > 0.
\end{equation}

\subsection{Lower bound on the mean value of $|v_1|$.}

Similarly to section \ref{se:lbav1}, we have:
\begin{lemma} \label{lem:lbv}
There exists $\delta > 0$ such that, for all $t$:

\begin{equation}
\int_{[0,1]} \left(\int_{\mathbb{R}^3} f\sqrt{\gamma^{-1}} |v_1| dv_1 dv_2 dv_3\right) d\theta> \delta.
\end{equation}
\end{lemma}
\begin{proof}
The proof of lemma \ref{lem:lbv_vla} is easily adapted to this setting.
\end{proof}

\subsection{Energy estimates}
We define $E(t)$ as the following energy integral:

\begin{equation}
E(t)=\int_{[0,1]} \frac{\nu_t}{t\sqrt{\alpha}}d\theta.
\end{equation}

We have:
\begin{lemma}$E$ admits a continuous extension to $t_f$. In particular $E$ is uniformly bounded on $(t_f,t_i]$.
\end{lemma}
\begin{proof}
As usual, we take the time derivative of $E$ and use the Einstein equations and the periodicity to simplify the resulting equations.
It follows that:

\begin{eqnarray}
\frac{dE}{dt}&=&-\int_{[0,1]} \Bigg(\frac{1}{2t^3 \sqrt{\alpha}} -\frac{k\sqrt{\alpha}e^{2\nu}}{t^3} \nonumber \\
&&+8\pi \intr{3} \bigg(\frac{f|v_0|}{t^2} +\frac{\alpha e^{2\nu}f L }{2t^4|v_0|}\bigg)\sqrt{\gamma^{-1}}dv_1dv_2dv_3 \Bigg) d\theta.
\end{eqnarray}
Since $k=-1$, we see that $E$ is increasing with decreasing $t$. 
Moreover from the last equation, the definition of $E$ and equation (\ref{ee:nut}), it follows that:

\begin{equation}
\frac{dE}{dt} \ge -\frac{4E}{t}
\end{equation}
and by integration of the last line, we obtain an upper bound for $E$ on $(t_f,t_i]$.
\end{proof}
\subsection{Estimate for $\sqrt{\alpha}e^{2\nu}$}
We have:
\begin{lemma} \label{lem:bveh} $\sqrt{\alpha} e^{2\nu}$ is uniformly bounded on $(t_f,t_i]$.
\end{lemma}
Proof: It follows from equations  (\ref{ee:nut}) and (\ref{ee:alphat}) that:
\begin{equation}
\partial_t ( \sqrt{\alpha} e^{2\nu}) \ge 0.
\end{equation}
We will use this bound in order to estimate the terms containing $\alpha e^{2 \nu}$ in the right-hand side of equation (\ref{ee:alphat}). This will follow from the next two lemmas.

\subsection{Estimate for $\int_{[0,1]}|\left(\sqrt{\alpha}e^{\nu} \right)_\theta|d\theta$}
Let $e^\beta=\sqrt{\alpha}e^{\nu}$. Equation (\ref{ee:nutheta}) can now be written as:

\begin{equation} \label{ee:betatheta}
\beta_\theta=-8 \pi \sqrt{\alpha} \intr{3} f v_1  \sqrt{\gamma^{-1}}dv_1 dv_2 dv_3.
\end{equation}

\begin{lemma} $\int_{[0,1]} |\beta_\theta| d\theta$ is bounded on $(t_f,t_i]$. In particular, there exists a bound independent of $t \in (t_f,t_i]$ on the difference between the maximum and the minimum of $\beta(t,.)$. \label{lem:betatheta}
\end{lemma}

\begin{proof}
From equation (\ref{ee:betatheta}), we have:

\begin{eqnarray}
|\beta_\theta | &\le& 8 \pi \sqrt{\alpha} \intr{3} f v_1 \sqrt{\gamma^{-1}}dv_1 dv_2 dv_3 \nonumber, \\
 &\le& 8 \pi \intr{3} f v_0\sqrt{\gamma^{-1}} dv_1 dv_2 dv_3, \nonumber \\\
&\le& \frac{\nu_t}{\sqrt{\alpha}},
\end{eqnarray}
where we have used the fact that $\sqrt{\alpha} |v_1| \le v_0$ from the mass shell relation to obtain the second line and equation (\ref{ee:nut}) to obtain the last line.

Dividing the last equation by $t$ and integrating over $[0,1]$ the last line, we obtain a bound on $\int_{[0,1]} |\beta_\theta| d\theta$ from the bounds on $t$ and $E$.
\end{proof}
\subsection{Control of $\alpha$ along special curves}
Similar to section \ref{sec:basc}, we now prove:
\begin{lemma} $\min_{S^1} \alpha(t,.)$ is bounded on $(t_f,t_i]$. \label{lem:bma}
\end{lemma}

\begin{proof}From the definition of $E$ and equation (\ref{ee:nut}), 

\begin{equation}
8\pi \int_{[0,1]} \intr{3} f \sqrt{\gamma^{-1}}|v_0| dv_1 dv_2 dv_3 d\theta \le t E(t).
\end{equation}
Since  $\sqrt{\alpha} |v_1| \le v_0$, we obtain:
\begin{equation}
\min_{[0,1]} (\sqrt{\alpha}) \int_{[0,1]} \intr{3} f|v_1| \sqrt{\gamma^{-1}}dv_1 dv_2 dv_3 d\theta \le \frac{t E(t)}{8\pi} \le A,
\end{equation}
for some constant $A$ depending on the bound on $E$.
However from lemma \ref{lem:lbv}, we have  $\delta \le \int_{[0,1]} \intr{3} f|v_1| \sqrt{\gamma}^{-1}dv_1 dv_2 dv_3$, for some $\delta > 0$. Therefore:

\begin{equation}
\min_{[0,1]} (\sqrt{\alpha}) \le A/\delta.
\end{equation}
\end{proof}

As in Corollary \ref{cor:controlalpha} of section \ref{sec:basc}, we obtain the following:
\begin{corollary} There exists $\bar{\theta}$ such that $\alpha(t,\bar{\theta})$ is bounded on $(t_p,t_i]$. \label{cor:bbt}
\end{corollary}

\subsection{Estimate on $e^{2 \beta}$}

\begin{lemma} $e^{2 \beta}=\alpha e^{2\nu}$ is uniformly bounded on $(t_f,t_i] \times [0,1]$.
\end{lemma}
\begin{proof}
This follows from corollary \ref{cor:bbt} and lemmas \ref{lem:bveh} and \ref{lem:betatheta} by an argument similar to the one given for the proof of lemma \ref{lem:seccontrol}.
\end{proof}

\subsection{Estimates for the support of $f$}
Let 
\begin{equation} \label{tr:u1w}
u_1=\sqrt{\alpha}v_1=\frac{\sqrt{\alpha}e^{\nu}}{\sqrt{t}}w
\end{equation}
 and define $\bar{u}_1$ by:  

\begin{equation}
\bar{u}_1(t)=\sup\left\{ |u_1| / \exists (\theta,L) / f\left(t,\theta,\frac{u_1}{\sqrt{\alpha}},L\right) \neq 0\right\}
\end{equation}

We have the following:
\begin{lemma}
$\bar{u}_1$ is uniformly bounded on $(t_f,t_i]$.
\end{lemma}
\begin{proof}

The characteristic equation for $u_1$ associated with the Vlasov equation written (\ref{eq:vlasov}) in terms of the coordinates $(t, \theta, u_1, L)$ gives:

\begin{equation} \label{eq:u1char}
\frac{d(u_1^2)}{ds}=\frac{\alpha_t}{\alpha}u_1^2+\frac{2\sqrt{\alpha}u_1}{v_0}\frac{e^{2\beta}}{t}\beta_\theta(1+L/t).
\end{equation}
The transformation (\ref{tr:u1w}) from $w$ to $u_1$ will avoid the difficulty arising from the term containing $\beta_\theta$ in equation (\ref{eq:vlasov}). Indeed, this term contains the factor $\sqrt{1+w^2+\frac{L}{t}}$ which depends in $w$ in a not completely trivial way. On the other hand, having $v_0$ at the denominator of the last term in the right-hand side of equation (\ref{eq:u1char}) will enable us to easily estimate this term.

Let us first estimate the factor $\frac{\alpha_t}{\alpha}$ appearing in the first term of the right-hand side of (\ref{eq:u1char}).
From equation (\ref{ee:alphat}) and the bounds on $e^\beta$ obtained previously, we have:

\begin{eqnarray}
\left| \frac{\alpha_t}{\alpha}\right| &\le& C+ 16\pi \alpha^{3/2}e^{2\nu} \intr{3} \frac{f\left( 1/t+L/t^2\right)}{|v_0|}\sqrt{\gamma^{-1}}dv_1dv_2dv_3, \nonumber \\
 &\le& C+ C' \sqrt{\alpha} \int^{\bar{u_1}}_{-\bar{u_1}} \int^{X}_{-X} \frac{f\left( 1/t+L/t^2\right)}{|v_0|} \frac{du_1}{\sqrt{\alpha}}\pi dL, \nonumber \\
&\le& C+ C''F\left(\frac{1}{t_f}+\frac{X}{t_f^2}\right) \int^{\bar{u_1}}_{-\bar{u_1}} \frac{du_1}{|v_0|}, \nonumber \\
&\le& C+A\int^{\bar{u}_1}_{-\bar{u}_1} \frac{du_1}{\sqrt{1+te^{-2\beta}u_1^2}},  \nonumber \\
&\le& C+A \left[ e^{\beta}t^{-1/2} \ln \left( u_1+\sqrt{e^{2\beta}/t+u_1^2}\right) \right]_{-\bar{u}_1}^{\bar{u}_1}, \nonumber \\
&\le& C+A' \left(\ln \left( \bar{u}_1+\sqrt{e^{2\beta}/t+\bar{u}_1^2}\right)+e^{-1}\right), \label{es:alphatalphav}
\end{eqnarray}
for some constants $C$ and $A'$.

We turn now to estimate the second term on the right-hand side of (\ref{eq:u1char}).
First note that the mass shell relation written in terms of $u_1$ reads as:

\begin{equation}
v_0=-\sqrt{\frac{\alpha}{t}e^{2\nu}+u_1^2+\frac{\alpha e^{2\nu}}{t^2} \gamma^{ab}v_a v_b}
\end{equation}
and thus, we have that $\frac{|u_1|}{|v_0|} < 1$.
Moreover, from equation (\ref{ee:betatheta}), we have that:

\begin{equation} \label{es:betahypv}
\sqrt{\alpha}\beta_\theta \le 8\pi^2 F X \bar{u}_1^2.
\end{equation}

Integrating (\ref{eq:u1char}) and using the estimates (\ref{es:betahypv}) and (\ref{es:alphatalphav}), we obtain an inequality of the form:

\begin{equation}
u_1^2(t) \le A +B\int_{t}^{t_i} u_1^2(s) \ln(1+\bar{u}_1^2)(s) ds+C\int_{t_i}^t \bar{u}_1^2(s) ds,
\end{equation}
for some positive constants $A$, $B$ and $C$. It follows from the last line, as in (\ref{es:flnf})-(\ref{es:flnfi}) that $\bar{u}_1$ is uniformly bounded.

\end{proof}
\subsection{Estimates for $\alpha$, $\beta$, $\nu$ and $\beta_\theta$}
\begin{lemma}
$\alpha$, $\beta$, $\nu$ and $\beta_\theta$ are uniformly bounded on $(t_f,t_i] \times [0,1]$.
\end{lemma}
\begin{proof}This follows easily from the Einstein equations since the right-hand sides of equations (\ref{ee:nut}), (\ref{ee:alphat}) and (\ref{ee:nutheta}) contain only quantities that have been shown to be bounded.
\end{proof}

\subsection{Estimates for the derivatives of $f$, $\alpha_\theta$, $\nu_\theta$ and higher order estimates}
This follows by standard arguments, which can be found for instance in \cite{mw:ast2v}.

\subsection{The conclusion}
Since all metric functions, the Vlasov field and all their derivatives have been shown to be uniformly bounded, the assumptions of Proposition \ref{pro:cc} have been retrieved. In particular, the maximal Cauchy development cannot have $t_0>0$, which concludes the proof of Theorem \ref{th:hyvla}.

\section{Comments and open questions} \label{se:coq}

\subsection{Weaver's estimate for Vlasov matter} 
The result of Theorem \ref{th:hyvla} was obtained in \cite{gr:csves, sbt:isss} under a small data assumption. The main difference in our analysis which enables us to remove this smallness assumption, is to use, following \cite{mw:ast2v}, the presence of the Vlasov field to obtain a lower bound on one of the matter terms (see lemma \ref{lem:lbv}). It would be interesting to see if this estimate could be applied in other geometries and what would be the consequences. 

Let us also note that if we couple the Einstein-Vlasov system to extra matter fields, a statement analogous to Lemma \ref{lem:lbv} would certainly be true if the extra matter fields satisfy the strong energy condition. For instance, the results of Theorems \ref{th:t2vla} and \ref{th:hyvla} can certainly be extended to include a massless scalar field.

\subsection{Theorem \ref{th:t2vac} and the hierachisation of the equations}
The proof of Theorem \ref{th:t2vac} is based on the recovery of the lower bound on the energy quantities $E_h$ and $E_g$. In the vacuum case, this lower bound is obtained directly from the monotonicity of $E_g$. However, this monotonicity is unstable to any perturbation in the setting of the problem, such as the introduction of matter or of a positive cosmological constant.

Our strategy has been to prove that, while $E_h$ is not necessarily monotone, one can recover a monotonocity for another energy, namely $E_A$, which controls $E_h$ from below and thus is sufficient to obtain the required lower bound on $E_h$. Since $E_A$ is the energy associated with the wave equation for $A$ only, while $E_h$ is associated for the system of equations for $(U,A)$, this shows that, in the contradiction setting that we have deployed, a certain hierarchy in the evolution equations appears, in the sense that one may first focus on the evolution equation for $A$ and extract information from it, which we then reintroduce in the whole system. 

Let us also note that not all estimates derived during the proof of Theorem \ref{th:t2vac} require the contradiction setting of section \ref{se:contrad}. In particular, in section \ref{se:lnalphatheta}, we have proven a new estimate for $T^2$-symmetric spacetimes which might be useful in a further study of these solutions.

\subsection{Antitrapped initial data}
One of the common features of $T^2$-symmetric and $k=-1$ surface-symmetric spacetimes is the antitrapping of the orbits of symmetry. This property arises from the positivity of the Hawking mass (exluding the flat case) and the fact that the orbits of symmetry have non-positive curvature. The positivity of the Hawking mass is itself a consequence of the topology of Cauchy surfaces and of the Einstein equations, especially the Raychaudhuri equations. The proofs of the positivity of the Hawking mass and of antitrapping for vacuum $T^2$ symmetry and for $k \le 0$ surface-symmetric spacetimes with Vlasov matter or with a massless scalar field were first obtained  in \cite{pc:u1u1} and \cite{ar:cssph}. In \cite{ar:cmc2ds}, the results on $T^2$-symmetry were extended to the non-vacuum cases where local $T^2$-symmetry only is assumed. In order to improve our understanding of the structure of cosmological singularities, it would be interesting to try to generalize these results. 
One might ask for instance the following question. Assume that $\Sigma$ is a compact Cauchy surface of a given spacetime satisfying the vacuum Einstein equations such that there exist a diffeomorphism $\phi$ between $\Sigma$ and $S^1 \times \mathcal{R}$ where $\mathcal{R}$ is a compact surface. 
Assume moreover that for every point $\theta \in S^1$,  $\phi^{-1}\left(\{\theta\}\times \mathcal{R}\right)$ has non-positive curvature. Is it then true that $\phi^{-1}\left(\{\theta\}\times \mathcal{R}\right)$ is necessarily trapped or antitrapped?

\subsection{Strong cosmic censorship in polarized $T^2$-symmetric spacetimes}
Theorems \ref{th:t2vla}, \ref{th:t2vac} and \ref{th:hyvla} complete our understanding of the value of $t_0$ for $T^2$-symmetric and surface-symmetric spacetimes, as can be observed in Table 2, and we should therefore focus our attention to the remaining, very difficult, open problems presented in Table 1. One of the first questions to consider is that of strong cosmic censorship for vacuum polarized $T^2$-symmetric spacetimes with $\Lambda=0$. While it is likely that the dynamics of these spacetimes are very different from those of general vacuum $T^2$-symmetric spacetimes, they are the simplest examples of vacuum inhomogeneous cosmological models where, writing the Einstein equations in areal coordinates, the constraint equations do not decouple from the evolution equations, as can be seen by removing the terms involving $A$ in (\ref{ee:nut2})-(\ref{ee:aev}).

\subsection{Future causal geodesic completeness of $T^2$-sym\-metric and $k=-1$ surface-symmetric spacetimes}
By the arguments of \cite{dr:iecs}, (non-flat) $T^2$-symmetric and $k=-1$ sur\-face-\-sym\-metric spacetimes are future inextendible. In the Gowdy case, where a complete understanding of the asymptotics has been obtained \cite{hr:wmegr}, and in the $k=-1$ surface-symmetric case with either small data \cite{gr:fgcvh} or $\Lambda>0$ \cite{tr:geabevpcc}, future geodesic completeness has also been proven. More generally, we have the following conjecture:

\begin{conjecture}
Let $(\mathcal{M},g)$ be the maximal Cauchy develop\-ment of $T^2$-sym\-metric or $k=-1$ surface-symmetric initial data in the vacuum or with Vlasov matter and with $\Lambda \ge 0$. Assume $(\mathcal{M},g)$ is non-flat. Denote by $t$ the area of the orbits of symmetry and orient $(\mathcal{M},g)$ by $\nabla t$.  Then $(\mathcal{M},g)$ is future causally complete.
\end{conjecture}

\subsection{The past boundary of $\tilde{\mathcal{Q}}$}
One might also consider the following question about the structure of singularities in $T^2$-symmetric or $k=-1$ surface-symmetric spacetimes. Let $\tilde{\mathcal{Q}}$ be the universal cover of the quotient by the group orbits of the maximal Cauchy development. It is possible to draw a Penrose diagram of $\tilde{\mathcal{Q}}$, by introducing bounded double null coordinates on $\tilde{\mathcal{Q}}$ and then regards $\tilde{\mathcal{Q}}$ as a bounded subset of $\mathbb{R}^{1+1}$. In the case of  vacuum non-flat $T^3$-Gowdy initial data with $\Lambda=0$, it is then a well known fact that  its past boundary is spacelike with respect to the causality of $\mathbb{R}^{1+1}$ and thus the Penrose diagram takes the following form:

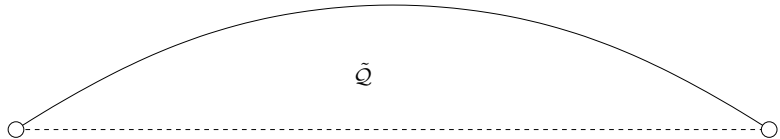
\[ \begin{picture}(0,0)%
\includegraphics{penrose_diagram.pstex}%
\end{picture}%
\setlength{\unitlength}{2565sp}%
\begingroup\makeatletter\ifx\SetFigFont\undefined%
\gdef\SetFigFont#1#2#3#4#5{%
  \reset@font\fontsize{#1}{#2pt}%
  \fontfamily{#3}\fontseries{#4}\fontshape{#5}%
  \selectfont}%
\fi\endgroup%
\begin{picture}(7366,1295)(2318,-1644)
\put(5626,-1111){\makebox(0,0)[lb]{\smash{{\SetFigFont{8}{9.6}{\rmdefault}{\mddefault}{\updefault}{\color[rgb]{0,0,0}$\tilde{\mathcal{Q}}$}%
}}}}
\end{picture}%
 \]
On the other hand, for the non-generic vacuum $T^2$-symmetric spacetimes\footnote{See the appendix of \cite{js:scct2pccm} for instance.} with $t_0>0$, the past boundary is null:
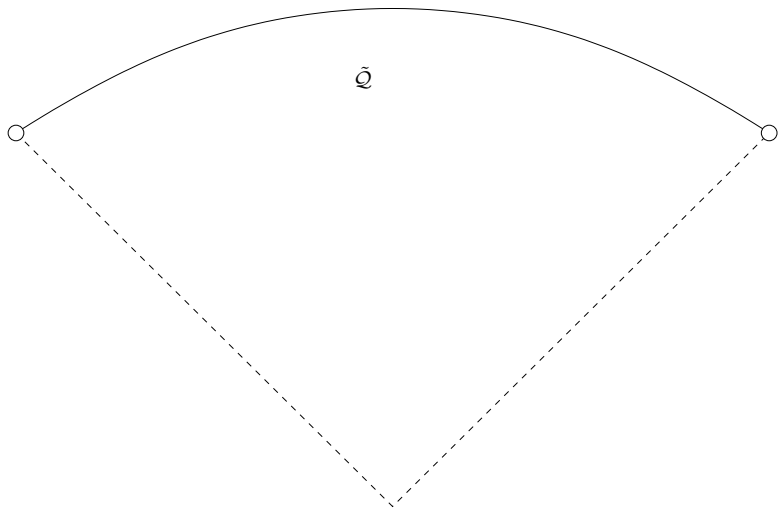
\[ \begin{picture}(0,0)%
\includegraphics{penrose_diagram2.pstex}%
\end{picture}%
\setlength{\unitlength}{2565sp}%
\begingroup\makeatletter\ifx\SetFigFont\undefined%
\gdef\SetFigFont#1#2#3#4#5{%
  \reset@font\fontsize{#1}{#2pt}%
  \fontfamily{#3}\fontseries{#4}\fontshape{#5}%
  \selectfont}%
\fi\endgroup%
\begin{picture}(7366,4824)(2318,-5173)
\put(5626,-1111){\makebox(0,0)[lb]{\smash{{\SetFigFont{8}{9.6}{\rmdefault}{\mddefault}{\updefault}{\color[rgb]{0,0,0}$\tilde{\mathcal{Q}}$}%
}}}}
\end{picture}%
 \]
It is natural to ask where the general case stands compared to these two particular cases, whether the past boundary is spacelike, null or neither spacelike nor null.

\section{Acknowledgements}
I would like to thank Mihalis Dafermos for his consistent support, his encouragements, his precious help and for introducing me to the problems addressed in this article. I would also like to thank John Stewart for much useful advice and Gustav Holzegel for many interesting discussions. The work necessary to prove Theorems \ref{th:t2vla} and \ref{th:hyvla} was conducted during my stay in MIT from October to December 2007 and I wish to thank the Department of Mathematics of MIT and the NSF for the help and funding I received during my stay. Finally, I wish to gratefully acknowledge funding from EPSRC.

\bibliography{bibliog}
\appendix
\section[Initial data and the constraint equations]{Initial data and constraint equations for the Einstein and Einstein-Vlasov systems} \label{ap:idev}
We present below the constraint equations of the Einstein-Vlasov system. To obtain the constraint equations in the vacuum case, it suffices to replace all matter terms (i.e.~all terms containing $\hat{f}$) by zero.

Recall that a smooth initial data set for the Einstein-Vlasov system is a quadruplet $(\Sigma,h,K,\hat{f})$ such that:
\begin{enumerate}
\item{$\Sigma$ is a smooth $3$-dimensional manifold,}
\item{$h$ is a smooth Riemannian metric on $\Sigma$,}
\item{$K$ is a smooth symmetric $2$-tensor on $\Sigma$,}
\item{$\hat{f}$ is a smooth function defined on the tangent bundle of $\Sigma$,}
\item{ $(\Sigma,h,K,\hat{f})$ satisfies the constraint equations:
\begin{eqnarray}
R^{(3)}-K_{ab}K^{ab}+(tr K)^2&=&16 \pi \rho+2\Lambda, \\
\nabla^{(3)}_a K_b^{\phantom{b}a}-\nabla^{(3)}_b (tr K)&=&8\pi j_b,
\end{eqnarray}
where $\nabla^{(3)}$ and  $R^{(3)}$ denote the Levi-Civita and the Ricci curvature scalar of $h$ and $\rho$ and $j_b$ are given by:

\begin{eqnarray}
\rho &=& \int_{\mathbb{R}^3} \hat{f} {(1+p^ap_a)}^{1/2}\sqrt{h}\hspace{1 pt}dp^1 dp^2 dp^3, \\
j_a &=& \int_{\mathbb{R}^3} \hat{f} p_a\sqrt{h}\hspace{1 pt}dp^1 dp^2 dp^3,
\end{eqnarray}
where it has been assumed in the above definitions that, if $\pi_{\Sigma}$ denotes the natural projection from $T\Sigma$ to $\Sigma$, then $(p^1,p^2,p^3)$ are global coordinates on $\pi_{\Sigma}^{-1}(x)$ for any $x \in \Sigma$.
}
\end{enumerate}
\section{Surface-symmetric spacetimes in areal coordinates} \label{ap:ssac}
We present in this appendix a change of coordinates and paramatrization of the metric which brings the metric (\ref{hy:metric}) from the usual parametrization:

\begin{equation} \label{def:im}
ds^2=-e^{2\mu(r,\theta)}dr^2+e^{2\lambda(r,\theta)}d\theta^2+r^2 \gamma_{ab}dx^a dx^b.
\end{equation}

We define the new time coordinate by $t=r^2$. The metric now takes the form:

\begin{equation}
ds^2=-\frac{e^{2\mu}}{4t}dt^2+e^{2\lambda}d\theta^2+t \gamma_{ab}dx^a dx^b.
\end{equation}
We can then define the functions $\alpha$ and $\nu$ by:

\begin{eqnarray}
e^{2\lambda} &=& \frac{e^{2\nu}}{t}, \\
e^{2\mu} &=& 4 \alpha e^{2\nu}.
\end{eqnarray}
in order to obtain the metric in the form (\ref{hy:metric}).

\section{From symmetric initial data to sym\-me\-tric\\ space\-times} \label{ap:sid}
We recall in this section that the symmetries of initial data are transmitted to the maximal Cauchy development. For the proofs in the vacuum case and a more exhaustive treatment of these questions, we refer the reader to the classical work of Chru\'sciel \cite{pc:ulsee}. We will write the theorems in the vacuum case for simplicity.

First, we recall that Killing data leads to Killing vector fields:

\begin{proposition}
Let $(\Sigma,h,K)$ be a vacuum initial data set for the Einstein equations. Assume that there exists a smooth vector field $Y$ such that:

\begin{equation}
\mathcal{L}_Y h = \mathcal{L}_Y K=0
\end{equation}
Let $(\mathcal{M},g)$ denote the maximal Cauchy development of $(\Sigma,h,K)$ as in the statement of the theorem of section \ref{se:mcd}  and let $\phi:\Sigma \rightarrow \mathcal{M}$ be the corresponding embedding. Then there exists a smooth vector field $X$ on $\mathcal{M}$ such that:

\begin{equation}
\mathcal{L}_X g=0, \quad X_{|\phi(\Sigma)}=\phi_*(Y)
\end{equation}
\end{proposition}

We have moreover the following:

\begin{proposition}
Let $(\Sigma, h, K)$ be a vacuum initial data set for the Einstein equations. Assume moreover that there exists a topological group $G$ acting smoothly by isometry on $(\Sigma, h, K)$ i.e.~a map $\phi$ such that:

\begin{eqnarray}
\phi: G \times \Sigma &\rightarrow& \Sigma \nonumber \\
(q,p) &\rightarrow&  \phi_q(p) \nonumber \\
\phi_g^*h=h,&& \phi_g^* K=K
\end{eqnarray}
Let $(\mathcal{M},g)$ denote the maximal Cauchy development of $(\Sigma, h, K)$ as in the statement of the theorem of section \ref{se:mcd} and let $i$ be the corresponding embedding of $\Sigma$ in $\mathcal{M}$. Then, there exists an action $\psi$ of $G$ on $\mathcal{M}$:

\begin{eqnarray}
\psi:  G \times \mathcal{M} &\rightarrow& \mathcal{M} \nonumber \\
(q,p) &\rightarrow& \psi_q(p)
\end{eqnarray}
such that, for all $q \in G$:

\begin{eqnarray}
\psi_q^*g=g, \quad \psi_q \circ i = i \circ \phi_g
\end{eqnarray}
\end{proposition}

\end{document}